\documentclass[twocolumn]{aastex63}
\usepackage{threeparttable}
\usepackage{graphicx}
\usepackage{ulem}
\hypersetup{linkcolor=cyan,citecolor=cyan,filecolor=cyan,urlcolor=cyan}

\def\nit{\hbox{N$_2$}}
\def\wat{\hbox{H$_2$O}}
\def\met{\hbox{CH$_4$}}
\def\nol{\hbox{CH$_3$OH}}
\def\ammon{\hbox{NH$_3$}}
\def\diox{\hbox{CO$_2$}}

\def\micron{\hbox{\,$\mu$m}}
\def\mum{\hbox{\,$\mu$m}}
\def\etal.{\hbox{\textit{et al.}}}

\def\ie{\hbox{\textit{i.e.}}}

\submitjournal{PSJ}

\shorttitle{Compositional study of TNOs}
\shortauthors{Fern\'andez-Valenzuela et al.}
\graphicspath{{./}{figures/}}

\begin{document}

\title{Compositional study of trans-Neptunian objects at $\lambda>2.2\,\mu$m\footnote{Accepted for publication on October 13, 2020. }}

\correspondingauthor{E. Fern\'andez-Valenzuela}
\email{estela@ucf.edu}

\author[0000-0003-2132-7769]{E. Fern\'andez-Valenzuela}
\affiliation{Florida Space Institute, University of Central Florida, 12354 Research Parkway, Partnership 1, Orlando, FL, USA}

\author[0000-0002-2770-7896]{\color{black} N. Pinilla-Alonso}
\affiliation{Florida Space Institute, University of Central Florida, 12354 Research Parkway, Partnership 1, Orlando, FL, USA}

\author{J. Stansberry}
\affiliation{Space Telescope Science Institute, Baltimore, MD, USA}

\author{J. P. Emery}
\affiliation{Department of Astronomy and Planetary Sciences, Northern Arizona University, Flagstaff, AZ, USA}

\author{W. Perkins}
\affiliation{Department of Earth and Planetary Sciences, University of Tennessee, Knoxville, TN, USA}

\author{C. Van Laerhoven}
\affiliation{University of British Columbia, Department of Physics and Astronomy, 6224 Agricultural Road, Vancouver, BC, Canada}

\author{B. J. Gladman}
\affiliation{University of British Columbia, Department of Physics and Astronomy, 6224 Agricultural Road, Vancouver, BC, Canada}

\author{W. Fraser}
\affiliation{NRC Herzberg Astrophysics, BC, Canada}

\author{D. Cruikshank}
\affiliation{NASA Ames Research Center, Moffett Field, CA, USA}

\author{E. Lellouch}
\affiliation{LESIA, Observatoire de Paris, Meudon, France}

\author{T. G. M\"uller}
\affiliation{Max-Planck-Institut f\"ur Extraterrestrische Physik, Center for Astrochemical Studies, Garching, Germany}

\author{W. M. Grundy}
\affiliation{Lowell Observatory, Flagstaff, Arizona, USA}

\author{D. Trilling}
\affiliation{Department of Astronomy and Planetary Sciences, Northern Arizona University, Flagstaff, AZ, USA}

\author{Y. Fernandez}
\affiliation{Department of Physics, University of Central Florida, Orlando, Florida, USA}

\author{C. Dalle-Ore}
\affiliation{SETI Institute, Mountain View, CA, USA ; NASA Ames Research Center, Moffett Field, CA, USA}



\begin{abstract}

Using data from the Infrared Array Camera on the Spitzer Space Telescope, we present photometric observations of a sample of 100 trans-Neptunian objects (TNOs) beyond 2.2 $\mu$m. These observations, collected with two broad-band filters centered at 3.6 and 4.5 $\mu$m, were done in order to study the surface composition of TNOs, which are too faint to obtain spectroscopic measurements. With this aim, we have developed a method for the identification of different materials that are found on the surfaces of TNOs. In our sample, we detected objects with colors that are consistent with the presence of small amounts of water and were able to distinguish between surfaces that are predominately composed of complex organics and amorphous silicates. We found that $86\%$ of our sample have characteristics that are consistent with a certain amount of water ice, and the most common composition ($73\%$ of the objects) is a mixture of water ice, amorphous silicates, and complex organics. $23\%$ of our sample may include other ices such as carbon monoxide, carbon dioxide, methane or methanol. Additionally, only small objects seem to have surfaces dominated by silicates. This method is a unique tool for the identification of complex organics and to obtain the surface composition of extremely faint objects. Also, this method will be beneficial when using the \textit{James Webb Space Telescope} for differentiating groups within the trans-Neptunian population.

\end{abstract}

\keywords{Trans-Neptunian objects (TNOs) -- TNOs: composition -- TNOs: surface -- Spectrosphotometric techniques -- Origin: solar system }



\section{Introduction}
\label{sec:introduction}

Trans-Neptunian objects (TNOs) are solar system objects whose heliocentric orbits have semi-major axes $a$ greater than that of Neptune and less than where the Oort clouds begin, i.e., $30.07<$a$<2000$ au \citep{Gladman2008}. Centaurs, with semi-major axes $a$ between those of Jupiter and Neptune, are a significant population of objects in the region between the giant planets; from a compositional point of view they are of great interest because these nearby objects are easier to study and believed to be derived from TNOs via perturbations by Neptune and the other giant planets \citep{Fernandez1980,Levison1997}. Therefore, both populations are studied together, understanding that there is a greater chance of recent modification of Centaur surfaces as they approach the Sun from the trans-neptunian region.

Due to their large heliocentric distances, all TNOs have surface temperatures that are low enough to retain water ice. Non-Centaur TNOs can also harbor highly volatile ices such as \diox, \met, CO and {\nit} in stable form, although the last three are likely only present on the TNO dwarf planets, which have enough mass to trap them gravitationally \citep{Schaller2007a,Pinilla-Alonso2020,Young2020}. In addition to these molecular ices, TNO surfaces are thought to be composed of refractory macromolecular complexes (termed tholins), derived via photolysis and radiolysis of ices \citep{Khare1993,Materese2014,Materese2015}, and silicates incorporated as grains during accretion from the solar nebula. The slightly- to extremely-red colors of some TNOs likely reflect the presence of these organic ``tholins'', although some silicates can also be slightly reddish in the visible spectral region.

Determining the relationships between the physical properties of TNOs to their dynamical classifications and orbital histories may help us understand the composition and thermal state of the solar nebula as a function of heliocentric radius \cite[e.g.,][]{Fernandez2020}. The cold-classical objects in the main belt (see section \ref{sec:dynamical_class}) are the only ones thought to still reside in their primordial orbits \cite[e.g.,][]{Brown2001,Van-Laerhoven2019}; TNOs in all other classes were likely perturbed greatly from their original orbits by interactions with the migrating giant planets \citep[e.g.,][]{Gladman2008,Petit2011,Morbidelli2020}. However, establishing the composition-dynamics relationship is difficult because TNOs are so faint (typical $m_{\rm V} \gtrsim 20$). Relatively few have near-IR colors, which could be diagnostic of composition, let alone near-IR spectra. The alteration of surface composition may also play a role in camouflaging links between dynamical class and composition \citep{Gil-Hutton2002}. This is more likely on Centaurs, which experience much higher temperatures and fluxes of solar UV and charged particles than the more distant TNOs. 

Multi-filter photometry has been used to study the composition of relatively large samples of TNOs, and has revealed some correlation between optical colors and size \citep[e.g.,][]{Fraser2012,Peixinho2015}, and between color and orbital inclination in the classical population \citep[e.g.,][]{Trujillo2002, Doressoundiram2002,Marsset2020}. Thermal data also revealed that the cold-classical objects also generally have higher albedos than the hot-classical objects \citep{Brucker2009,Lacerda2014,Vilenius2014}. Near-IR colors potentially offer better discrimination of composition, but only $\sim25\%$ of all ground-based data for TNOs are at these wavelengths \citep[i.e, from $\lambda\sim1$ to $\sim2.5\,\mu$m; e.g.,][]{Hainaut2012}. \cite{Fraser2012} and \cite{Fraser2015} reported on the visible and NIR observations of 100 small TNOs and Centaurs using the \textit{Hubble} WFC3, finding two clusters in visible vs.~near-IR color, and evidence of ubiquitous \wat\ ice. To date, spectra of various quality have been collected for only about 75 TNOs and Centaurs total \citep[e.g.,][and references therein]{Barkume2008,Brown2008,Barucci2011,Barucci2020}. Based on those data, it appears that {\wat} is the most common ice detected on TNOs' surfaces. 

These past studies have achieved limited success in part because they are limited by the sensitivity of ground-based facilities, and in part because only the relatively weak near-IR bands of the components of interest could be studied. The ices and complex organics (sometimes we also refer to the latest as only organics) that we expect to find on TNOs and Centaurs have their fundamental absorption bands beyond 2.2 $\mu$m, necessitating space-based observations due to the atmospheric interference and low brightness presented by this populations. Here we summarize a collection of 3.6 and 4.5\mum\ observations of TNOs made with the Infrared Array Camera, IRAC \citep{Fazio2004}, on the \textit{Spitzer Space Observatory} (hereafter, Spitzer). These data are consistent with absorption bands due to \wat, organic tholins, and volatile ices, and the presence of silicates. They also reveal the range of geometric albedos of TNOs at 3 -- 5 $\mu$m, which will be important for planning observations at those wavelengths using the \textit{James Webb Space Telescope (JWST).} JWST can in principle provide high-quality spectra for hundreds of TNOs from 1 -- 5 \mum.


\section{Spitzer/IRAC observations and data reduction}
\label{sec:observations}

Spitzer, with an aperture of 85 cm, was launched in August 2003 into an Earth-trailing, heliocentric orbit \citep{Werner2004}. Its Infrared Array Camera (IRAC), a broad-band imager with four channels, has a field of view (FoV) of $5.2'\times5.2'$ and an image scale of $\sim1.2$ arcsec pixel$^{-1}$ \citep{Fazio2004}. The four broad-band channels are centered at 3.6, 4.5, 5.8, and $8.0\,\mu$m with full width at half maximum of 0.68, 0.87, 1.25, and $2.52\,\mu$m, respectively. Spitzer's instruments were originally cooled by liquid helium, but this cryogen was exhausted in May 2009. Thereafter, Spitzer entered its so-called ``warm mission'' during which only the 3.6 and 4.5 $\mu$m channels of IRAC are operational. 

From 2005 to 2016, we observed TNOs and Centaurs during different Spitzer observational cycles\footnote{Specifically GO2, GO4, GO6, GO7, GO8 and GO12 cycles, with program ID 20769, 40389, 60155, 70115, 80116, and 12012, respectively.}, and included all dynamical classes. In this work we present the results from a sub-sample that includes detached, classical and resonant types (see section \ref{sec:dynamical_class}). Data are available for download from the Spitzer Heritage Archive\footnote{\url{http://sha.ipac.caltech.edu/applications/Spitzer/SHA/}}. Even though some objects were observed in 5.8 and $8.0\,\mu$m, only the 3.6 and 4.5 $\mu$m data are reported and analyzed here. Note that for these cold ($\sim40$ K) and distant objects (beyond 30 au) objects the flux at these wavelengths is only reflected light, with no thermal emission from their surfaces. A summary of observations is provided in table \ref{tb:observations}, which includes: object designation, date of observation, Spitzer-object distance ($\Delta$), heliocentric distance ($r_{\rm H}$), phase angle ($\alpha$), flux at 3.6 and 4.5 \micron\ ($F_{3.6}$, and $F_{4.5}$, respectively), and geometric albedo at 3.6 and 4.5 \micron\ ($p_{3.6}$, and $p_{4.5}$, respectively).

Each object was observed twice with time intervals between several hours to several days, with the aim of having two different measurements at different locations relative to the background star fields, while keeping the object within the same FoV (movement of the object was around $30''$). This provides several benefits: identification of the object by its motion against background stars, straightforward and accurate background subtraction, and increased probability of at least one good measurement in the case of a field star obscuring the object \citep{Emery2007}. Each observation consists of nine or more dithered frames, which allows image correction for effects such as bad pixels, latent images from previous observations, and stray light from bright objects in or just off the frame. 

IRAC data frames were pre-processed by the Spitzer Science Center (SSC) automated pipeline for dark, bias, and flat field corrections, and flux calibrations. Corrections for IRAC specific artifacts such as column and array pulldown, muxbleeding and stray light contamination, are done for individual frames, if necessary. The two observations per object were used as background frames for each other in order to remove diffuse flux and most of the contribution from stars in the field. Corrections (available from the SSC) for pixel solid angle variations and array location dependent photometric variations were applied to the frames.

We performed synthetic aperture photometry in order to calculate the flux from the object, as outlined in the IRAC Instrument Handbook\footnote{\url{http://irsa.ipac.caltech.edu/data/SPITZER/docs/spitzermission/}}. We used four combinations of aperture radius and background annulus as shown in table \ref{tb:photometry_parameters} \citep{Emery2007}. Aperture corrections for each combination are given in the IRAC Instrument Handbook. Color corrections are calculated for each broad-band channel assuming a solar spectral slope through each passband \citep{Smith1974}. All aperture/annulus combinations and all frames in an observation are averaged together for each channel and recorded as final fluxes. Uncertainties of 1$\sigma$ are reported. The uncertainties account for photon counting statistics, deviation among the dithered frames, and deviation among the aperture/annulus combinations. The absolute calibration of all IRAC channels is accurate to $\sim3\%$ \citep{Reach2005}.

A visual inspection of each frame, the average frame, and the background subtracted frame for each observation was conducted to assess success of the observation. If objects were not discernible by eye in the observation, an upper limit calculated from the background flux is presented. The background flux is the mean of the repeated aperture photometry process for 50 random center points, removing outliers outside $\pm2\sigma$, within a radius of 50 pixels of the location of the object predicted by its ephemeris.

\begin{table}
\caption{\label{tb:photometry_parameters} Parameters used to perform the synthetic aperture photometry.}
\centering
\begin{tabular}{ccc}
Aperture    &   Annul radius    & Annul width \\
(pix)       &       (pix)       & (pix)     \\
\hline
2           &       2           &       4       \\
2           &       10          &   10          \\
3           &       3           &       4       \\
3           &       10          &   10  \\
\hline
\end{tabular}
\end{table}

With the aim of ensuring that our measurements are reliable, the objects had to satisfy the following criteria in order to be selected for the final analysis: 

\begin{enumerate}
\item Not contaminated by a background star or image artifact. 
\item Clearly visible in the average frames.
\item Greater than $3\sigma$ detection.
\end{enumerate}

For any objects selected for final analysis that have $<3\sigma$ detection, we used a $3\sigma$ upper limit calculated from the background flux. From the total sample, 100 objects satisfied these criteria and were analyzed in this work. All measured fluxes with errors are provided in table \ref{tb:observations}.


\section{Supporting data}
\label{sec:data_sample}

\begin{figure}
\includegraphics[width=\columnwidth]{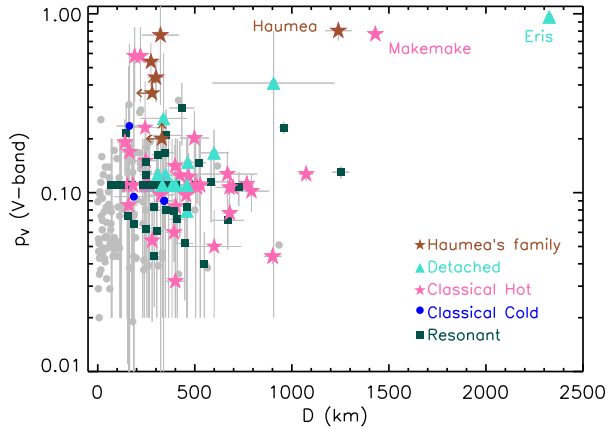}
\caption{\label{fig:physical_properties} Diameter vs.~geometric albedo in $V$-band from the ``TNOs are Cool'' sample (grey circles). Colored symbols denote objects of our sample \cite[note that some of them have diameters assuming the median albedo calculated from][]{Mueller2020}. Green squares show resonant objects, turquoise triangles show detached objects, pink stars and blue circles represent hot and cold classical objects, respectively. Brown stars represent Haumea's family members \cite[the brown arrows indicate lower limits][]{Brown2007b,Snodgrass2010}. Only dwarf planets and Haumea's family members have albedos over 40\%, which belong to detached and hot classical groups. The remaining objects have albedos under 40\%.}
\end{figure}

\subsection{Dynamical classification}
\label{sec:dynamical_class}
The dynamical classes of the objects in the sample, which are included in table \ref{tb:physical_properties_compilation}, were determined using the numerical procedure detailed in \cite{Gladman2008}. 

These classifications are categorized as either secure or insecure. Given the observed on-sky location of each TNO over time, we find the best fit orbit as well as the highest- and lowest- semi-major axis orbits that are consistent with the observations\footnote{For each candidate orbit, we consider the deviation between the predicted position from that orbit versus the observed astrometry. To be consistent with observations, the worst predicted versus observed astrometric position must be no more than 1.5 times the worst from the best fit orbit, and the root mean square of these deviations must be no more than 1.5 times that of the best fit.  The selected extremal orbits are then those with the lowest and highest semi-major axis.}. These three orbits are used as starting points for 100 Myr numerical simulations. If all three ``clones'' show the same orbital behaviour in these simulations, the classification is considered secure. If not, it is insecure and additional future observations are still required to establish the classification; 
in this case we give the best-fit classification.

Each of the dynamical classifications are designed to capture orbital behavioral characteristics, particularly to describe the manner in which each TNO is (or is not) interacting with Neptune. TNOs which are experiencing active gravitational scattering by Neptune are classified as {\it scattering} objects; we consider a TNO to be scattering if the semi-major axis changes by $>1.5$ au over the 100 Myr simulation. {\it Resonant} objects are TNOs in mean-motion resonance with Neptune (resonance with Uranus are searched for, but are only known for Centaurs). We diagnose resonance by determining if a resonant angle is librating as opposed to circulating \cite[see][for more details]{Gladman2008,Khain2018}. TNOs that are not strongly interacting with Neptune are classified as either {\it detached} or {\it classical} objects; if the TNO is not scattering or resonant and has an initial eccentricity above 0.24, then it is called ``detached''; otherwise, it is considered ``classical''. 
Classical objects are then subdivided, based on their position relative to two major TNO resonances, as {\it inner} ($a<39.4$~au, where the $3:2$ resonance is located), {\it main} ($39.4<a<47.0$ au), and {\it outer} ($a>47.0$~au). Orbital inclination $i$ is not part of the dynamical classification because the TNO $i$ structure is complex (varying with $a$), and causes confusion if an overly simplistic
$i$ cut is used. In particular, modern understanding shows that the so-called 'cold' component \citep{Brown2001} appears to be present  only in the main classical belt between $42.5<a<47.0$ au \citep{Petit2011}, and should be located by using free inclinations $i_{\rm free}$ corrected for secular effects and thus measured with respect to the local Laplace pole; \citet{Van-Laerhoven2019} show that the cold component then has an impressively small ``inclination width'' of $<2^{\circ}$, and that in the main belt choosing objects with $i_{free}<4^{\circ}$ gives a reasonable separation, where the majority of the objects on either side of this boundary belong to only one of the two otherwise-overlapping hot and cold components. At other semi-major axes, the ``cold classical'' component appears not to exist and objects with low inclinations are simply the low-$i$ tail of the hot component.

In this manuscript, we analyzed those that did not have or are not having planetary encounters. In other words, we excluded scattering objects, which will be analyzed separately with Centaurs in a future paper. In summary, from the 101 objects that satisfied the criteria from the above section: 40 are hot classical, 4 are cold classical, 41 are resonant, and 15 are detached. 

\subsection{Albedos and diameters}

Different physical properties were used for our albedo calculations (see section \ref{sec:results}). Radii and visible geometric albedos where compiled from the database of the ``TNOs are cool'' project\footnote{\url{http://public-tnosarecool.lesia.obspm.fr}} \citep{Mueller2009}, which is, to date, the most complete and accurate database of these properties for TNOs \citep{Mueller2020}. Figure \ref{fig:physical_properties} shows diameters and albedos for resonant objects (green squares), detached objects (turquoise triangles), and hot and cold classical objects (pink stars and blue circles, respectively). The ``TNOs are cool'' sample is shown in grey circles. To maintain consistency, we also adopted the corresponding absolute magnitudes used in the ``TNOs are cool'' project.

Not all our objects were observed by the ``TNOs are cool'' project; in those instances, we used $11\pm9\%$, the median albedo calculated from the ``TNOs are cool'' database.  Coupled with the absolute magnitude specified by the Minor Planet Center (\href{https://www.minorplanetcenter.net}{MPC}), we calculate the radius ($R$) using the equation:
\begin{equation}
\label{eq:radius}
R=C\,p^{-1/2}10^{-H/5},
\end{equation}
where $C$ is a constant depending on the observed wavelength (i.e., 1329 km for $V$-band), $p$ and $H$ are the geometric albedo and the absolute magnitude of the object, in the same photometric band, respectively. The same procedure was also applied to three Haumea family members that only have upper limits in their $V$-band albedos \citep{Vilenius2018}, specifically, 1995~SM$_{55}$, 1996~TO$_{66}$, and 1999~CD$_{158}$. For these three objects, we calculated a median value using the albedos from Haumea family members ($p_{\rm V, Haumea}=0.58\pm0.26$), as the median value for the TNO population is not representative of the family. Table \ref{tb:physical_properties_compilation} shows a compilation of the physical properties used in this work.

\begin{figure}
\centering
\gridline{\fig{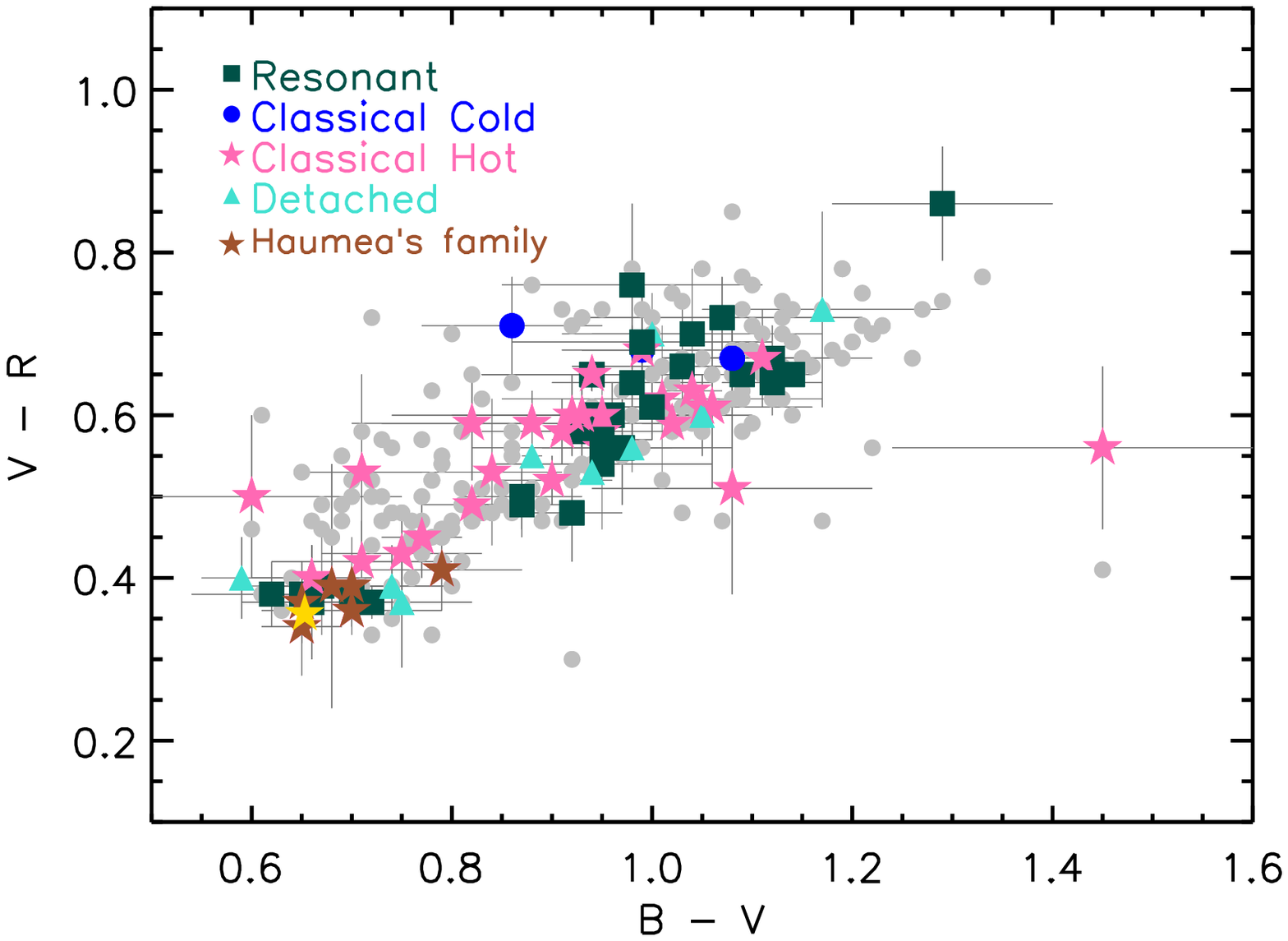}{\columnwidth}{(a)}}
\gridline{\fig{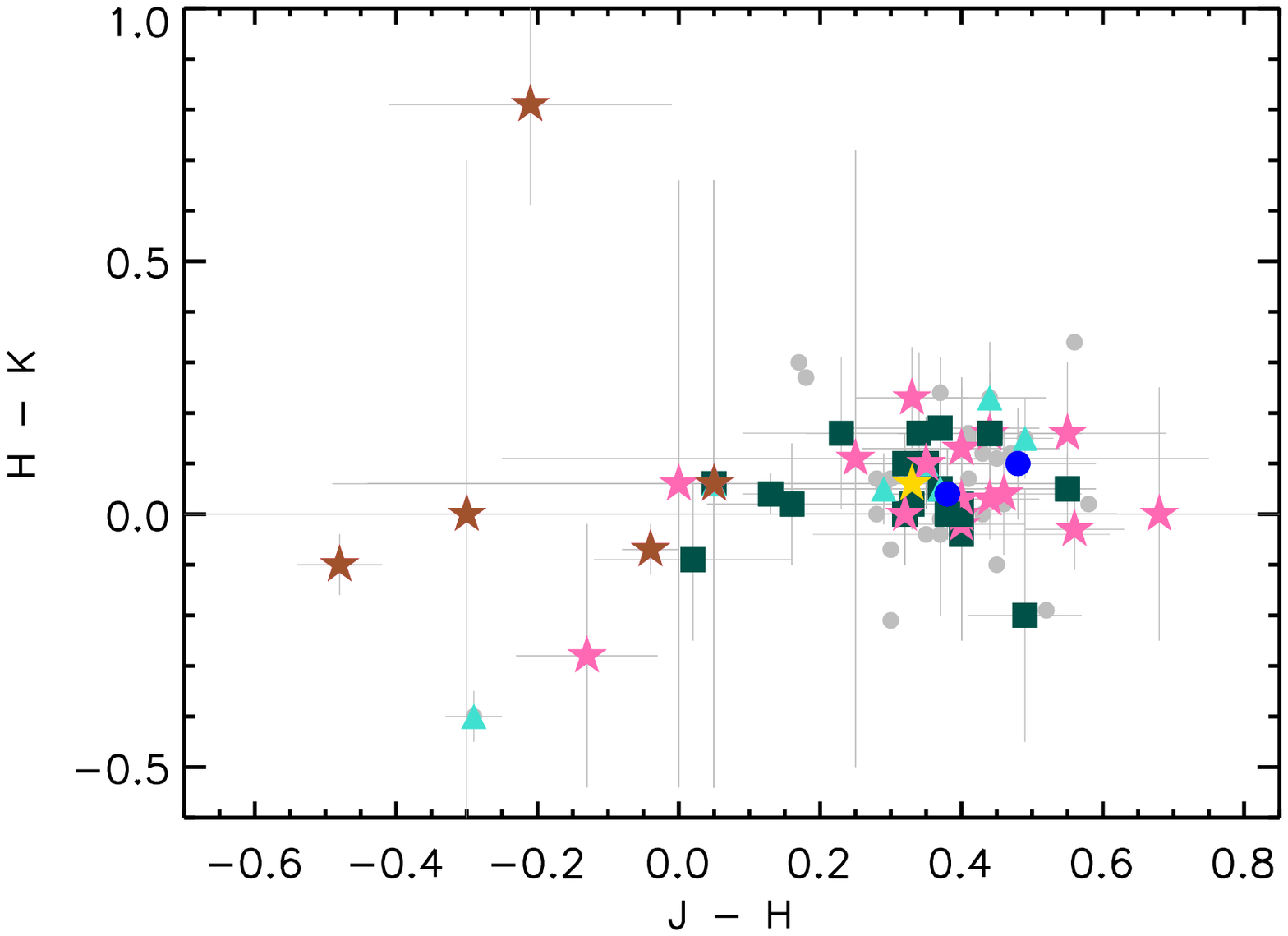}{\columnwidth}{(b)}}
\caption{\label{fig:sample_colors} (a) $B-V$ versus $V-R$ color-color plot for our sample using colors and symbols to represent different populations. Grey circles represent data from \cite{Peixinho2015}. (b) $J-H$ versus $H-K$ color-color plot for our sample using colors and symbols to represent different populations. Grey circles represent the sample from \cite{Fulchignoni2008}. In both panels: the same scale as in \cite{Doressoundiram2008} was used in order to compare the diversity of our sample; green squares show resonant objects, triangles show detached objects, pink stars and blue circles represent hot and cold classical objects, respectively. Solar colors are indicated by a yellow star.}
\end{figure}

\subsection{Visible and near-infrared photometric data}
Ground-based observations for VNIR photometric data were compiled from published literature using the following process: 
\begin{itemize}
\item [$-$] Visible colors were taken from \cite{Peixinho2015}, and near infrared (NIR) colors from \cite{Fulchignoni2008}, with some of the NIR colors completed using data from \cite{Belskaya2015}, MBOSS, \cite{Tegler2016}, among others (see table \ref{tb:colors_compilation}).
\item [$-$] For objects with no color available in the above references, we used the average colors published by \cite{Fulchignoni2008} as a function of their taxonomy \cite[given in][]{Belskaya2015}.
\item [$-$] Some exceptions were made during the process: Because the NIR colors available in the literature for Makemake and Quaoar have  very large errors, we decided to extract their colors using VNIR spectra published in the literature. We also used the measured spectrum of 2002 TX$_{300}$ \citep{Licandro2006} and synthesized its NIR colors because they are unavailable in the literature.
\end{itemize}

As can be seen in figure \ref{fig:sample_colors}, our sample exhibits the full diversity of colors reported by different surveys \citep[e.g.,][]{Barucci1999,Delsanti2001,Delsanti2004,Delsanti2006,Boehnhardt2002,Doressoundiram2001,Doressoundiram2002,Doressoundiram2005,Doressoundiram2007,DeMeo2009,Perna2010}. Table \ref{tb:colors_compilation} provides a compilation of colors for our sample. The reference for each object and any exceptions to the process to select the reference are included in the table.

\subsection{Spectroscopic data}

Spectra were also compiled as a baseline to understand our sample and to assess whether our compilation of colors that are translated to spectrophotometric measurements (i.e., geometric albedo versus wavelength) are in agreement with the spectra. We followed the work performed by \cite{Barucci2011} regarding surface composition of TNOs using VNIR spectra. We used this work as a reference for water and other ices detection because it provides a homogeneous analysis of spectra for, at least, some of our targets. However, because this review was published eight years ago, we also searched into more recent papers in order to find spectra published for other objects. 

Although we have not performed any calculation to measure bands, we followed the criteria of \cite{Barucci2011}, in which three categories are defined: objects with clear detection of absorption bands reported by the authors are considered ``sure detections'', objects with some indications of absorption bands reported by the authors are considered ``tentative detections'' and objects with no indications of absorption bands reported by the authors are considered ``no detections''.

Literature regarding the spectra used in this work are cited in sections \ref{sec:results} and \ref{sec:other_cc}, where each object is independently analyzed. In general, there is good agreement for the whole sample, as demonstrated in appendix \ref{ap:plot_individual_objects}. There exists only 5 exceptions in which the spectrum and the spectrophotometric measurements do not completely agree (discussed in section \ref{sec:analysis}).


\section{Reflectance and color indices}
\label{sec:reflectance_color_indices}

\subsection{Reflectance}

Fluxes in the IRAC wavelengths are converted into geometric albedos in order to be combined with ground-based data and to plot as spectrophotometric measurements. The geometric albedo ($p_{\lambda}$) at wavelength $\lambda$, is given as follows:
\begin{equation}
\label{eq:albedo_equation}
p_\lambda=\frac{F_\lambda r_{\rm H}^2 \Delta^2}{F_{\odot,\lambda}\Phi R^2},
\end{equation}
where $F_{\lambda}$ is the measured flux at wavelength $\lambda$; $F_{\odot,\lambda}$ is the solar flux at 1 au at the same wavelength; $r_{\rm H}$ is the heliocentric distance in au; $\Delta$ is the Spitzer-centric distances in km; $R$ is the radius of the object in km and $\Phi$ is the phase correction factor. We have calculated the phase correction factor by $\Phi=10^{\frac{-\beta\alpha}{2.5}}$; where $\alpha$ is the phase angle with respect to Spitzer, and $\beta=0.14^{+0.07}_{-0.03}$ mag deg$^{-1}$ is the median of phase coefficients tabulated by \cite{Alvarez-Candal2016}, where we have excluded values that are negative or greater than 0.5 as these are likely due to uncharacterized rotational light-curve effects\footnote{The median value using the complete sample was $\beta=0.1\pm0.1$ mag deg$^{-1}$, which produces an error of the same order of magnitude than the $\beta$ value; for that reason, for that reasons, outliers were excluded first.}.

Colors from the literature were also converted into geometric albedo by means of the equation:
\begin{equation}
\label{eq:albedo_color}
p_{\rm R}=p_{\rm V}10^{\frac{(V-R)-(V-R)_{\odot}}{2.5}},
\end{equation}
where $p_{\rm R}$ is the albedo in $R$-band, $p_{\rm V}$ is the albedo in $V$-band, $(V-R)$ is the color of the object and $(V-R)_{\odot}$ is the color of the Sun. Analogous relations are used for the other colors, in our particular case: $(B-V)$, $(V-I)$, $(V-J)$, $(V-H)$ and $(V-K)$. As a result, spectrophotometric measurements have been incorporated for each object (see appendix \ref{ap:plot_individual_objects}), allowing us to analyze the surface composition using the widest possible wavelength range (see section \ref{sec:results}). 

Note that, from Eq. (\ref{eq:albedo_equation}), the IRAC albedo depends on the radius $R$ explicitly and, therefore, one may expect that the large relative uncertainties in the measurements of the radius of known objects would translate into large relative uncertainties in the IRAC albedos. However, Eq. (\ref{eq:radius}) provides a workaround, as one can check explicitly that combining both Eqs. (\ref{eq:albedo_equation}) and (\ref{eq:radius}) results into an expression for the ratio of the ground-based $V$-band geometric albedo and IRAC albedo, $p_V/p_\lambda$ that does not depend explicitly on the radius $R$ nor $p_V$ (in other words, in the resulting combined equation only the relative albedo plays a role). Hence, if one focuses on the relative albedos (i.e., on the absorption), the effect of large uncertainties in $R$ is neutralized, provided, of course, that $R$ and $p_V$ are constrained to satisfy Eq. (\ref{eq:radius}). Table \ref{tb:results} shows median and average values for the geometric albedos at 3.6 and 4.5~\mum\ of our sample, while tables \ref{tb:observations} and \ref{tb:albedos_compilation} show the geometric albedos obtained for each object at IRAC and VNIR wavelengths, respectively. Note that measurements at 3.6 and 4.5 $\mu$m were taken consecutively one after the other. The total exposure times (including all sub-exposures) were no longer than 40 min at 3.6 $\mu$m. Considering that the faster rotation period for a TNO known to date is 4 hours (Haumea), with a median value of $\sim8$ hours \citep{Thirouin2012}, large effects due to rotational variability between both filters are very unlucky.

\begin{table*}
    \centering
      \caption{Median and average values for the geometric albedos and colors obtained in this work. Abbreviations are defined as follows: geometric albedo at 3.6 and 4.5 $\mu$m ($p_{3.6}$, $p_{4.5}$, respectively).  
    \label{tb:results}}
    \begin{tabular}{ccccccc}
    \hline
         &$p_{3.6}$& $p_{4.5}$ & $3.6-4.5$  & $V-3.6$ & $J-3.6$ & $K-3.6$ \\
          &        &             & (mag)     & (mag)  & (mag)   & (mag) \\
                  \hline
    Median       & $0.12\pm0.01$  &$0.15\pm0.03$  & $0.2\pm0.1$  & $0.0\pm0.1$  & -$0.9\pm0.1$  & $-0.9\pm0.1$ \\
    Average       & $0.16\pm0.05$  &$0.21\pm0.05$  & $0.3\pm0.2$  & $0.1\pm1.0$  & -$0.7\pm0.3$  & $-0.8\pm0.6$ \\
    \hline
    \end{tabular}

\end{table*}

\subsection{Color indices}

In order to combine our IRAC/Spitzer measurements with existing apparent magnitude measurements at shorter wavelengths,
we define colors, expressed as magnitude differences, as:

\begin{eqnarray}
\nonumber m_1 - m_2 = -2.5 \log\left(\frac{F_1/S_1}{F_2/S_2}\right) \\
= -2.5 \log\left(\frac{F_1}{F_2}\right)+2.5\log\left(\frac{S_1}{S_2}\right),
\end{eqnarray}
where $F_{\rm n}$ is the flux from the target in a given band, and $S_{\rm n}$ is the solar flux in that band (with $\rm n=1,2$). In what follows we denote these colors as ``V~-~3.6~\mum'', ``J~-~3.6~\mum'', ``K~-~3.6~\mum'', and ``3.6~\mum~-~4.5\mum''.
Because these colors are corrected for the intrinsic solar color, they can be related to the albedos in the bands, for example:
\begin{equation}
3.6\,\mu {\rm m}-4.5\,\mu {\rm m}=-2.5\log\frac{p_{3.6}}{p_{4.5}},
\end{equation}
where $3.6\,\mu {\rm m}-4.5\,\mu$m is the color from IRAC/Spitzer measurements, $p_{3.6}$ is the geometric albedo at $3.6\,\mu$m and $p_{4.5}$ is the geometric albedo at 4.5 $\mu$m. Other colors using IRAC/Spitzer broad-band and ground-based measurements can be obtained using the same equation. A compilation of the different colors used in the analysis is shown in table \ref{tb:colors_Spitzer_compilation}.

\section{Results}
\label{sec:results}
Below we present our results in two ways: (1) by comparing the color indices we have measured to synthetic color indices computed from synthetic reflectance models for pure substances and binary- and trinary mixtures of those substances, and (2) by presenting the measured reflectance spectra (i.e., spectrophotometric measurements) of all objects along with measured visible -- near-IR spectra found in the literature, or spectral models if available (see appendix \ref{ap:plot_individual_objects}). 

\subsection{Measured Colors, and synthetic spectra}

\begin{figure*}
\gridline{\fig{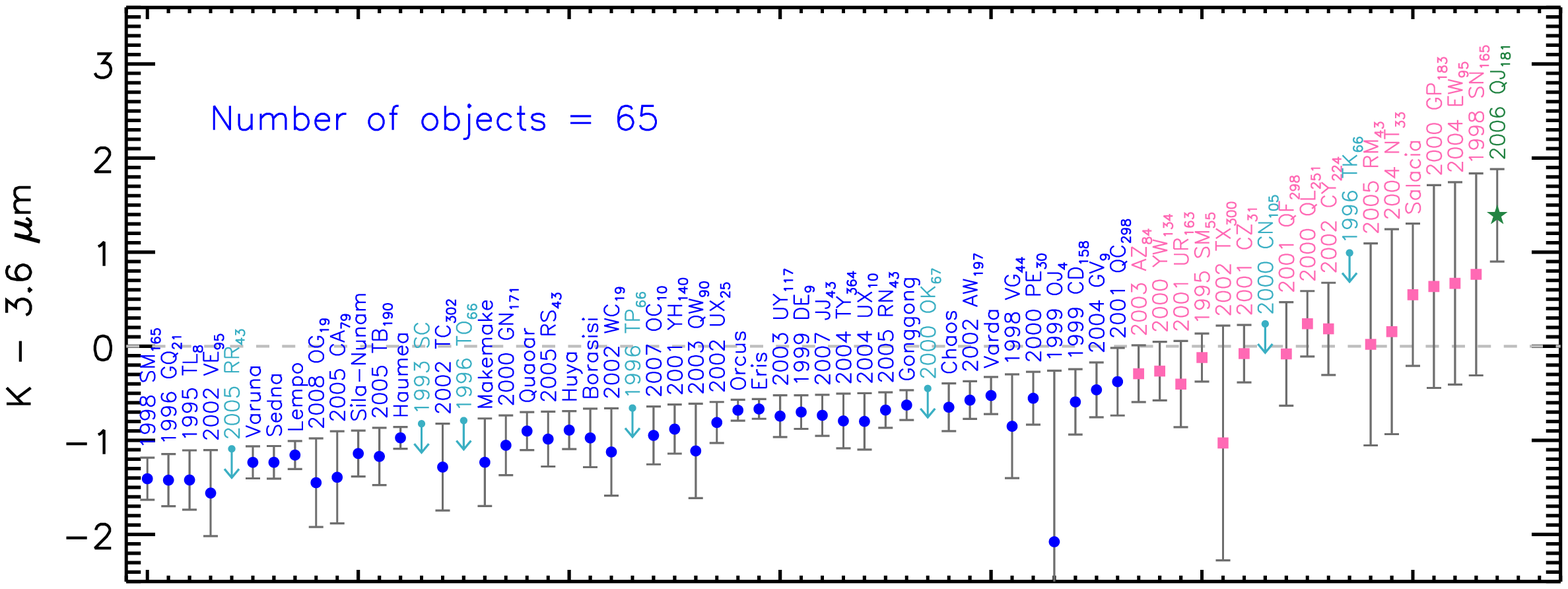}{\textwidth}{(a)}}
\gridline{\fig{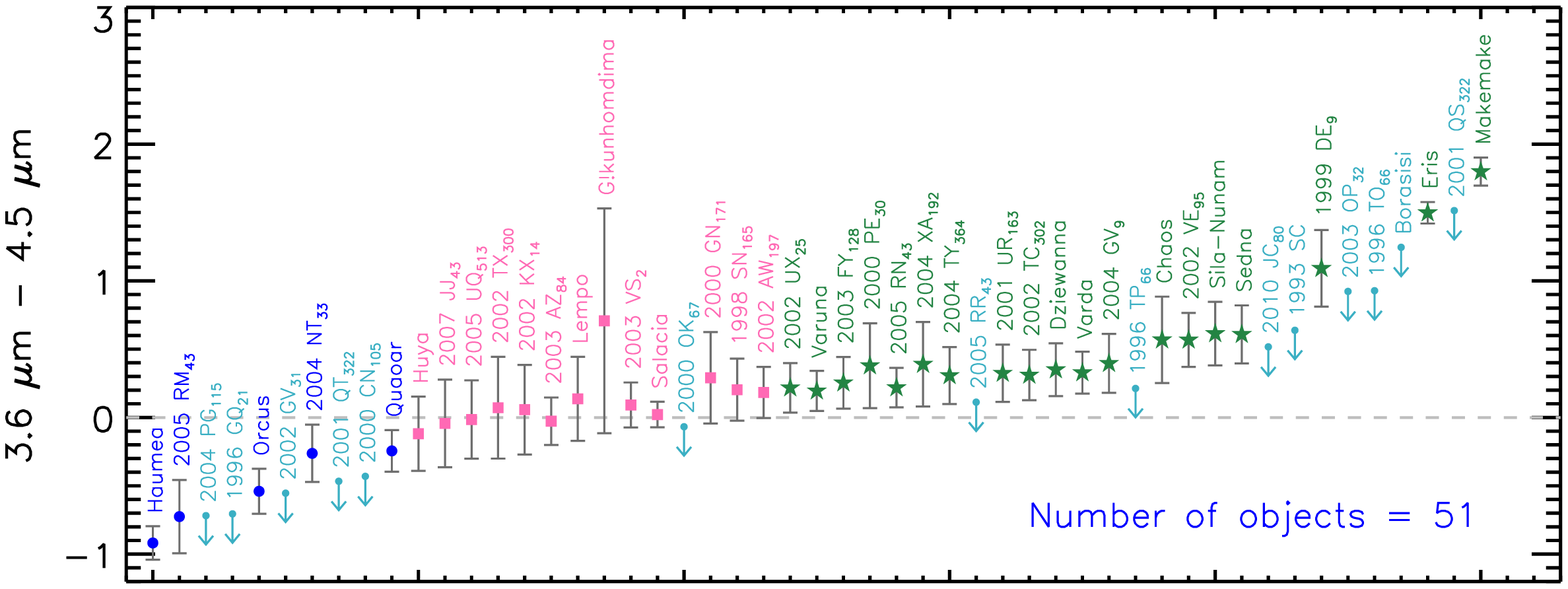}{\textwidth}{(b)}}
        
\caption{$K$-band vs.~IRAC color indices for our entire sample, plotted in order of increasing color index. Negative (positive) values indicate absorption in the longer (shorter) wavelength; zero indicates neutral (\ie, solar) colors (horizontal dashed line). Note that not all the objects of our sample presented measurements at 4.5 $\mu$m or published data in the $K$ band, which drops the number of objects to 65 and 51 for $K-3.6\,\mu$m (upper panel) and $3.6\,\mu{\rm m}-4.5\,\mu{\rm m}$ (bottom panel), respectively. There are also a few objects that have only upper-limits in both IRAC bands, so those are also not plotted in the bottom panel. \textbf{Error bars give $1-\sigma$ uncertainties while arrows give $3-\sigma$ upper limits.} Objects with 1-$\sigma$ independent probability of presenting absorption at 3.6 $\mu$m (upper panel) and 4.5 $\mu$m (bottom panel) are represented by blue circles and are labeled in blue, objects with 1-$\sigma$ independent probability of \textbf{no} presenting absorption are plotted as green stars and labeled in green. Objects that error bars overlap both regions are represented by pink squares and labeled in pink. Object for which upper limits have been determine are represented by turquoise arrows and are labeled in turquoise. The range of TNO colors in these bands ($\gtrsim 2$mag) significantly exceeds that seen in the visible \cite[$\lesssim 1$mag; see figure \ref{fig:sample_colors} and e.g.,][]{Peixinho2015,Schwamb2019}. \label{fig:color_index}}
\end{figure*}

\begin{figure*}
\gridline{\fig{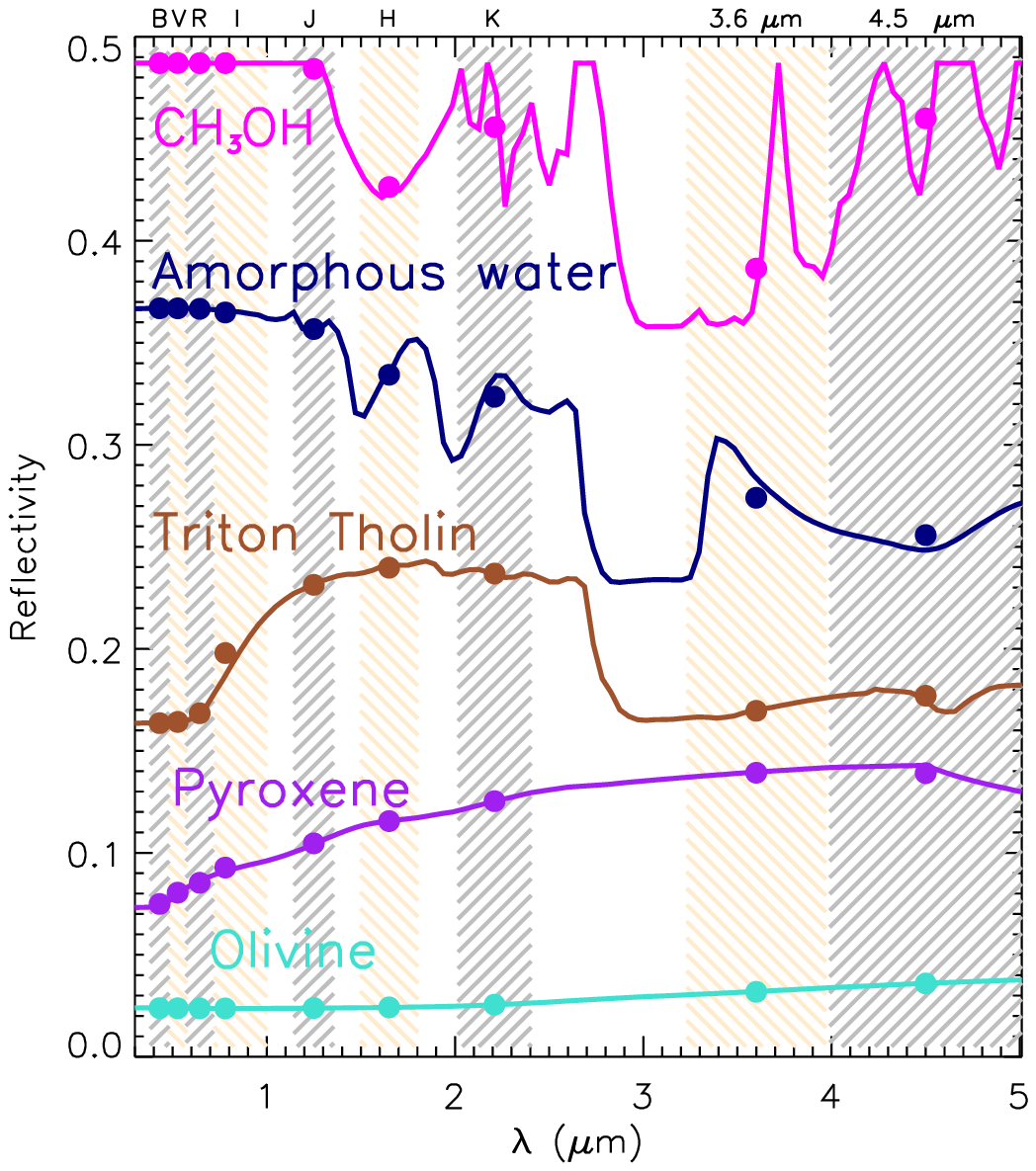}{0.5\textwidth}{(a)}
          \fig{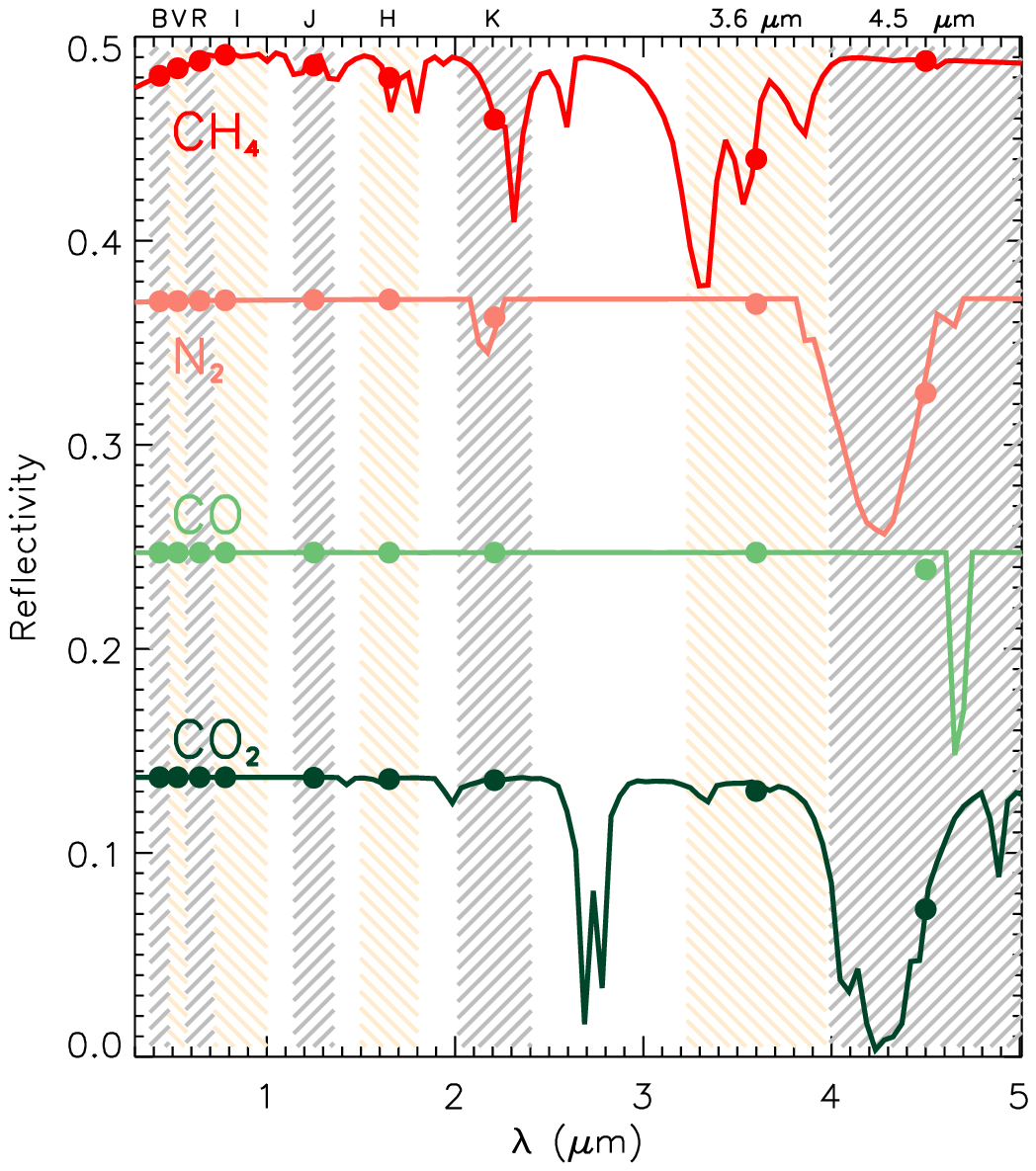}{0.5\textwidth}{(b)}
          }
\caption{\label{fig:sinthetic_models} Synthetic reflectance spectra of some pure materials expected or known to exist on the surfaces of small TNOs and Centaurs (left) and additional molecular ices found on TNO dwarf planets (right). The particle size was 10 \mum\ for all materials except for \nit, where the particle size is 10 cm. Vertical grey and brown shaded bars indicate VNIR and IRAC/Spitzer filter bands, as labeled along the top axis. The bandwidth is the full width of the band at 50\% of the average in-band transmission. Dots represent the spectrophotometric measurements as given by the convolution of each synthetic spectrum with the filters. For clarity, reflectance have been shifted by an arbitrary offsets as follows: olivine (0.02), pyroxene (0.07), Triton tholin (0.16), amorphous \wat\ (0.23), \nol\ (0.35), \met (0.355), N$_2$ (0.235), CO (0.11), and CO$_2$ (no offset).}
\end{figure*}

Figure \ref{fig:color_index} shows the measured $K$ vs.~IRAC color indices for all of our targets. Note that not all the objects of our sample presented measurements at 4.5 $\mu$m or published data in the $K$ band, which drops the number of objects to 66 and 52 for $K-3.6\,\mu$m and $3.6\,\mu{\rm m}-4.5\,\mu{\rm m}$, respectively. From the upper panel in figure \ref{fig:color_index} we see that, from the 65 objects with $K-3.6\,\mu$m, 48 present $1\sigma$ independent probability of having an absorption at $3.6\,\mu$m ($74\%$ of the sample), with 44 objects representing the $1\sigma$ compound probability (i.e., $68\%$). Only one object present $1\sigma$ independent probability of not having absorption at $3.6\,\mu$m, with 2 objects representing the $1\sigma$ compound probability (i.e., $3\%$). Additionally, and as shown in the bottom panel, with a total of 51 objects, 11 objects present $1\sigma$ independent probability of having an absorption at $3.6\,\mu$m ($22\%$ of the sample), with the same number of objects representing the $1\sigma$ compound probability. From both panels we can appreciate that the range of colors is significantly larger than seen in the visible -- near-IR wavelengths \citep[e.g.,][]{Fulchignoni2008,Peixinho2015,Schwamb2019}, suggesting the potential of using the IRAC colors to constrain TNO composition in ways that have previously been impossible. Figure \ref{fig:sinthetic_models} shows synthetic spectral models for some materials typically found, or expected to be present, on TNOs and Centaurs. The figure illustrates how much stronger the absorption bands of most of these materials are at the IRAC wavelengths than in the near-IR. This is because the longer-wavelength absorptions are associated with fundamental molecular vibration frequencies, while those short-ward of 2.5\mum\ are weaker overtone bands, which explains the strong color diversity of TNOs seen in figure \ref{fig:color_index}. Other colors are also important for constraining composition, and help establish continuum levels (particularly true for $J$, $H$ and $K$-band). In addition to the colors discussed above, we also use  $V$-3.6\mum\ and $J$-3.6\mum\ colors in the following analysis. Table \ref{tb:colors_Spitzer_compilation} shows our compilation of all the colors for our targets (including the aforementioned and VNIR colors) used in this work.

\begin{table*}
\centering
\caption{\label{tb:synthetic_models} References of the optical constants for the synthetic models used in this work.}

\begin{tabular}{lcc}
\hline
Model         & Temperature (K)/Phase &   Reference \\
\hline\hline
\nol          &   90   &    Robert Brown$^{\star}$ \\
\met          &   40 &  \cite{Grundy2002}\\
\nit          &   21   &  \cite{Quirico1997}\\
CO            &  21   & \cite{Quirico1997}\\
\diox         & 150  &  \cite{Hansen1997}\\
\wat          & 120~amorphous &\cite{Mastrapa2009}\\
\wat          &150~crystalline &\cite{Mastrapa2009}\\
Triton Tholin &  Not available    & \cite{Imanaka2012}\\
Titan Tholin  & $\simeq 290$ & \cite{Khare1984}\\
Pyroxene 5: Mg$_{0.7}$Fe$_{0.3}$SiO$_3$    & amorphous &\cite{Dorschner1995}\\
Pyroxene 8: Mg$_{0.4}$Fe$_{0.6}$SiO$_3$    & amorphous &\cite{Dorschner1995} \\
Olivine 1     & amorphous &\cite{Dorschner1995} \\
Olivine 2     & amorphous &\cite{Dorschner1995} \\
\hline
\end{tabular}
\begin{tablenotes}\footnotesize
\centering
\item $^{\star}$See Appendix in \cite{Cruikshank1998}.
\end{tablenotes}
\end{table*}

\subsection{Spectrophotometric measurements}
Our compilation of visible--near-IR and IRAC spectrophotometric measurements for each object are plotted in terms of geometric albedo in appendix \ref{ap:plot_individual_objects}. The measurements were converted to albedo as described in section \ref{sec:reflectance_color_indices}. When a previously-published visible--near-IR spectrum of an object is available, it is also plotted to allow comparison to our IRAC results. The figures are ordered by provisional designation in ascending order, followed by the named objects in alphabetical order. The albedo spectra are shown for all objects, regardless of whether we found $J$- and/or $K$-band photometry in the literature. The objects lacking those measurements are excluded from the color index analysis we present below.

Inspecting each albedo spectrum in appendix \ref{ap:plot_individual_objects}, we found that seven of our objects do not show absorption at 3.6 and 4.5 $\mu$m with respect to the $K$-band. Referencing figure \ref{fig:sinthetic_models}, it seems plausible that such objects have surfaces dominated by amorphous silicates. Those objects are: 1998 SN$_{165}$, 2000 GP$_{183}$, 2000 QL$_{251}$, 2002 CY$_{224}$, 2004 EW$_{95}$, 2006 QJ$_{181}$ and Salacia, which have positive $K$~-~3.6\mum\ colors in figure \ref{fig:color_index}. The remaining objects have absorption identification in at least one of the two IRAC/Spitzer bands. These absorption identifications are related to either ices (\wat, \met, \nol,...) or complex organics, as explained below.

\section{Analysis}
\label{sec:analysis}
\subsection{Synthetic Color Indices}
\label{sec:Synthetic_color_indices}

\begin{figure}
\centering
\includegraphics[width=\columnwidth]{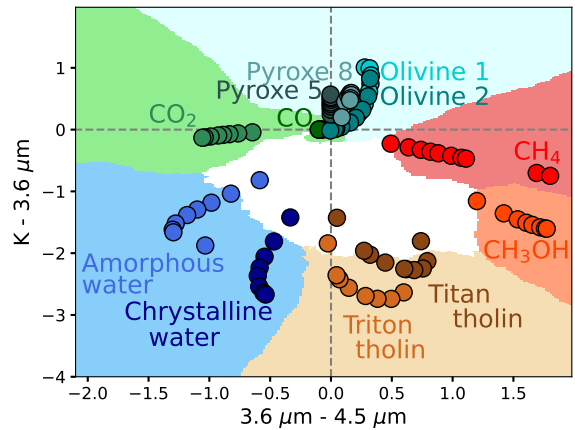}
\caption{ ``Compositional-clock'': color-color diagram for $K-3.6\,\mu$m versus $3.6-4.5\,\mu$m synthetic colors for pure materials (each plotted with a different color and symbol, and labeled). Colors were synthesized from spectral models using grain sizes from 10 -- 100 {\mum} in 10 {\mum} increments (\met\ also included 400 and 500 \mum). The larger the grain, the deeper the absorption band produced, and therefore grains of 10 \mum\ are nearest to the origin of the plot.}
\label{fig:clock_purematerial}
\end{figure}

In order to interpret the colors of our objects in terms of composition, we computed synthetic spectral models for the substances shown in figure \ref{fig:sinthetic_models} and table \ref{tb:synthetic_models}, using a range of grain sizes for each component. We convolved the synthetic spectral models with the Johnson-Cousins-Bessell standard filter system \citep{Johnson1964,Bessell1988,Bessell2005}, and the IRAC broadbands \citep{Fazio2004}, to derive synthetic photometry and colors for the synthetic spectra. Figure \ref{fig:clock_purematerial} illustrates the results for the $K-3.6\,\mu{\rm m}$ and $3.6\,\mu{\rm m}-4.5\,\mu{\rm m}$ color indexes (circles) for the pure materials as a color-color diagram. There are distinct regions that can be attributed to most of the materials we consider, with the color vs.~grain-size trends extending approximately radially in different directions from the origin for each material. Because of this layout, we informally call this diagram the ``compositional clock''. While most of the materials result in color trends at different `times' on the clock, \diox, CO and \nit, occupy nearly the same region (for clarity within the diagram, we did not represent the \nit\ in figure \ref{fig:clock_purematerial}), although for a given color the relevant grain size would be orders of magnitude larger for CO or \nit\ than for \diox. While the trends for these pure materials are simple, minor changes to the model assumptions, such as having a distribution of grain sizes or mixtures of different materials complicate these trends. In order to plot the region of influence for each pure material, we have applied the K-Nearest neighbors (KNN) method \citep{Hastie2001,James2013}, as implemented in {\ttfamily scikit learn} \citep{Pedregosa2011}, with $K=15$, and assigning weights proportional to the inverse of the distance from the query point. The method will classify each coordinate in the graphic considering the 15 nearest points provided by the models, filling the empty regions of the diagram. The selection of K was made by inspection of the results provided by different values. Small values of k will be too moldable leading to unstable decision boundaries, and large values of K will have smoother decision boundaries which mean lower variance but increased bias (also, computationally expensive). A good estimation is made from the square root of the total number of models (in our case, we have 207 points of synthetic materials, i.e., K $\sim$ 15). We inspected values of K between 10 and 25, which produced similar results, and therefore we chose 15 as a good compromise. Each colored region is dominated by a different material (water, complex organics, methanol, methane, silicates, and supervolatiles--CO and CO$_2$), while the white region is dominated by different mixtures of those materials, as explained bellow. 

\begin{figure*}[h!]
\gridline{\fig{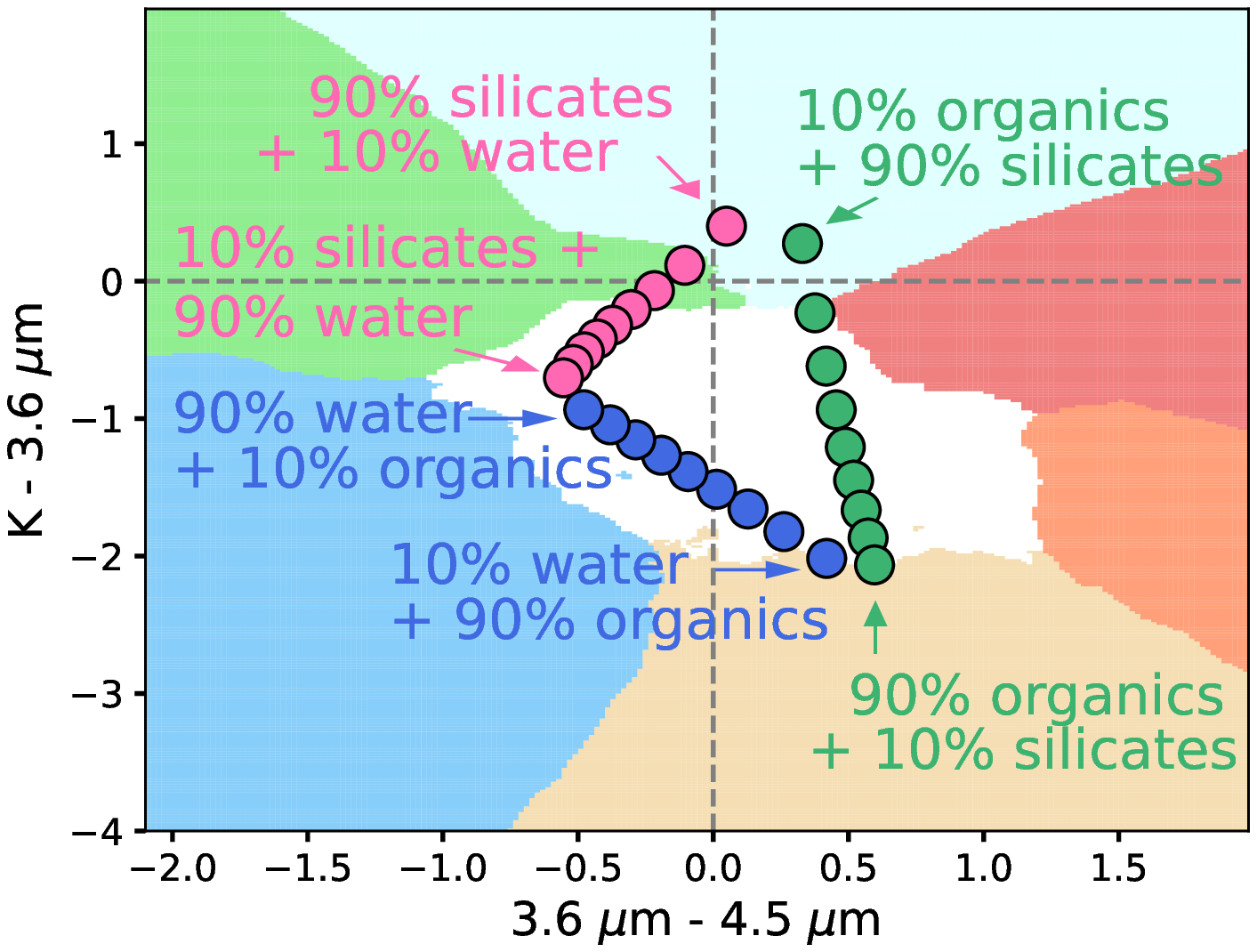}{0.5\textwidth}{(a) \footnotesize{Mixtures of two components.}}
          \fig{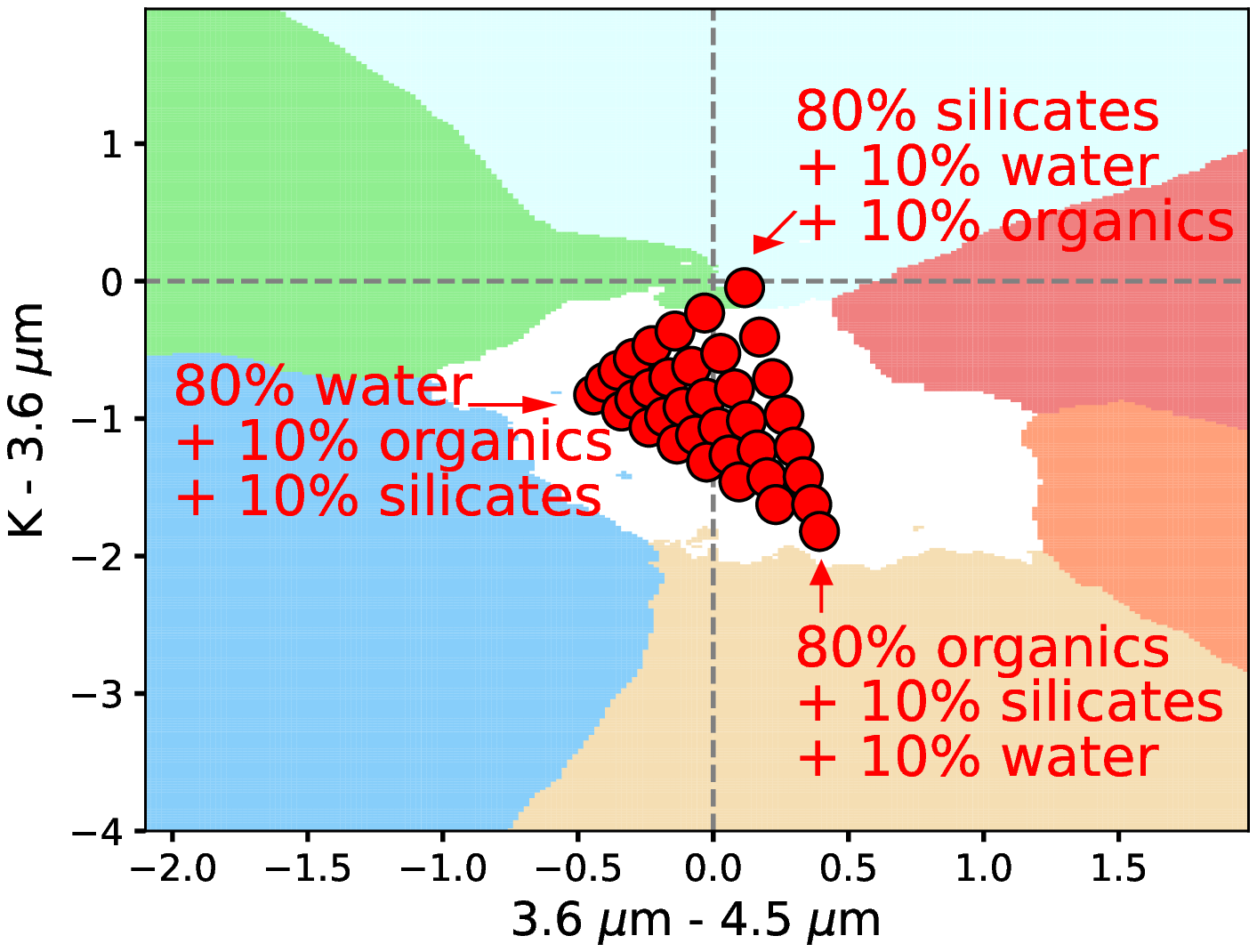}{0.5\textwidth}{(b) \footnotesize{Mixtures of three components.}}
          }
    \caption{\label{fig:mixtures} Same color-color space as in figure \ref{fig:clock_purematerial}, but showing trends for intimate mixing between two (left, panel a) and three (right, panel b) components. Here we have chosen amorphous \wat\ ice, Triton tholin, and olivine 1 as representative materials, and we show only the 10\mum\ grain size results. Models including crystalline and amorphous {\wat} have similar relative positions in the diagram. For comparison, we plotted the same colored regions as in figure \ref{fig:clock_purematerial}, to indicate those regions dominated by pure materials.}

\label{fig:mixture_models}
\end{figure*}

Surfaces of TNOs are very unlikely to be dominated by pure materials, so we now explore synthetic colors for intimate combinations between spectral models for three of our representative components (i.e., olivine 1, Triton tholin and amorphous \wat), with each component having a single grain size. The intimate mixtures have been carried out using Hapke scattering models \citep{Hapke1981,Hapke1993}. The results are shown in figure \ref{fig:mixture_models} for mixing between each pair of components (a) and between all three components (b). If additional combinations of grain sizes and/or mixtures between more than three components are considered, the region covered in the color-color diagram would expand, but still be bounded by the colors of the pure components. The figure shows that surfaces with significant fractions of multiple components can have colors that deviate from the trends seen for the pure materials in figure \ref{fig:clock_purematerial}, but also illustrates that the colors for such surfaces are still confined to certain regions of the diagram, and therefore can constrain the composition in useful ways, and exclude the presence of some components.


\subsection{Compositional interpretation of the TNO colors} 
\label{sec:test}

In order to explore the utility of the synthetic colors and color-color diagram discussed above, we now plot the measured colors for our targets and include the composition regions and color trends shown in figure \ref{fig:clock_purematerial} and figure \ref{fig:mixture_models}. To start, we focus on 12 objects with relatively well-understood spectral properties, shown in figure \ref{fig:test} (a). (Albedo spectra for these objects are given in appendix \ref{ap:plot_individual_objects}.) Fortunately, the colors of these objects span much of the color-color diagram, providing a fairly complete sample for testing the predictions based on the compositional-clock approach. The colors of all 12 objects appear to be roughly consistent with the predictions based on synthetic colors for the pure substances and simple mixtures between those substances. To obtain the proportion of each material for each object, we have implemented a routine that interpolate using the KNN method \citep{Hastie2001,James2013}. For each object (or observational point), we calculate the euclidean distance to search for the K-nearest points given by the synthetic models. The selection of K was made following the same procedure explained in section \ref{sec:Synthetic_color_indices}, which lead to K = 15. Then, we average the different proportion of each material for each point to provide an interpolation for each object. To calculate the errors in the proportions, we obtained the uppermost and lowest value of each of the two colors considering their error bars, which provides four new points for each object (these are arranged in the form of a cross around the central value). We applied the K-nearest method to calculate the proportion of different materials for each of those four new points. We obtained the difference between the proportion of each material given for the object and the proportion of each material given the four new points. For a more conservative perspective, we chose as the error the biggest difference from those obtained from opposite points in the cross shape.
 
\begin{figure*}
\gridline{\fig{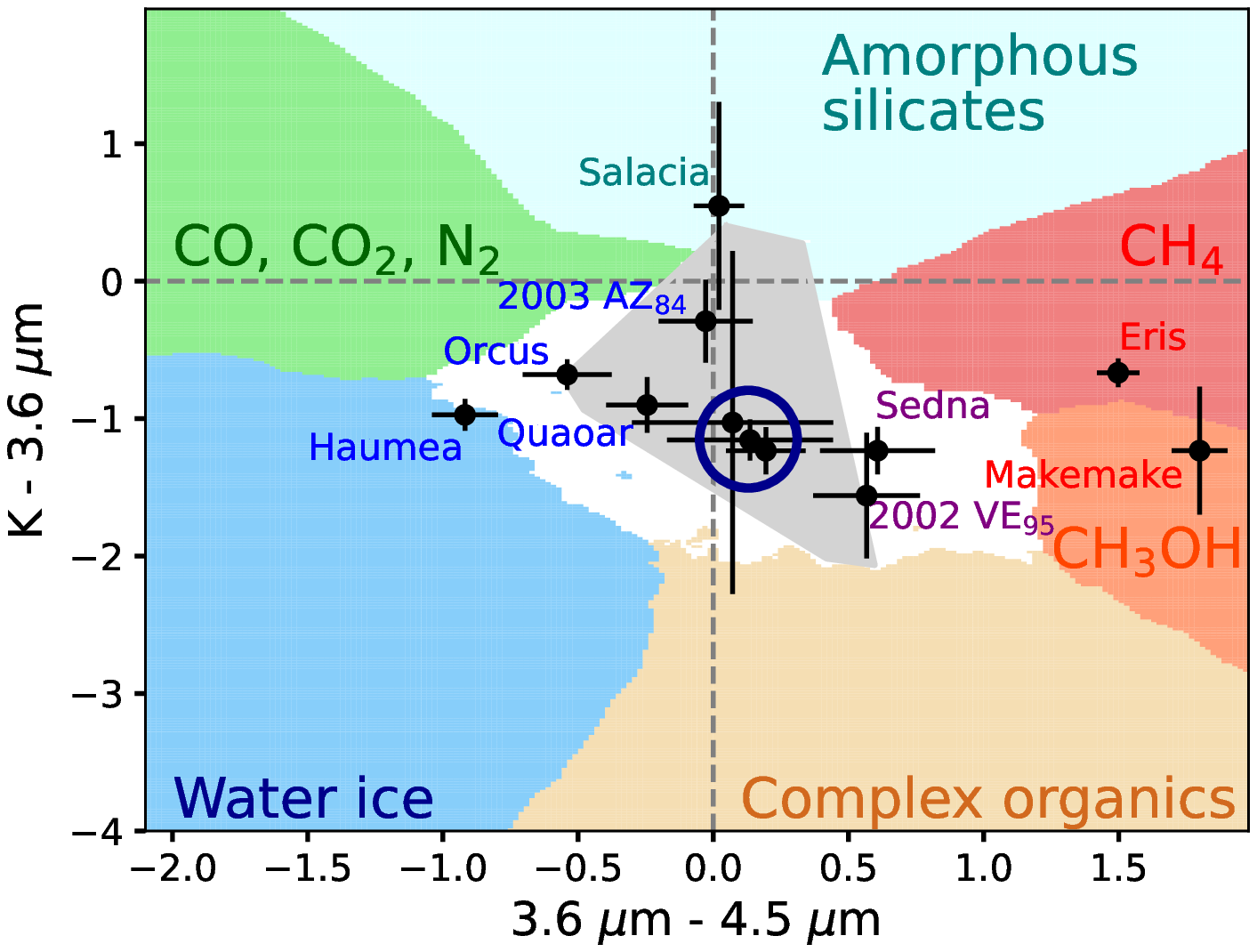}{0.5\textwidth}{(a) \label{fig:testing_CC}}
\fig{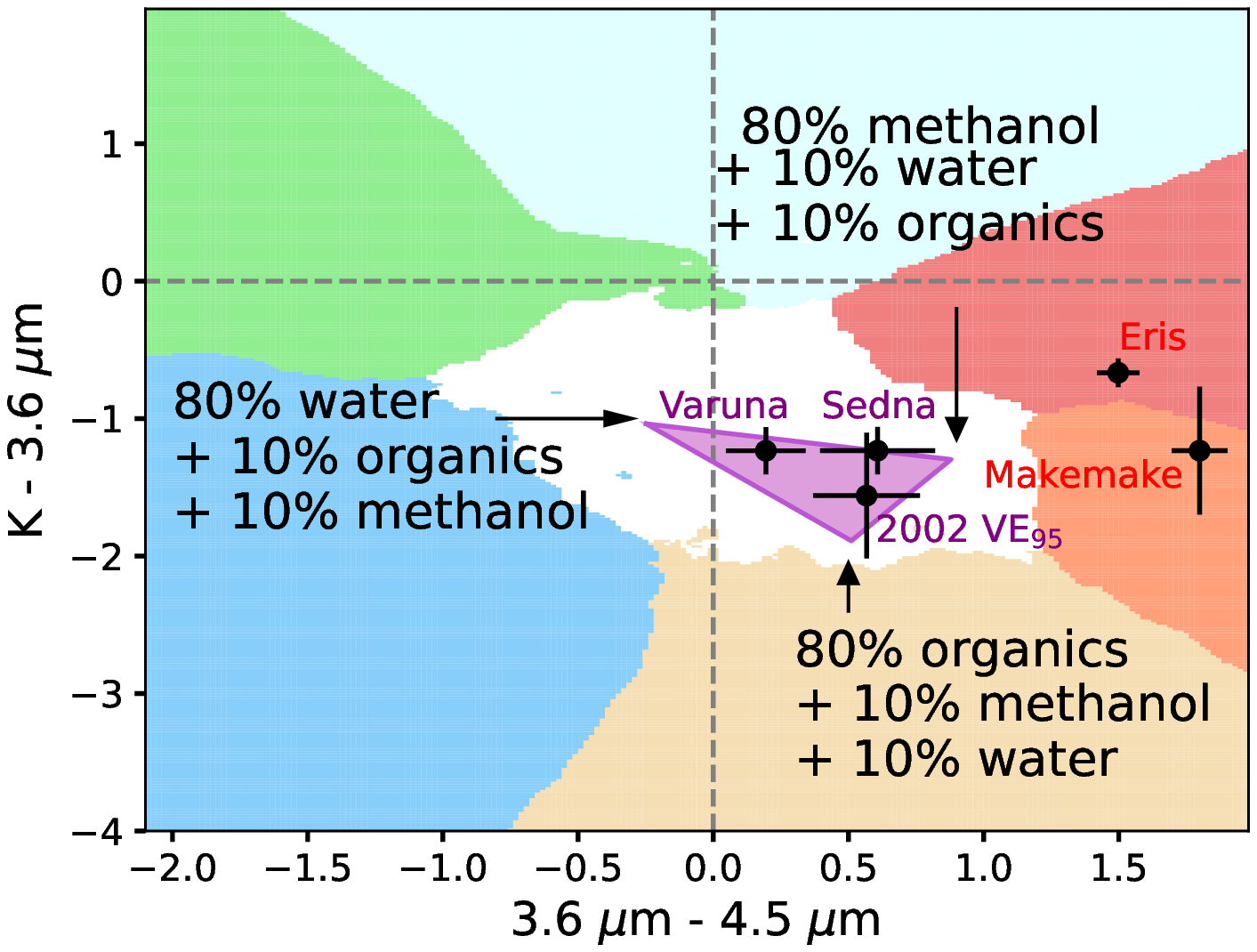}{0.5\textwidth}{(b) \label{fig:testing_CC_methanol}}
           }
 \caption{Color-color diagram, similar to figure \ref{fig:clock_purematerial}, including the shaded regions for to pure materials, and with measured colors of 12 TNOs with relatively well-characterized compositions from VNIR spectroscopic observations. Left: Target names are given in colors indicating whether the spectrum has previously been characterized as having \wat\ (blue), \met\ (red), or \nol\ (purple). Silicates (turquoise) has not being detected in Salacia, however, its flat spectrum and low albedo is consistent with our results shown in this diagram of a surface composition dominated by silicates. Objects surrounded by a blue circle are 2002~TX$_{300}$, Lempo and Varuna (from left to right). The grey polygon is bounding the regions shown in figure \ref{fig:mixture_models} for binary and ternary models of \wat, Olivine 1 and Triton tholin (i.e., \wat, silicates, and organics). Right: the colors for Varuna, Sedna and 2002~VE$_{95}$ are compared to ternary mixing models of amorphous \wat, Triton tholin and \nol\ with grain sizes of 10\mum.}
    
 \label{fig:test}
\end{figure*}

\textbf{Haumea} plots squarely in the \wat-ice pure region, as expected based on previous studies \citep[e.g.,][]{Trujillo2007,Merlin2007,Pinilla-Alonso2009}. We obtain a composition of $80\pm\%5$ \wat, $10\pm10\%$ silicates, $10\pm10\%$ organics.

\textbf{Quaoar} has been found to have large amounts of water ice in its surface according to absorption bands at 1.5, 1.65 and 2.0 \micron\ \citep{Jewitt2004,Schaller2007b,Guilbert2009,Dalle-Ore2009}. Another band was found at 2.2 \micron, which has produced some discussion among the different authors, suggesting that this band could be either due to ammonia hydrate (\ammon$\cdot$\wat) or \met. The model by \cite{Dalle-Ore2009} also included photometric data at wavelengths 3.6 and 4.5 \micron\ from Spitzer. They claim a surface composition of $\sim40\%$ \wat, $\sim10\%$ \met, and $\sim50\%$ complex organics. They also fit a different model with including up to $20\%$ \nit. Taking into account those percentages of water and complex organics, Quaoar should appear in a region in the color-color diagram slightly more downward and rightward than depicted. Considering our method, we obtain a composition of $60\pm10\%$ \wat, $20\pm10\%$ silicates, and $20\pm10\%$ complex organics. Note that for simplicity we have not included mixtures with supervolatiles; nonetheless, if there exists {\nit} on its surface, which overlaps the CO region in the $K-3.6$ {\micron} vs.~$3.6\,\micron-4.5\,{\micron}$ diagram, the data point for Quaoar can be displaced up and to the left, towards its position shown in figure \ref{fig:test} (a). The water percentage is consistent with the models provided by \cite{Dalle-Ore2009}, and the variation in the other materials could be due to the inclusion of \met\ and \nit\ on their mixtures. Overall, its position is consistent with the models provided by \cite{Dalle-Ore2009}.

\textbf{Orcus} has been studied by several authors, and most of them agree in a composition with a low percentage of water \citep{Fornasier2004b,deBergh2005,Delsanti2010}. Others have fit models with larger amounts of water, for instance, \cite{Trujillo2005} imposed an upper limit of $50\%$ water ice and \cite{Barucci2008} modeled the spectrum with $\sim40\%$ water ice. \cite{Demeo2010} also modeled the spectrum of Orcus with a larger amount of water (up to $70\%$), claiming that models from earlier papers \citep[i.e.,][]{Fornasier2004b,deBergh2005} and \cite{Trujillo2005} did not include the albedo of Orcus, which was published by \cite{Stansberry2008}. \cite{Demeo2010} also discussed that the difference between their model and the model in \cite{Barucci2008} may be due to a different blue component used to fit the data. Other ices, such as \met, \ammon\ and, C$_2$H$_6$, have been proposed for Orcus \citep{Trujillo2005,Delsanti2010}, and the presence of \diox\ is hypothesized to not exceed $5\%$ \citep{Demeo2010}. Our data, which include the $V$-band albedo, is consistent with a large amount of water on the surface of Orcus, as can be seen in figure \ref{fig:test} (a), our method produces a composition of $70\pm10\%$ \wat, $20\pm10\%$ silicates, and $10\pm10\%$ complex organics. Smaller amounts of water would be also consistent with our measurements if we include other volatiles (e.g., \diox, N$_2$) in the mixture.

Another object that has its surface distinctly dominated by water ice is \textbf{2002 TX$_{300}$}, which is part of Haumea's family \citep[see its spectrum in appendix \ref{ap:plot_individual_objects} and][]{Licandro2006,Barkume2008}. We obtain a composition of $30\pm30\%$ \wat, $30\pm50\%$ silicates, and $40\pm50\%$ complex organics, which is consistent with the spectroscopic measurements considering the error bars. However, we think that given the high proportion of water detected through its spectrum \citep[see e.g.,][]{Licandro2006}, our method does not seem to be very accurate for this specific object. This could be due to the extremely wide absorption bands produced at 3.6 and 4.5 \micron{} by the water, that could overlap one another, producing uncertainty on the photometric measurements when using wide pass-bands as those used in this work (see figure \ref{fig:sinthetic_models}). Also, the presence of \nol\ could move its position rightward as explain in section \ref{sec:methane}. (Note that NIR colors of this object were extracted from its spectrum, for which $p_{\rm V}$ is needed; therefore the large uncertainty on $p_{\rm V}$ translates into large uncertainty on the NIR colors).

Spectral models for \textbf{2003 AZ$_{84}$} have included $17-44\%$ of water and small amounts of organic compounds \citep{Barkume2008,Guilbert2009,Barucci2011}. In the color-color diagram it plots in a region intermediate between \wat-dominated and silicate-dominated colors, with a composition of $30\pm10\%$ \wat, $60\pm20\%$ silicates, and $10\pm10\%$ complex organics. If organics are present, these will not be greater than $20\%$, which is consistent with its slightly steeped spectral slope.

The spectrum of \textbf{Lempo} has been modeled with different proportions of \wat-ice, complex organics and silicates \cite[$5-35\%$, $10-65\%$ and, $0-85\%$, respectively;][]{Dotto2003,Merlin2005,Barkume2008,Guilbert2009,Protopapa2009}. The position of this object in the color-color diagram corresponds to a composition of $30\pm20\%$ \wat, $20\pm10\%$ silicates, and $50\pm10\%$ complex organics, in agreement with the proportions above mentioned. This is also supported by the $3.6\,\mu{\rm m} - 4.5\,\mu{\rm m}$ versus $3.6\,\mu{\rm m}$ and the $V-3.6\,\mu{\rm m}$ versus $J-3.6\,\mu{\rm m}$ diagrams (see section \ref{sec:other_cc}).

\textbf{Eris} and \textbf{Makemake} both have spectra dominated by \met\ and spectral models suggest particle sizes as large as 20 mm for Eris \citep{Licandro2006b,Merlin2009}, and 1 cm for Makemake \citep{Licandro2006c,Brown2007}, and both appear well inside the methane region. Note that Makemake is at the boundary between methane and methanol, where these classification methods do not provide accurate results. Also note that, comparing with figure \ref{fig:clock_purematerial}, its position is closer to points corresponding to larger methane particles than to those corresponding to methanol particles. Since it is clear that they have a composition dominated by pure methane and that the grain size is playing an important role, given their position in the compositional clock, for these two objects we have applied the KNN method to obtain the grain size of this material. For this calculation, we took only methanol models and added 6 models with larger grain sizes (specifically, 0.4, 0.5, 1, 1.5, 2, 2.5, 5, 10 mm). We did not include larger sizes because the position of the objects in the diagram clearly indicated sizes between 0.4 and 1 mm. In total we had 15 models from 0.01 to 10 mm to perform the KKN method, for which we chose K = 5 following the same explanation given in section \ref{sec:Synthetic_color_indices}. We obtain a particle size of $0.2\pm0.1$ and $1\pm0.4$ mm for Eris and Makemake, respectively. Even though our models result into particle sizes smaller than those obtained by the spectroscopic models, they manifest the necessity of using larger particle sizes. The difference between the measured grains sizes and those considered here can be explained due to the response of methane molecules at 3.6 and 4.5 \micron, which is higher than in the visible, producing wider absorption bands.

\textbf{Salacia} is an object with a very flat spectrum \citep{Pinilla-Alonso2008,Schaller2008}, for which no \wat-ice\ or other absorption features have been documented. \cite{Pinilla-Alonso2020} suggest that this object has a highly processed surface cover by a mixture of carbon and amorphous silicates. Its position within the silicate region of the color-color diagram results in a proportion of $10\pm20\%$ \wat, $90\pm20$ silicates and no organics, consistent with the lack of features that has been detected before. 

\textbf{Sedna} and \textbf{2002 VE$_{95}$} are objects for which water ice has been detected in their spectra, yet both objects appear well away from the \wat-ice region in the color-color diagram. However, their locations are clarified if we consider ternary mixtures of \wat, complex organics, and \nol, as shown panel b, figure \ref{fig:test}. Because \met\ and \nol\ occupy a similar region in the compositional clock, a similar location for each object is displayed if we use \met\ instead of \nol, which makes it impossible to distinguish between them using this method. However, we can indicate the presence of \met\ and/or \nol\ on the surface of the objects, and confirm their existence using VNIR spectroscopy, as \met\ and \nol behave quite different at those wavelengths (see figure \ref{fig:sinthetic_models}). Sedna's spectrum has been modeled by \cite{Emery2007} using VNIR spectroscopy and Spitzer measurements at 3.6 and 4.5\micron. They found the best model was given by a mixture of $50\%$ \met, $25\%$ complex organics, $15\%$ \wat\ and, $10\%$ \nit, which agrees with Sedna's position in the color-color diagram, that corresponds to a composition of $25\pm10\%$ \wat-ice, $50\pm10\%$ \met, and $25\pm10\%$ complex organics. The spectrum of 2002~VE$_{95}$ has been modeled using $12-13\%$ \wat, $10-12\%$ \nol, $64-78\%$ complex organics, and $0-11\%$ silicates \citep{Barucci2006,Barkume2008}. This is consistent with 2002~VE$_{95}$'s position in the color-color diagram, which corresponds to $20\pm10\%$ \wat, $40\pm10\%$ \nol, and $40\pm20\%$ organics.

\textbf{Varuna} has been found to have absorption bands related to water ice \citep{Licandro2001,Barkume2008}. Also, \cite{Lorenzi2014} fit its spectrum using two different mixtures: one composed of $25\%$ water, $25\%$ silicates, $35\%$ complex organics, and $15\%$ carbon; and a second one composed of $20\%$ water, $25\%$ silicates, $35\%$ complex organics, $10\%$ carbon and $10\%$ of \met. The position of this object in the compositional clock results in different proportion of water depending on whether the mixture is a combination of water-silicates-organics or water-methanol-organics. For the former the resulting composition is $30\pm10\%$ \wat-ice, $20\pm10\%$ silicates, and $50\pm10\%$ organics, while for the latter the resulting composition is $50\pm10\%$ \wat-ice, $20\pm10\%$ methane, and $30\pm10\%$ organics. A combination between both mixtures is in agreement with the spectroscopic models.

Based on the test cases presented, we find that overall the color-color diagram provides compositional information that is consistent with that derived from visible to near-IR spectral fits for well-characterized TNOs. Appendix \ref{ap:sample_composition} presents an individually exploration to understand the composition, based on IRAC data, for the large numbers of objects in our sample, that lack detailed characterization in the visible and near-IR wavelengths.

\section{Other color-color diagrams}
\label{sec:other_cc}

\begin{figure*}[h!]
\gridline{\fig{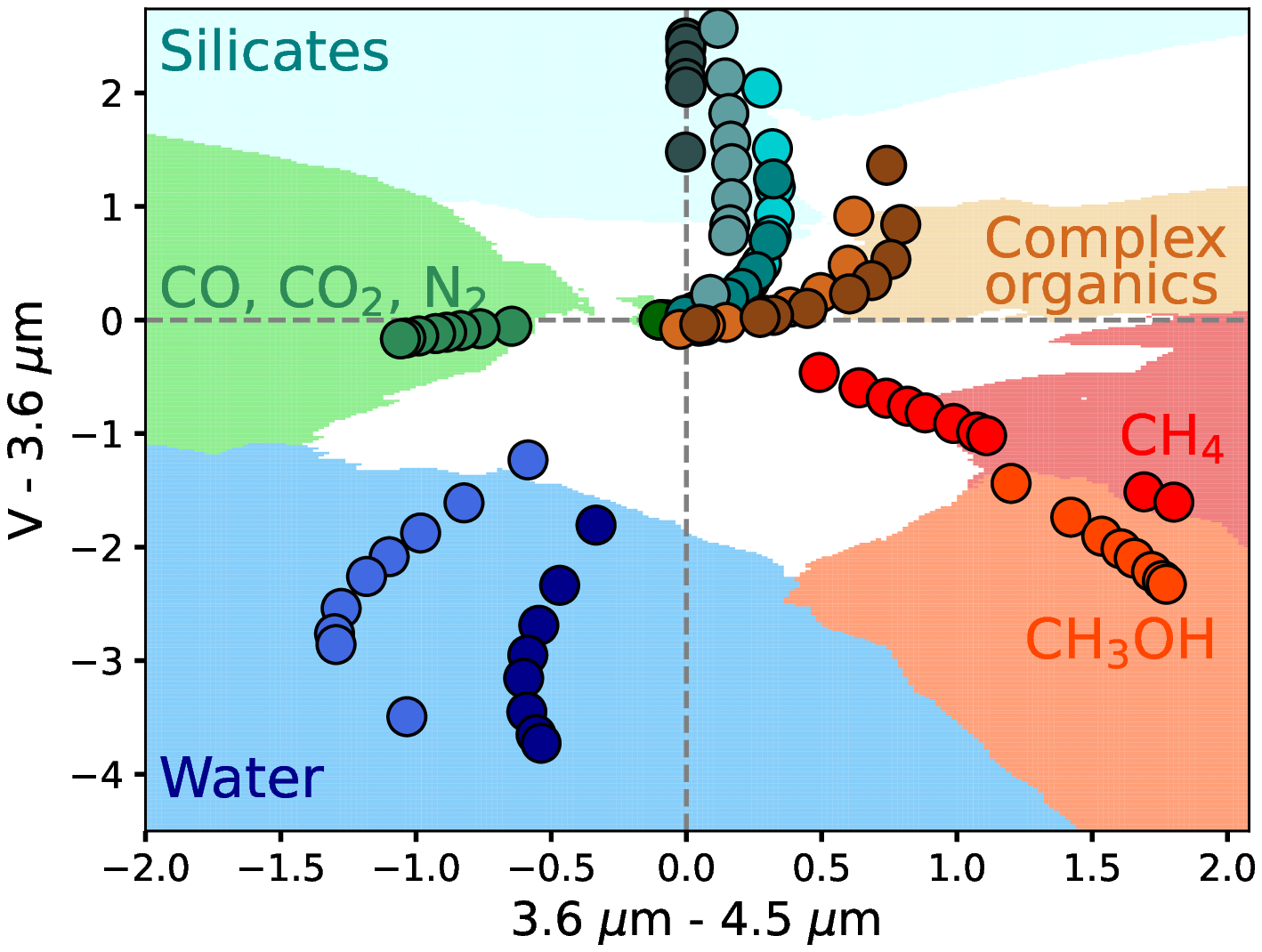}{0.4\textwidth}{(a) Pure materials. }
          \fig{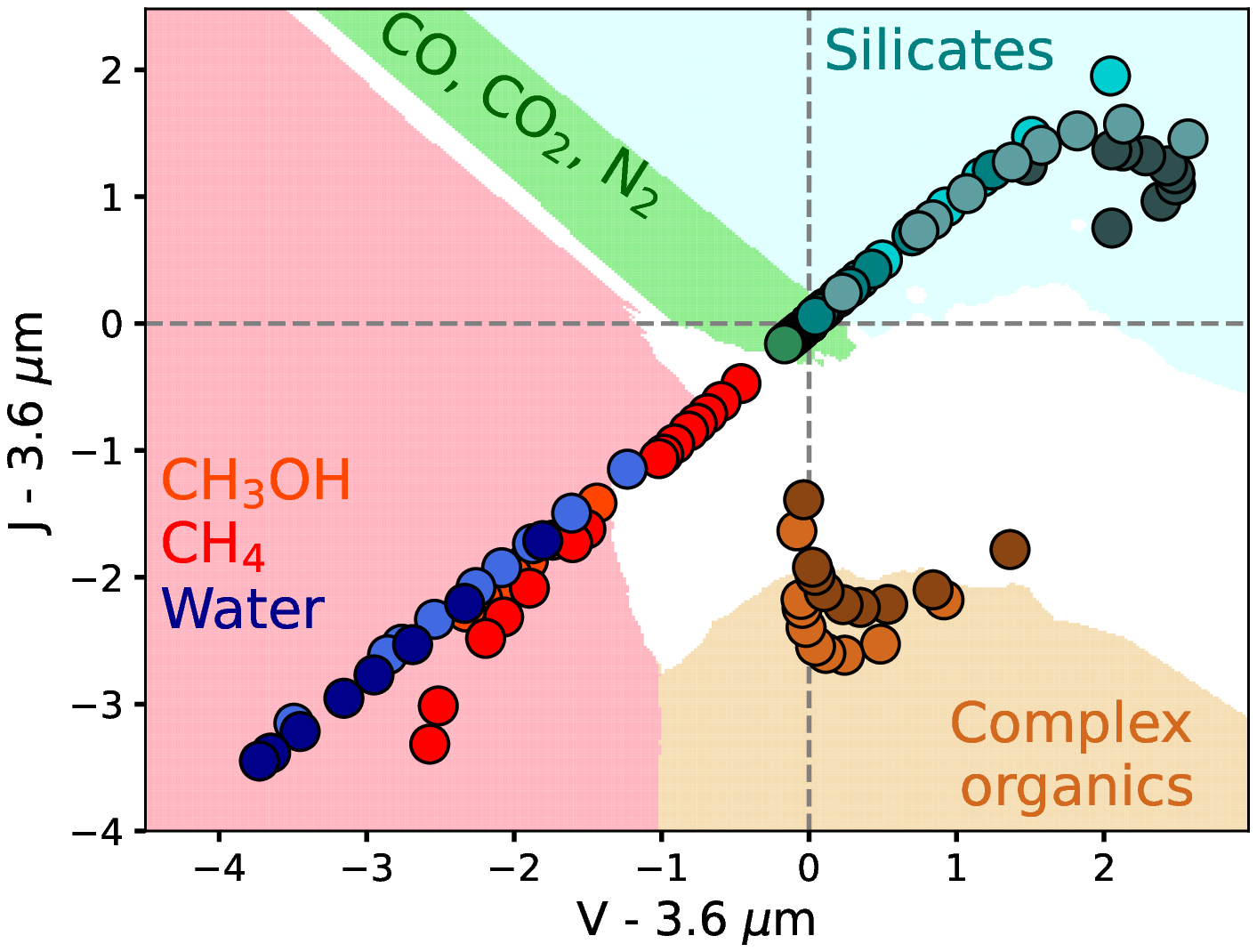}{0.4\textwidth}{(b) Pure materials. }}
    
     \gridline{ \fig{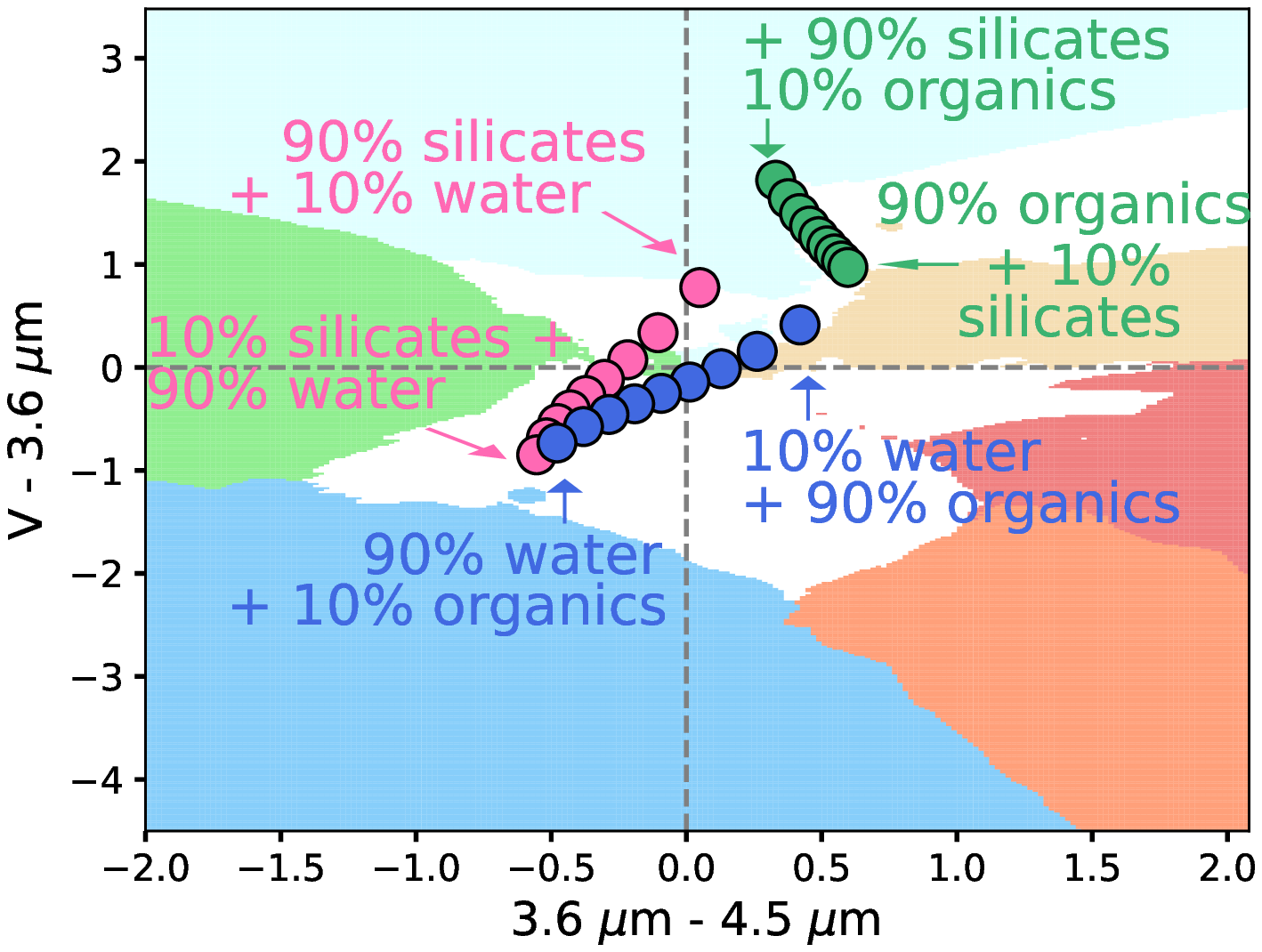}{0.4\textwidth}{(c) Mixture of two components.}
	\fig{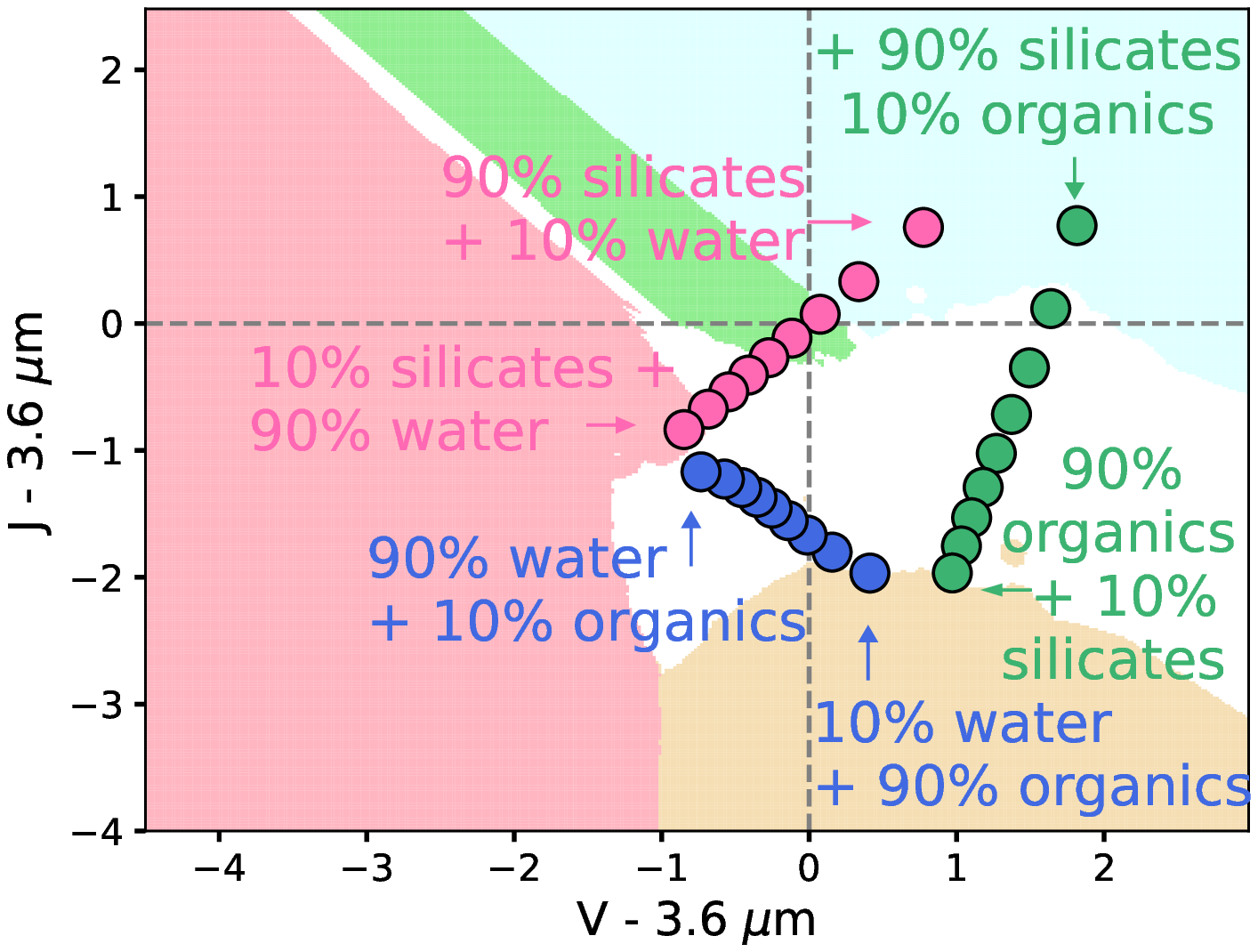}{0.4\textwidth}{(d) Mixture of two components.}
}
     \gridline{         \fig{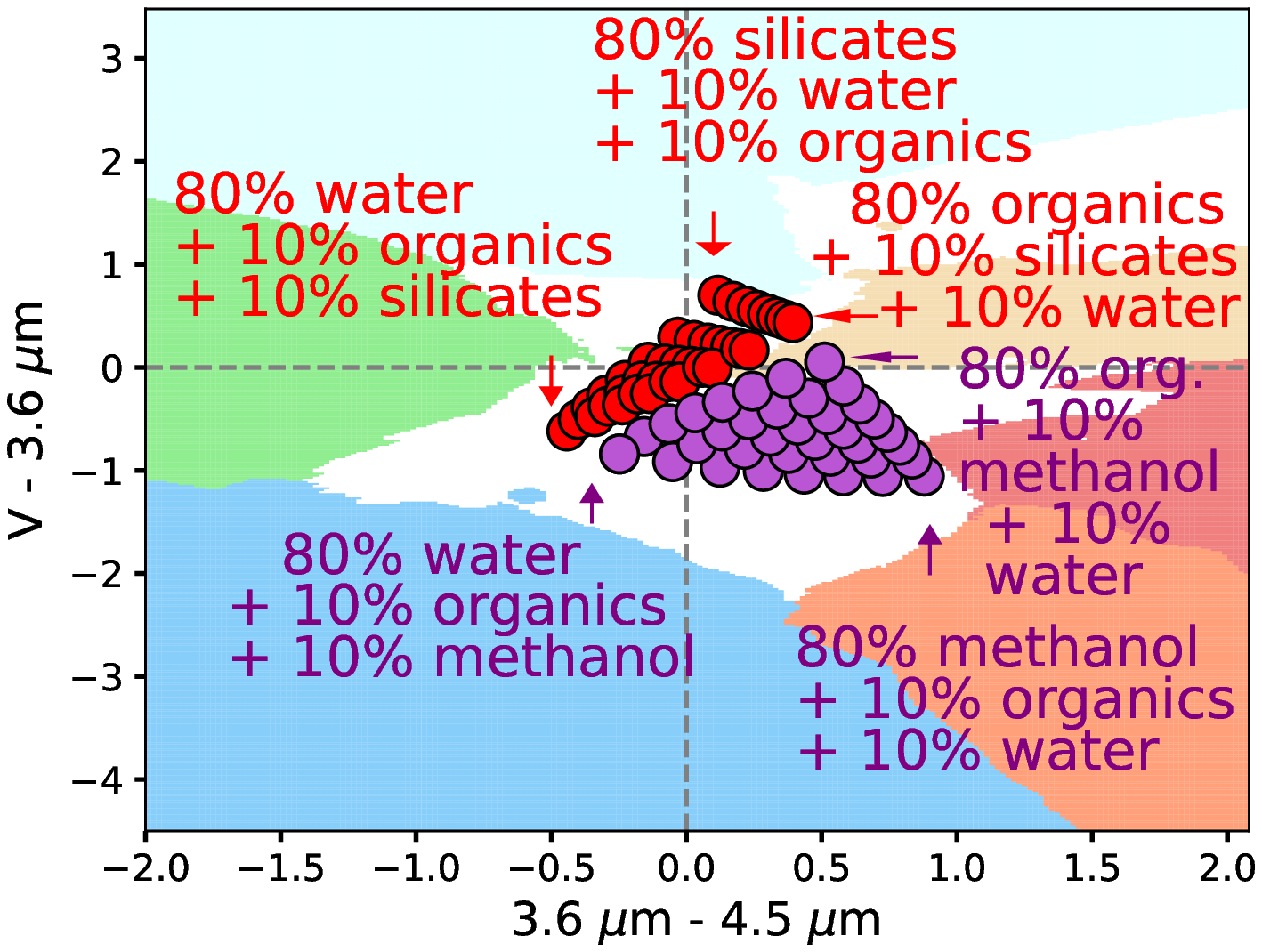}{0.4\textwidth}{(e) Mixture of three components.}
        \fig{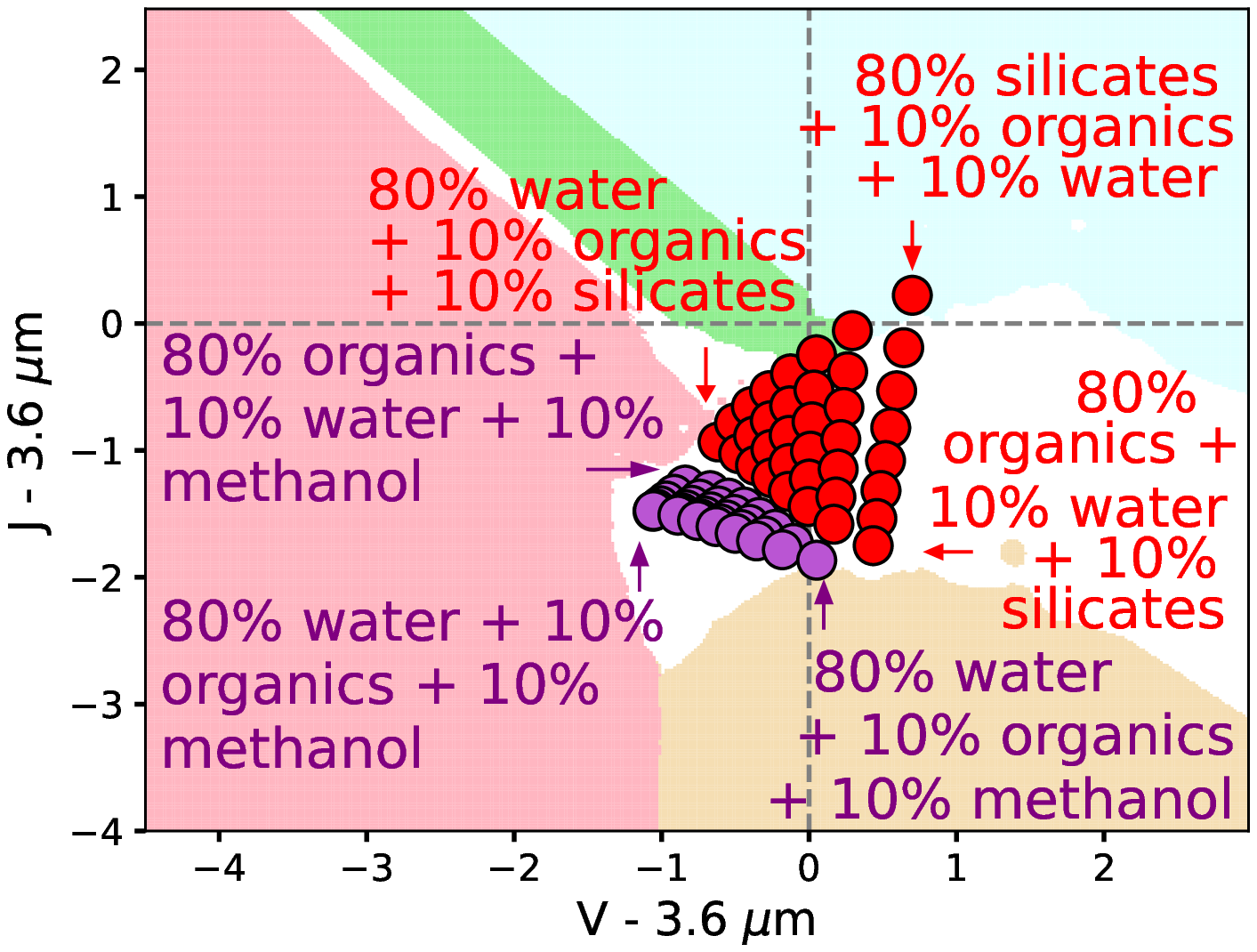}{0.4\textwidth}{(f) Mixture of three components.}}

      \caption{\label{fig:otherdiagrams}
 Left panels: $3.6\,\mu{\rm m}-4.5\,\mu{\rm m}$ versus $V-3.6\,\mu{\rm m}$ diagrams for pure, mixtures of two components and mixtures of three components (panels a, c, and e, respectively). shaded regions corresponds to a different material as indicated in panel (a) and equivalent to the compositional clock (figure \ref{fig:clock_purematerial}). Right panels: $V-3.6\,\mu$m versus $J-3.6\,\mu$m diagrams for pure, mixtures of two components and mixtures of three components (panels b, d, and f, respectively). In this case, since water, \met, and \nol\ occupy the same region, the pink shaded region correspond to those three material as indicated in panel (b).}
\end{figure*}

We have built other color-color diagrams in order to verify and/or identify the different components that dominate the surface of our sample (see figure \ref{fig:otherdiagrams}). The result of these diagrams are explained similarly to the compositional clock.

Of special interest is the diagram $3.6-4.5\,\mu$m versus $V-3.6\,\mu$m. Synthetic models of the components plotted in the compositional clock can be seen in figure \ref{fig:otherdiagrams}, left panels, where pure materials, mixtures of two components, and mixtures of three components are plotted. The region of influence of each material has been plotted using the KNN method as explained in section \ref{sec:Synthetic_color_indices}. The inconvenience of this diagram is that the organic materials occupied a very similar region to the silicates, so we can not distinguish between them. Also, comparing the three panels it can be seen that the mixtures with high percentages of silicates and/or organics overlap the regions occupied by models of pure materials, which makes that region unuseful for our purpose. This is noticeable, for instance, when using the K-nearest neighbor method, which produces blurred results in that region. However, there are two advantages. The first one is that, because there are more objects observed with visible colors than near infrared colors, we can analyze a larger sample than using the compositional clock. The second is that some organic materials change their position with respect to the compositional clock, therefore we are able to distinguish between complex organics and {\met}, and {\nol} compounds more easily. 

In figure \ref{fig:otherdiagrams}, right panels depict models by the colors $V-3.6\,\mu$m and $J-3.6\,\mu$m. As in the other cases, this figure shows models of pure materials, mixtures of two components, and mixtures of three components (panels b, d and f, respectively).  In this diagram, models with higher proportions of silicates are in the upper right quadrant and organic materials appear in a completely different region than the other materials. Meanwhile, super volatiles are nearly indistinguishable and \wat, \met, and {\nol} share the same region in the lower left quadrant. For that reason, we have applied the KNN considering one region for \wat, \met, and \nol\ models (pink shaded region), while the rest of the colors are equivalent to the other diagrams. This diagram is specially suitable for the identification of complex organics.

At VNIR wavelengths (up to 2.2 $\mu$m) complex organics and amorphous silicates present a very similar behaviour, with no absorption bands \citep{Cruikshank2005}. Therefore, we can only claim indications of objects with higher proportions of complex organics than silicates and vice-versa depending on the slope of the spectrum, i.e., objects with abrupt slope will be model using complex organics, while for objects with less abrupt slopes, the spectrum will be model by a combination of silicates and complex organics \citep{Emery2004}. However, the different behavior of silicates and complex organics at IRAC wavelengths (figure \ref{fig:sinthetic_models}) enables separation of these two materials in the compositional clock and the $J-3.6\,\mu$m vs $V-3.6\,\mu$m diagram (panels b, d, and f in figure \ref{fig:otherdiagrams}), allowing the identification of complex organics (due to absorption bands beyond 2.2 $\mu$m), and demonstrating that the wide band-pass of IRAC at 3.6 and 4.5 $\mu$ are a powerful tool to identify, for the first time, what the coloring agent that produces redness on TNOs is.

\section{The surface composition of our sample}
\begin{figure*}
\centering
\includegraphics[width=0.9\textwidth]{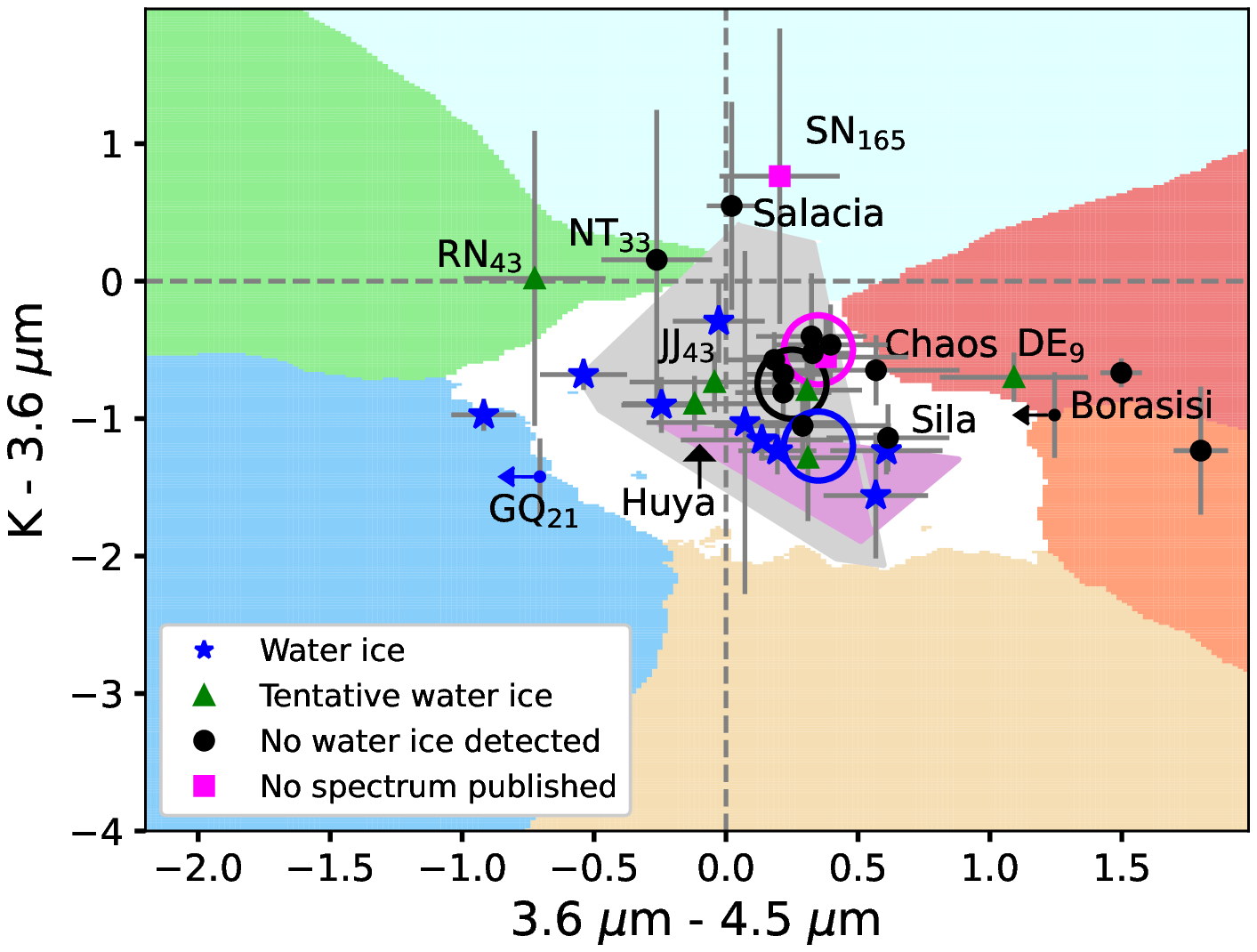}
\caption{\label{fig:water_detections} Compositional clock plotting observational data. For comparison between spectroscopic measurements and our results, symbols and colors represent the presence or lack of water detected in spectroscopic data as found in the literature: Blue stars show objects for which water have been identified before using published spectra up to $2\,\mu$m. Green triangles represent objects with tentative detection of water ice using published spectra up to $2\,\mu$m. Black circles represent objects with no identification of water ice in their published spectra. Pink squares show objects for which no spectra have been published. Pure materials are represented by shaded colored ovals that correspond to figure \ref{fig:clock_purematerial}. The shaded regions correspond to pure materials as labeled in figure \ref{fig:test} (a). The white region then is dominated by the combination of all mixture of different proportions, as explained in figures \ref{fig:mixture_models} and \ref{fig:test} (b). The grey shaded polygon is representing the binary and ternary models of different proportions of \wat-silicates-ornganics plotted in figure \ref{fig:mixture_models}. The purple triangle is representing models of different proportions of \wat-\nol-organics. Objects within the black circle are the following: 2002~AW$_{197}$, 2005 RN$_{43}$, 2002 UX$_{25}$, and 2004 TY$_{364}$ (from top to bottom). Objects within the blue circle are: 2000 GN$_{171}$ and 2002 TC$_{302}$ (from top to bottom). Objects within the pink circle are the following: Varda, 2000 PE$_{30}$, 2001 UR$_{163}$, and 2004 GV$_{9}$.} 
\end{figure*}

We have demonstrated the consistency of our photometric measurements with the presence of \wat-ice and other materials such as {\met}, {\nol}, complex organics and amorphous silicates for objects that present clear signature of those materials in their VNIR spectra (up to 2.0 $\mu$m). Our idea within the following subsections is to focus on the different materials that can be detected using the IRAC colors.

\subsection{Water ice}
\label{sec:detection_water}

\begin{figure*}
\gridline{\fig{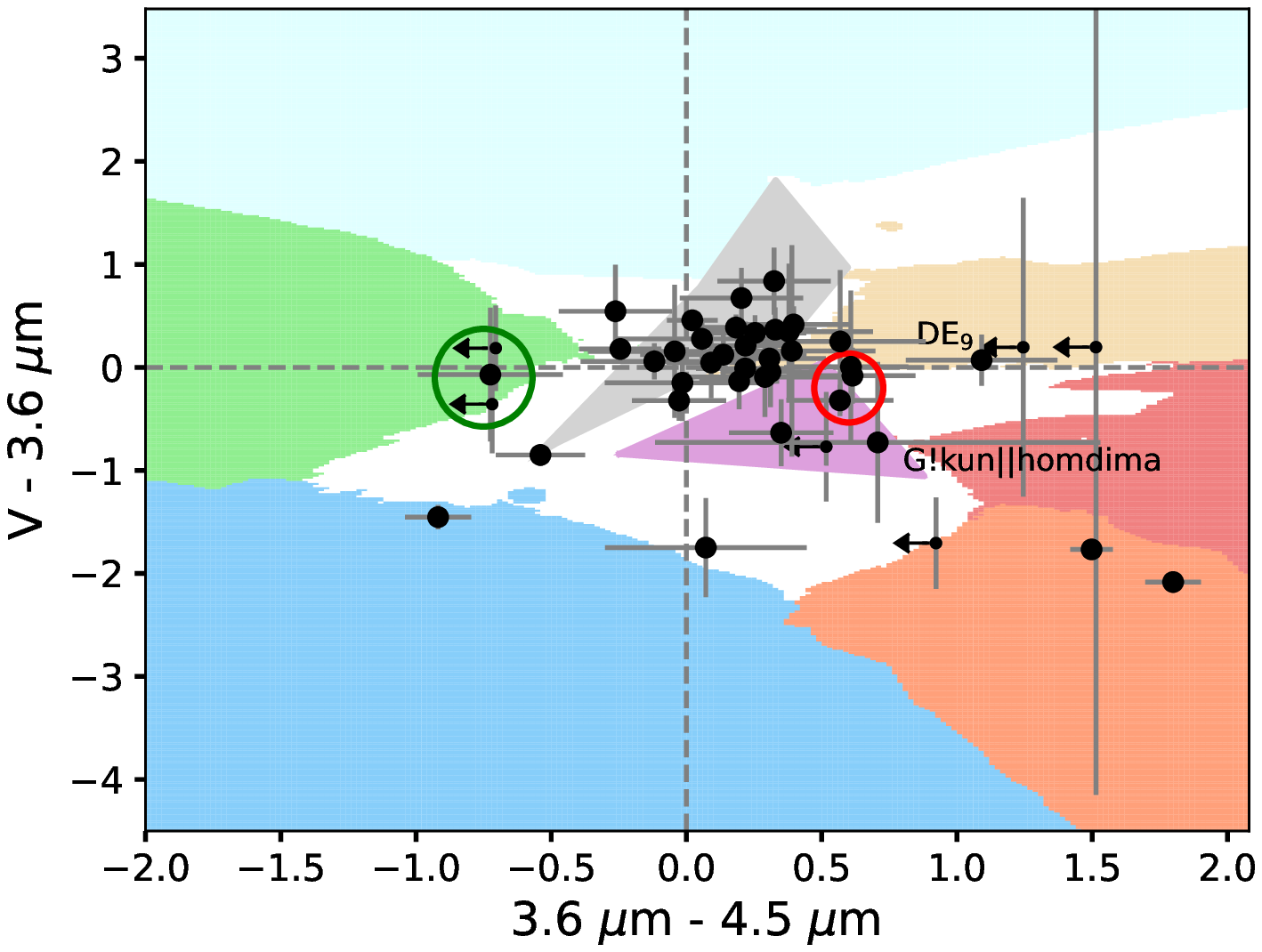}{0.5\textwidth}{(a)}
         \fig{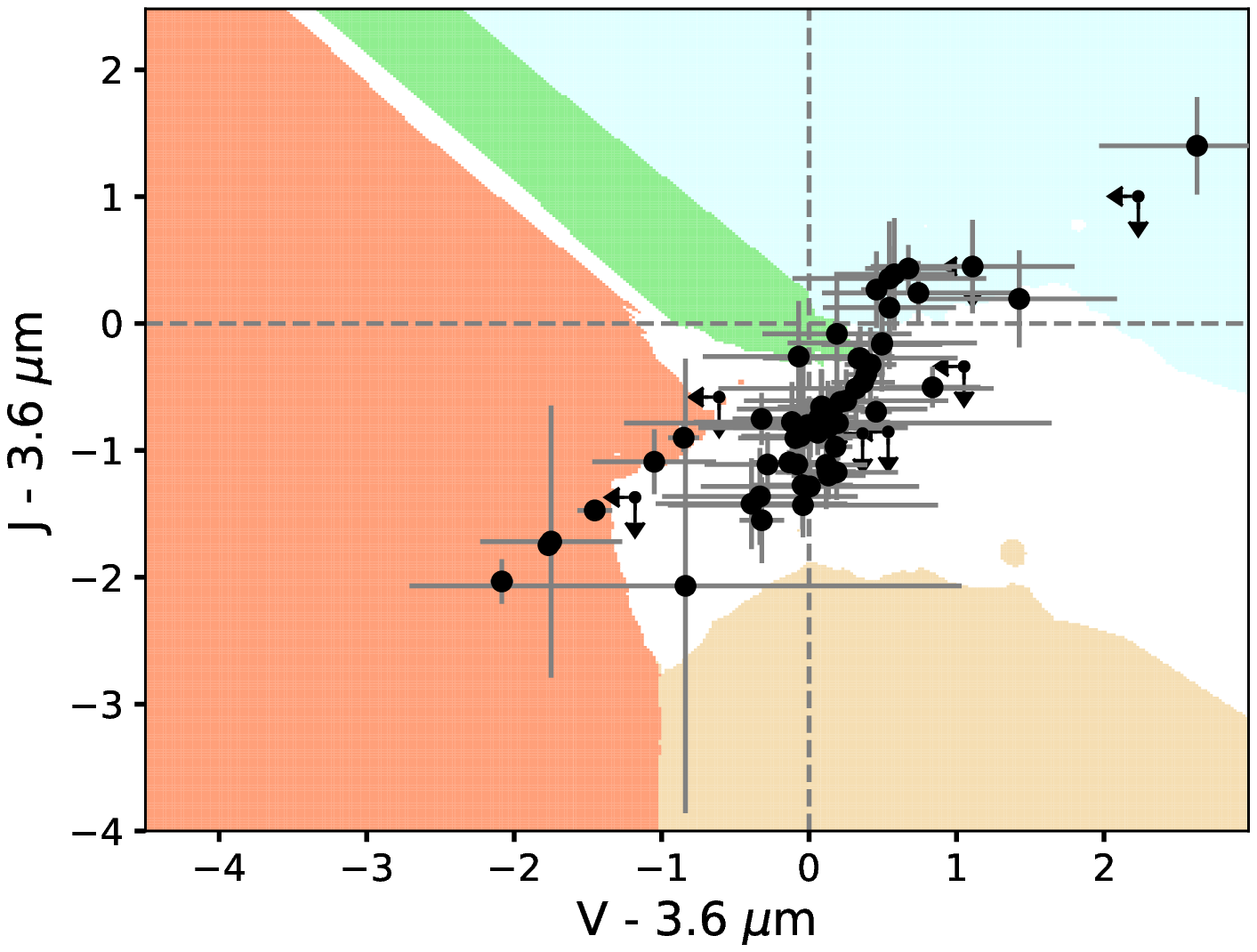}{0.5\textwidth}{(b)
                }}
  \caption{\label{fig:todos}(a) Diagram $3.6-4.5\,\mu$m versus $V-3.6\,\mu$m, with black points representing our sample. Objects within the red circle are Sedna, Sila-Nunam, and 2002~VE$_95$ (from top to bottom). Objects within the green circle are 1996~GQ$_{21}$, 2005~RM$_{43}$, and 2004~PG$_{115}$ (from top to bottom). shaded regions corresponds to a different materials as indicated in figure \ref{fig:otherdiagrams} (a) and equivalent to the compositional clock (figure \ref{fig:clock_purematerial}). (b) Diagram $V-3.6\,\mu$m versus $J-3.6\,\mu$m. In both diagrams, the shaded regions are the same explained in figure \ref{fig:otherdiagrams}. The grey shaded polygon is representing the binary and ternary models of different proportions of \wat-silicates-ornganics plotted in figure \ref{fig:mixture_models}. The purple triangle is representing models of different proportions of \wat-\nol-organics.}

\end{figure*}

In figure \ref{fig:water_detections} we plotted our sample as a function of the detection of water ice in the VNIR spectra published in the literature \cite[up to 2.0 $\mu$m; e.g.,][]{Barucci2011,Licandro2001,Lorenzi2014}. Blue stars represent objects for which water detection is already known from the spectra, and were used to test our method (see section \ref{sec:test}). This includes Sedna and 2002 VE$_{95}$, for which \wat\ has been detected using spectroscopic studies and, as we explained at the end of section \ref{sec:test}, the presence of \nol\ move them rightward within the diagram. One object we have not yet discussed is 1996 GQ$_{21}$ because we only obtained an upper limit at 4.5\micron{} (a more detailed discussion can be found in appendix \ref{ap:sample_composition}). However, this limit constrains the region in which the object is localized within the diagram, eliminating the possibility of having other ices but water. 

Objects with tentative detection of water in the VNIR spectra are represented by green triangles. As can be seen, with the exception of 2005 RM$_{43}$ and 1999 DE$_9$, all of them fall within regions where \wat\ has to be part of the composition. The position of 2005 RM$_{43}$ in figure \ref{fig:water_detections} is quite interesting, as this region is the one dominated by CO$_2$. However, since the errorbars are quite large, we have considered a mixture of organics-silicates-water to obtain a composition, resulting in $50\pm40\%$ \wat-ice, $50\pm40\%$ silicates, and no organics. Large amounts of \wat\ should be detected by VNIR spectroscopy, which is not the case. The presence of CO$_2$ in the mixture could be placing the object in that region of the diagram while decreasing the amount of water. Additionally, as we point out in section \ref{sec:methane}, other color-color diagrams support the possibility of this object having CO$_2$. On the other hand, the presence of \wat\ for 1999~DE$_9$ has been discuss by several authors with no clear agreement. We conclude that the position of this object within the diagram is not consistent with the presence of \wat\ but \nol\ on its surface \cite[see also][]{Jewitt2001}. Considering a mixture of \wat-\nol-organics, the resulting proportion for this object is $20\pm10\%$ \wat, $60\pm10\%$ \nol, and $20\pm10\%$ organics. These two objects are deeply discussed in section \ref{sec:methane}.

Black circles in figure \ref{fig:water_detections} represent objects for which no \wat\ has been reported before. Our results show that 2004 NT$_{33}$ and the groups formed by 2000 GN$_{171}$, 2002 AW$_{197}$, 2002 UX$_{25}$, 2004 TY$_{364}$, 2005 RN$_{43}$, Varda, 2001 UR$_{163}$, and 2004 GV$_9$ marked by a black rectangle and a pink circle, are consistent with $\sim20\%$ \wat-ice within its composition. The resulting proportions for each object are given in table \ref{ap:sample_composition}.

For those objects that have published spectra in the literature, we are able to say that our results are in agreement with the spectra (see appendix \ref{ap:sample_composition} for an individual explanation on each object). In summary, IRAC colors are highly sensitive to the presence of \wat, thus using the $3.6\,\mu{\rm m}-4.5\,\mu{\rm m}$ diagram, with a total of 30 objects (32 if counting the upper limits), 26 objects (86\%) present errors bars within $1\sigma$ independent probability consistent with the presence of \wat\ on their surface, with  22 (73\%) representing the $1\sigma$ compound probability.

This conclusion is also consistent with the other two color-color diagrams (figure \ref{fig:otherdiagrams}). In figure \ref{fig:todos}, we have plotted the shaded regions corresponding to each pure material (as described in figure \ref{fig:otherdiagrams}), with the white region representing binary and trinary mixtures, and the sample of objects represented by  black circles. From panel (a), we obtain that 30 of a total of 37 objects (81\%) present errors bars within $1\sigma$ independent probability consistent with the presence of \wat\ on their surface, with  26 (70\%) representing the $1\sigma$ compound probability.  From panel (b), we obtain that 50 of a total of 59 objects (85\%) present errors bars within $1\sigma$ independent probability consistent with the presence of \wat\ on their surface, with 36 (61\%) representing the $1\sigma$ compound probability. 

\subsection{Complex organics}

As we have been discussing, one of the potential of IRAC using the compositional clock diagram is that, in order to obtain colors that occupied the center of the diagram, it is required to include complex organics (e.g., tholins) within our models. Tholins have been used for modeling the spectra of different objects but have not been detected before, as they do not produce absorption band at the VNIR spectra. Our method provide a high level of confidence that the surface of most objects within our sample is composed by mixture that include complex organic materials such as tholins \cite[e.g.,][]{Khare1993,Materese2014,Materese2015}. For instance, in figure \ref{fig:water_detections}, we obtain that 80\% of the sample present error bars within $1\sigma$ independent probability consistent with the presence of organics material on their surface composition with the same percentage (80\%) representing the compound probability. From figure \ref{fig:todos} (b), we obtain that $90\%$ of the sample present error bars within $1\sigma$ independent probability consistent with the presence of organics material on their surface with 63\% representing the compound probability. We do not include statistics from figure \ref{fig:todos} (a), since both organics and silicates occupy similar regions, and does not provide a clear separation between them. Our preferred statistics are the ones provided by the compositional clock ($K-3.6\,\mu{\rm m}$ versus $3.6\,\mu{\rm m}-4.5\,\mu{\rm m}$ diagram), since it is the one that maximize the range difference on colors depending on the material.

Specifically in figure \ref{fig:water_detections}, the groups marked by a black rectangle and a pink circle require a percentage between $10-60\%$ of complex organics. All of them present an absorption identification at 3.6 $\mu$m with respect to 4.5 \micron\ and an abrupt spectral slope in the visible (see figure \ref{fig:color_index} and appendix \ref{ap:plot_individual_objects}, with the exception of 200 PE$_{30}$, which has no spectrum published). Huya, Quaoar, and 2007 JJ$_{43}$ require between $10-40\%$ of complex organics; however, this amount of complex organics might be hide in their spectra due to the large amount of \wat-ice, which flattens the spectral slope.

\subsection{Supervolatiles, \met, and/or \nol}
\label{sec:methane}

\begin{figure*}
\centering
\includegraphics[width=\textwidth]{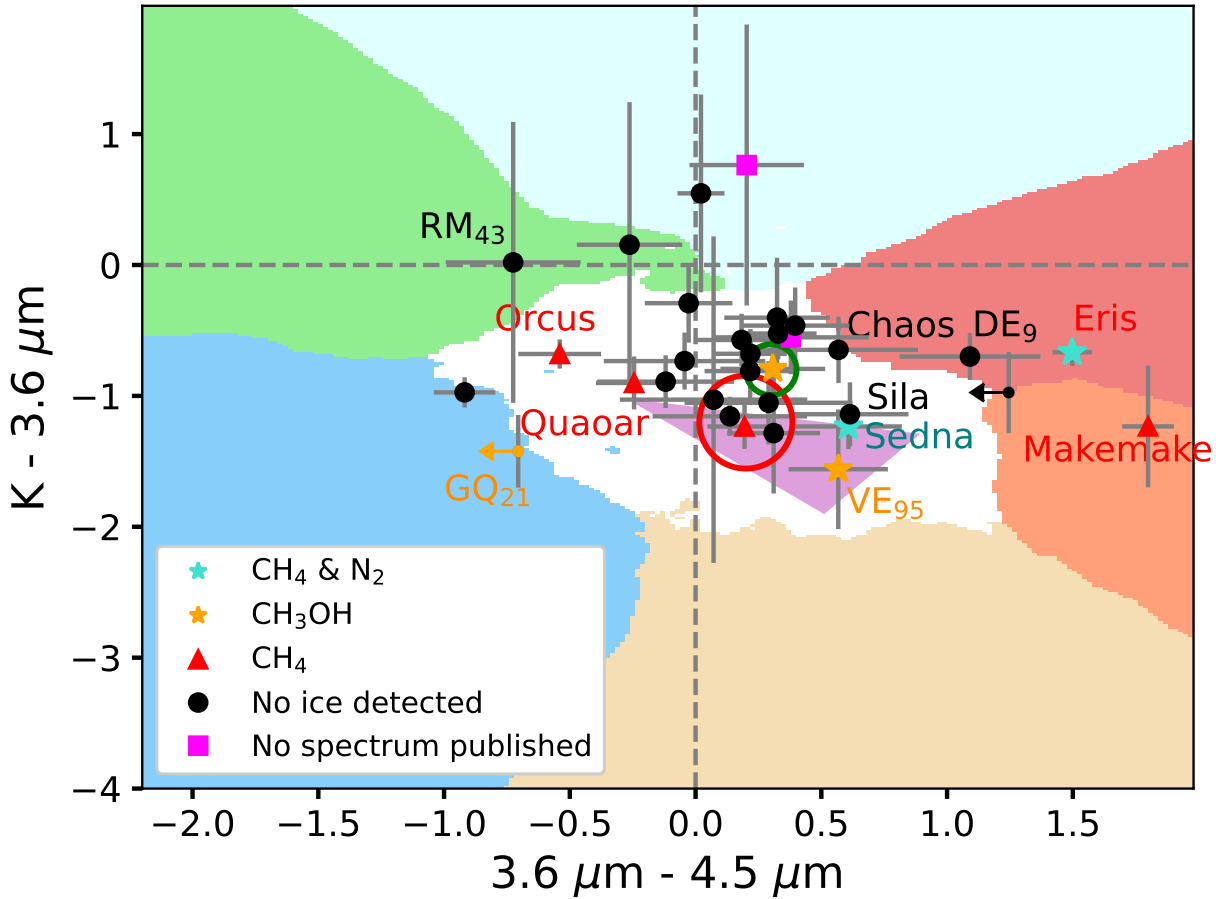}
\caption{\label{fig:ice_detections} Compositional clock with the sample highlighting objects for which bands related to ices have been identified in spectroscopic measurements up to 2.3 $\mu$m. Turquoise stars depict objects that have CH$_4$ and, N$_2$ in their spectra. Orange stars represent objects with possible detection of CH$_3$OH. Red triangles represent objects with {\met} detections. Black circles represent objects with no identification of any ice in their spectra. Pink squares show objects for which no spectroscopic data have been published. The orange star surrounded by the green circle is representing 2004 TY$_{364}$. Objects within the red circle are as follows: 2002 TX$_{300}$, Lempo, Varuna, 2000 GN$_{171}$ and 2002 TC$_{302}$ (from left to right). The red, orange, and green shaded regions represent the regions of influence of synthetic models of different grain sizes of pure methane, methanol, and supervolatiles (\nit, CO,...), respectively. The white region represents the region of influence of synthetic models with different proportions of mixtures of \wat-\met-organics, and \wat-silicates-organics. The purple triangle is bounding the theoretical points of different proportions of \wat-\nol-organics.}
\end{figure*}

This section is dedicated to the identification of compositions consistent with the presence of supervolatiles, \met, and/or \nol. Figure \ref{fig:ice_detections} shows our sample highlighting the regions dominated by \met, \nol, \diox, CO and \nit, and mixtures that contain these components. Turquoise stars show objects with detection of both, CH$_4$ and N$_2$, as found in the literature from VNIR spectra. The two objects, Sedna and Eris, with this composition were discussed in section \ref{sec:test} where we explained the good agreement between the results from spectra and ours. Red triangles represent those objects for which CH$_4$ has been detected from their spectra. As discussed in section \ref{sec:test}, Makemake and Eris, both have detection of methane with grain sizes over 0.4 mm in their spectra and appear in a region not only dominated by this component but also in which a larger particle size is necessary. The other two objects, Quaoar and Orcus, with \met\ detection in their spectra are located in a region that seems to contradict this detection. In these specific cases, the large amount of \wat-ice is hiding the detection of \met\ using this method, and therfore we do not exclude \met\ as part of their composition. Additionally for Orcus, its position also suggest the presence of \diox\, which has been suggested by \cite{Demeo2010}.

Objects represented by an orange star are those for which {\nol} has been tentatively suggested. For 2002~VE$_{95}$ the detection of \nol\ is in clear agreement with its position in figure \ref{fig:ice_detections}. As explained in section \ref{sec:test}, we obtained a composition that include $40\pm20\%$ of complex organics considering a mixture of \wat-\nol-organics. The position of 2004~TY$_{364}$ is not as clear. Considering models with a mixture of \wat-\nol-organics, we obtain a proportion of $50\pm10\%$, $30\pm10\%$, and $20\pm10\%$, respectively. However, such amount of water should be noticeable in its spectrum, and that is not the case. On the other hand, considering a mixture of \wat-silicates-organics, we obtain a proportion of $20\pm10\%$, $50\pm10\%$, and $35\pm10\%$, respectively. The later is more in agreement with its spectra, although a combination between both mixture would be a very luckily situation. The other two interesting objects in this figure are Sila-Nunam, and 1999 DE$_9$, whose position is consistent with the presence of \met\ or \nol. Due to the small size of the binary system \cite[around 300 km;][]{Vilenius2012,Lellouch2013}, it is unlikely that these objects, Sila and Nunam, possess \met\ on its surface, being more realistic to think that their position in figure \ref{fig:ice_detections} is due to \nol. We obtain a composition of $30\pm10\%$ \wat, $50\pm10\%$ \nol, and $20\pm10\%$ organics for Sila-Nunam and $20\pm10\%$ \wat, $60\pm10\%$ \nol, and $20\pm10\%$ organics for 1999 DE$_9$. A more detailed explanation is given in appendix \ref{ap:sample_composition}.

A composition including \met\ and \nol\ on the surfaces of Sedna and 2002~VE$_{95}$, respectively, is consistent also with their position in in figure \ref{fig:todos} (a), as explained in section \ref{sec:test}. 

There are three objects objects (Lempo, 2000 GN$_{171}$, and 2002 TC$_{302}$) depicted by black points within the red circle that are very close or within the purple triangle, for which their position could indicate the presence of \nol. However, for them to have \nol\ and be located on that region, they would also require a high percentage of \wat\, which should have been detected on their spectra. Specific proportions for each of them are given in appendix \ref{ap:sample_composition}. On the contrary, the object 2002 TX$_{300}$, which is right next to the the red circle, is known for having large amounts of water on its surface \citep{Licandro2006}. Considering a mixture of \wat-\nol-organics, we obtain a composition of $60\pm30\%$ \wat, $20\pm10\%$ \nol, and $30\pm30\%$ organics. Therefore, the existence of \nol\ on the surface of 2002~TX$_{300}$ could be displacing this object to the right of the diagram.

Finally, as we mentioned in section \ref{sec:detection_water}, in figure \ref{fig:ice_detections}, 2005 RM$_{43}$ shares the region occupied by CO$_2$. Although the error bars are large and could place this object out of this region, the diagram $V-3.6\,\mu{\rm m}$ versus $J-3.6\,\mu{\rm m}$ also support this interpretation (see panel a in figure \ref{fig:todos}), where there are four objects that share the region occupied by CO$_2$: 1996 GQ$_{21}$, 2002 GV$_{31}$, 2004 PG$_{115}$ and 2005 $RM_{43}$. Although three of these objects have measurements with only upper limits in the $3.6\,\mu{\rm m}-4.5\,\mu$m color, these limits constrain the objects to regions dominated by supervolatiles (\diox, CO, \nit) and/or water. On the other hand, the error bars for the $V-3.6\,\mu$m color constrain the region to one dominated by the supervolatiles. Our measurements suggest the possibility of these objects containing CO$_2$ and \wat. A more detailed explanation and specific proportions considering a mixture of \wat, silicates and organics are given in appendix \ref{ap:sample_composition}. 

Summarizing, seven of 30 objects within the compositional clock (Eris, Makemake, Sila-Nunam, Sedna, 2002 VE$_{95}$, 2004 TY$364$, and 2002 TX$_{300}$), 32 including those with upper limits, are consistent with a composition that includes \nol\ and/or \met on their surfaces. Considering other color-color diagramas (figure \ref{fig:todos}) a total of four objects are consistent with a composition that includes \diox.

\subsection{Silicates}

In the compositional clock, the region dominated by the amorphous silicates (see figure \ref{fig:water_detections}) is occupied by two objects: Salacia and 1998 SN$_{158}$. Salacia was discussed in section \ref{sec:test}, for which we conclude a surface clearly dominated by silicates. 1998~SN$_{165}$'s measurements are also consistent with a surface dominated by amorphous silicates. We obtain a proportion of $90\pm20\%$ silicates, and $10\pm10\%$ \wat, and no organics. 

\begin{figure}
\gridline{          \fig{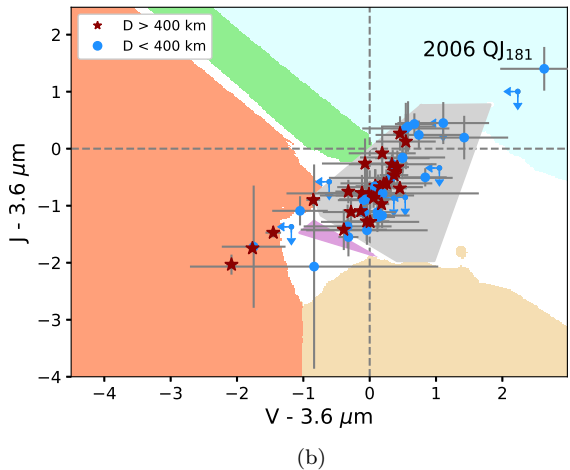}{\columnwidth}{(b)}
                }
  \caption{\label{fig:vch1_jch1}Diagram $V-3.6\,\mu$m versus $J-3.6\,\mu$m as a function of the size. Objects with diameters smaller than 400 km are represented by brown circles, while objects larger than 400 km are represented by red stars. Shaded regions are equivalent to those in figure \ref{fig:todos} (b).}
\end{figure}

In this regard, we also found interesting information representing the $V-3.6\,\mu{\rm m}$ versus $J-3.6\,\mu{\rm m}$ diagram when dividing the sample between small objects (those with diameters, $D$, smaller than 400 km) and large objects (those with $D>400$ km, see figure \ref{fig:vch1_jch1}). In this specific diagram, where there is a total of 28 objects over 400 km, only one object presents $1\sigma$ probability of having $J - 3.6\,\mu{\rm m}>0$, or, in other words, being dominated by silicates \citep[namely, Salacia, which has a diameter of $\sim900$ km;][]{Fornasier2013}. This translates into a $4\%$ probability of an object over 400 km presenting $J-3.6\,\mu{\rm m}>0$. While there are 6 objects with $D<400$ km (from a total of 38) with colors consistent with a surface dominated by silicates: 2000~GP$_{183}$, 2000~QL$_{251}$, 2001~CZ$_{31}$, 2001~QJ$_{181}$, 2002 CY$_{224}$, and 2004 EW$_{95}$. For all of them, we obtain over $80\%$ of silicates on its surface, considering a mixture of \wat, silicates and organics (see detailed information in appendix \ref{ap:sample_composition}). In fact, the spectrum of 2004 EW$_{95}$ was studied by \cite{Seccull2018}. They demonstrated that its composition is ``consistent'' with a C-type asteroid and the spectrum present a clear feature produced by hydrated, iron-rich silicates. This result provides validity for these specific colors to indicate surface consistence with compositions dominated by silicates.

\section{Diagrams by dwarf planets and Haumea's family}

\begin{figure*}
\gridline{\fig{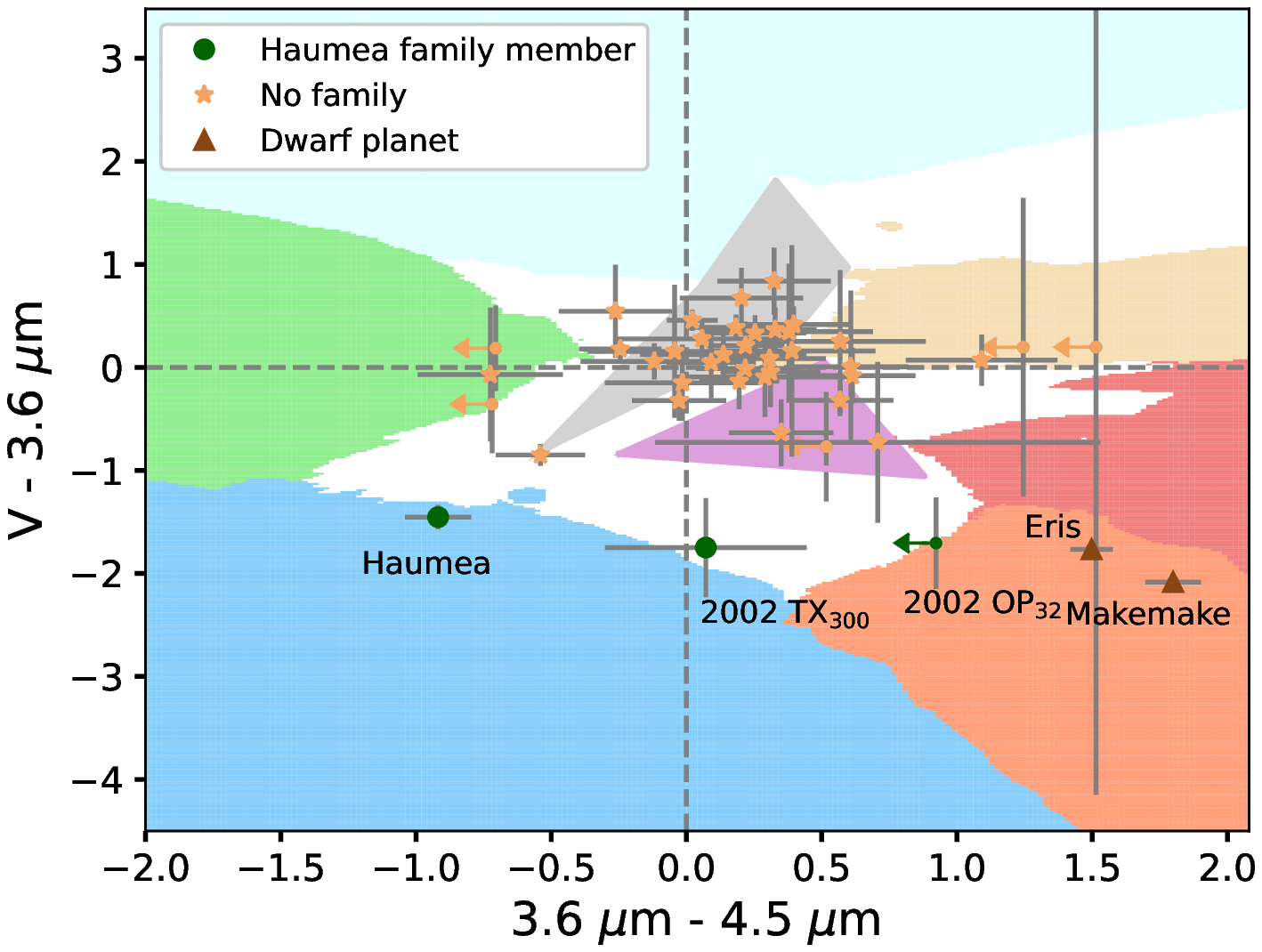}{0.5\textwidth}{(a)}
          \fig{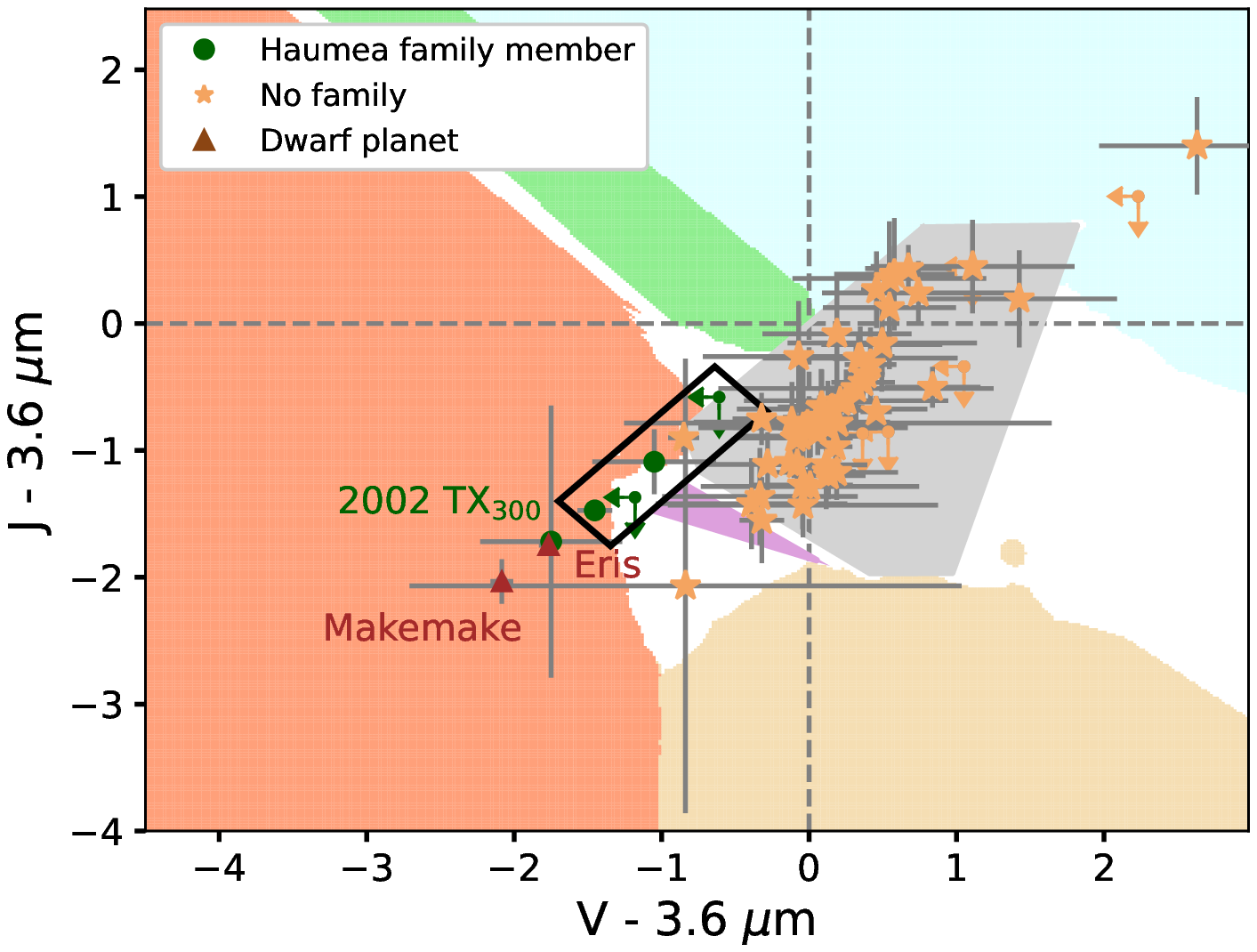}{0.5\textwidth}{(b)
                }}
  \caption{\label{fig:ch1ch2_vch1_data} (a) Diagram $3.6-4.5\,\mu$m versus $V-3.6\,\mu$m. (b) $V-3.6\,\mu$m versus $J-3.6\,\mu$m. In both panels, we represented the sample as a function of Haumea's family members and dwarf planets \cite[confirmed through dynamical models and spectroscopic measurements;][]{Brown2007b,Snodgrass2010} and dwarf planets Objects within the black rectangle are as follows: Haumea, 2005~RR$_{43}$, 1995~SM$_{55}$, Orcus, and 1996 TO$_{66}$ (from left to right). In both panels, shaded regions are equivalent to those explained in figure \ref{fig:todos}.}

\end{figure*}

However, Panel (a) in figure \ref{fig:ch1ch2_vch1_data} shows the $V-3.6\,\mu$m vs $3.6-4.5\,\mu$m diagram and indicates whether the objects are dwarf planets, Haumea family members, or neither of these two classifications. This diagram reveals that dwarf planets and Haumea family members \cite[confirmed through dynamical models and spectroscopic measurements;][]{Brown2007b,Snodgrass2010} appear segregated from the rest of the TNO population (including those non-Haumea family members that show H$_2$O), in regions dominated by {\wat}, {\met} or {\nol} and in agreement with the previous knowledge about their composition. 

In panel (b) of figure \ref{fig:ch1ch2_vch1_data}, these objects appear again in a detached region from the rest of the sample. The dwarf planets, Makemake, Eris and Haumea and 2002~TX$_{300}$ are clearly located over the region dominated by pure {\met} or {\wat} materials. The location of 1995~SM$_{55}$ is not as clear, however, it is detached from the rest of the population. Measurements for 2005 RR$_{43}$ and 1996 TO$_{66}$ are upper limits in both axes; however, these limits constrain the surface composition to models with high percentage of water, in agreement with their published spectra \citep{Brown1999,Pinilla-Alonso2007,Barucci2011}. 

Due to the lack of measurements for Haumea family members either at infrared wavelength or at 4.5 $\mu$m, we are not able to use the diagram $K-3.6\,\mu$m vs $3.6-4.5\,\mu$m (Haumea and 2002 TX$_{300}$ are the only two objects within the family that provide such a combination of measurements). 

New observations of objects that have been dynamically identified as part of the family could be also ``spectroscopically'' confirmed if they fall in the same region of both diagrams. 

\section{Diagrams by taxonomic classifications}
\label{sec:taxonomy}
\begin{figure*}
\gridline{\fig{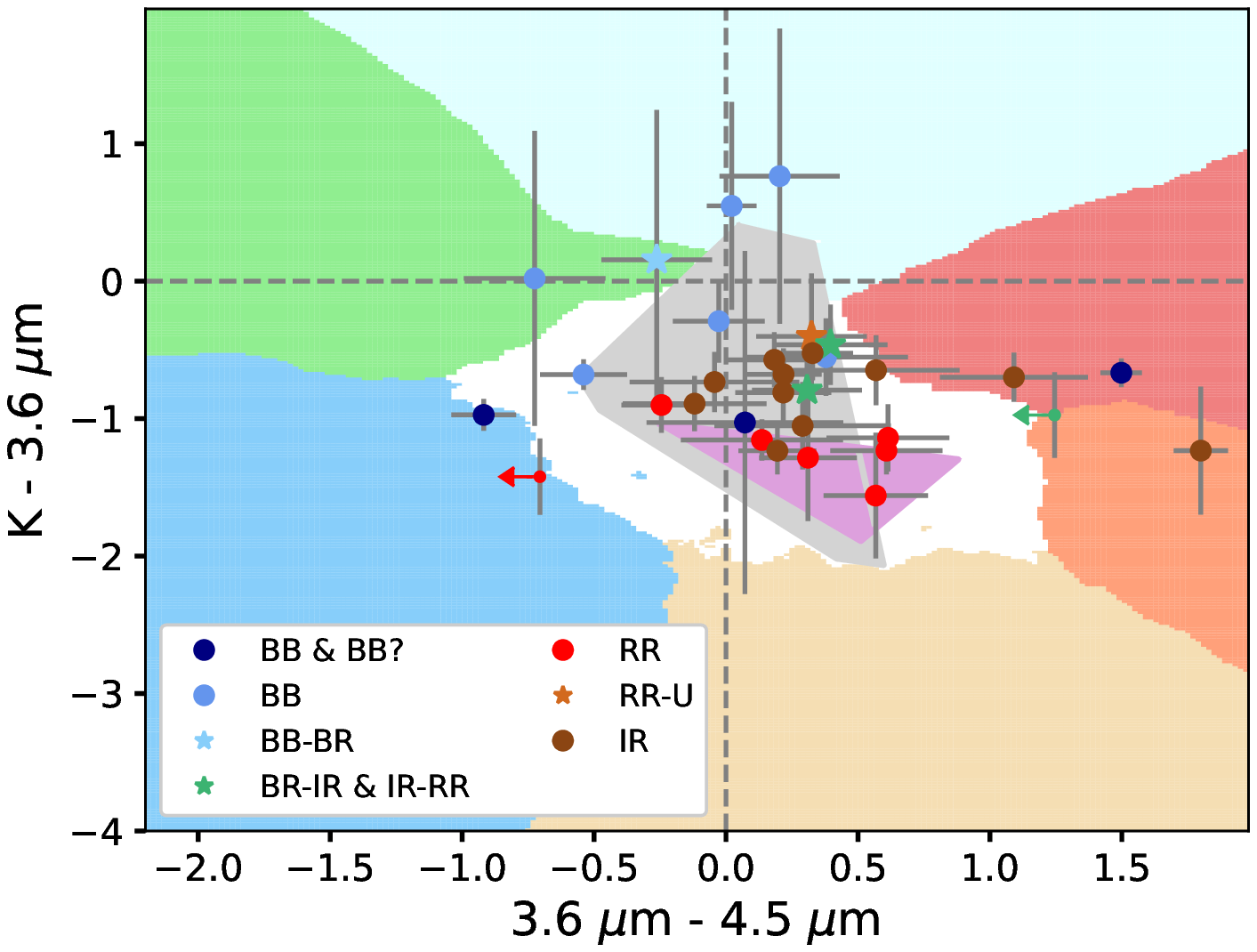}{0.5\textwidth}{(a)}
\fig{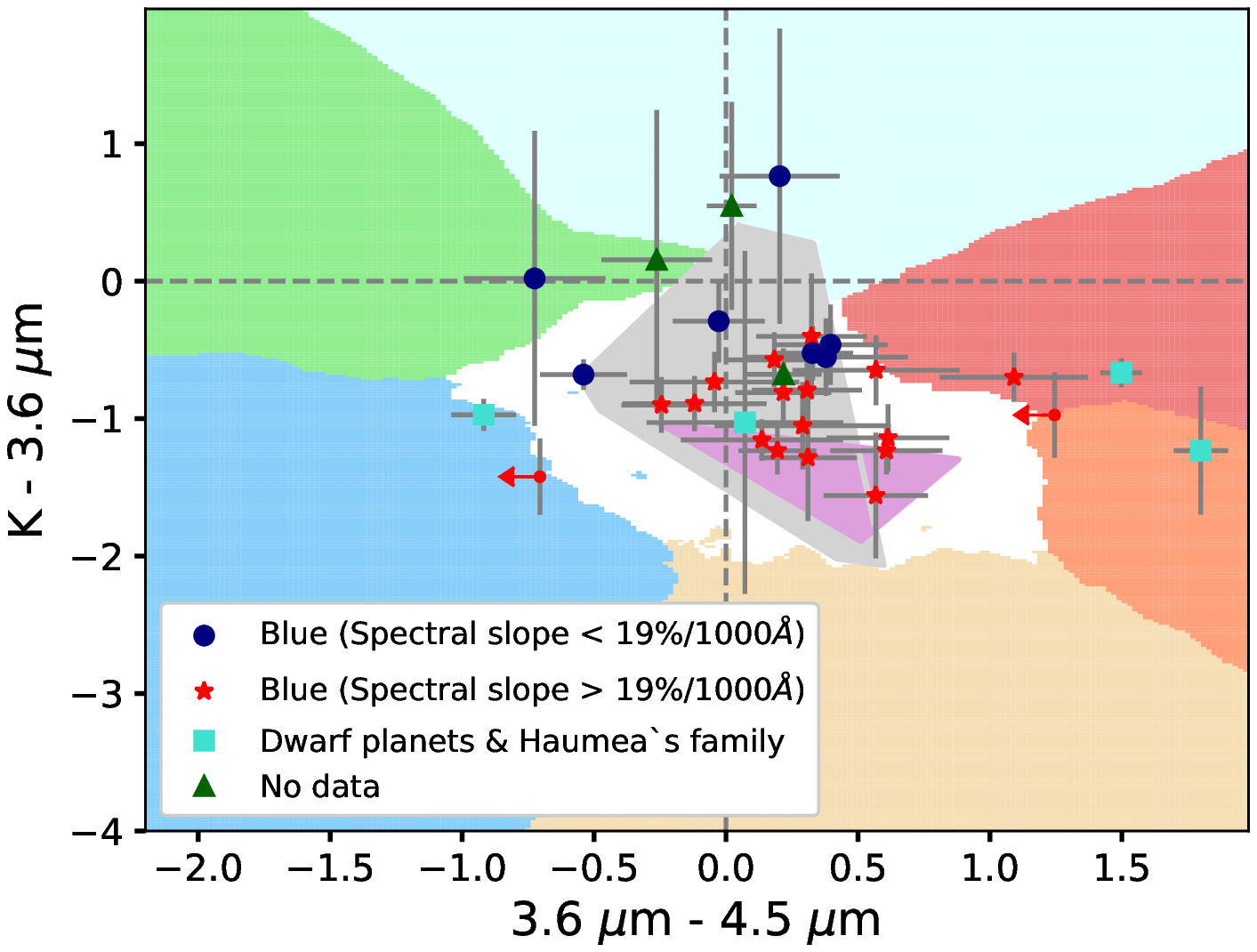}{0.5\textwidth}{(b)}}
\gridline{\fig{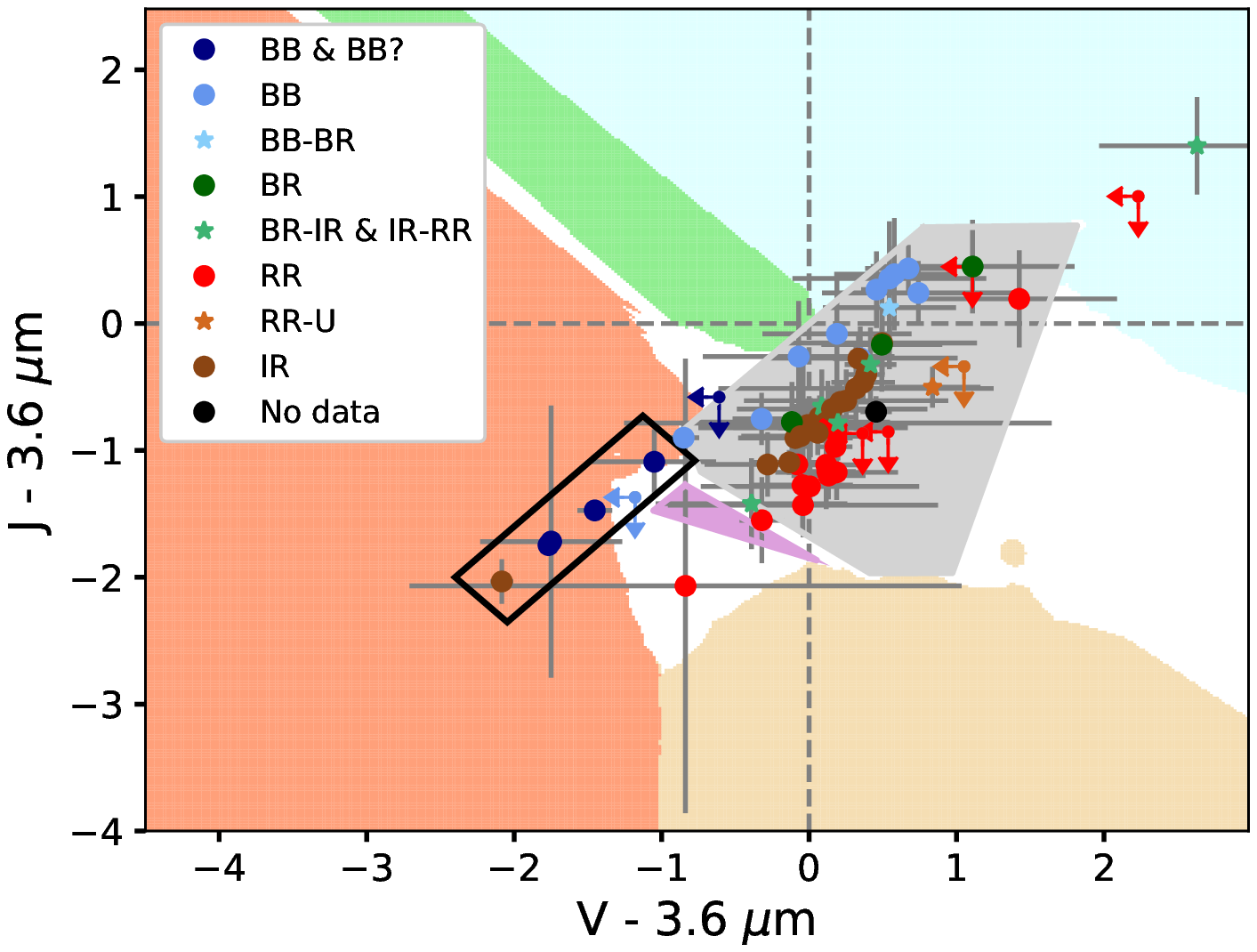}{0.5\textwidth}{(c)}\fig{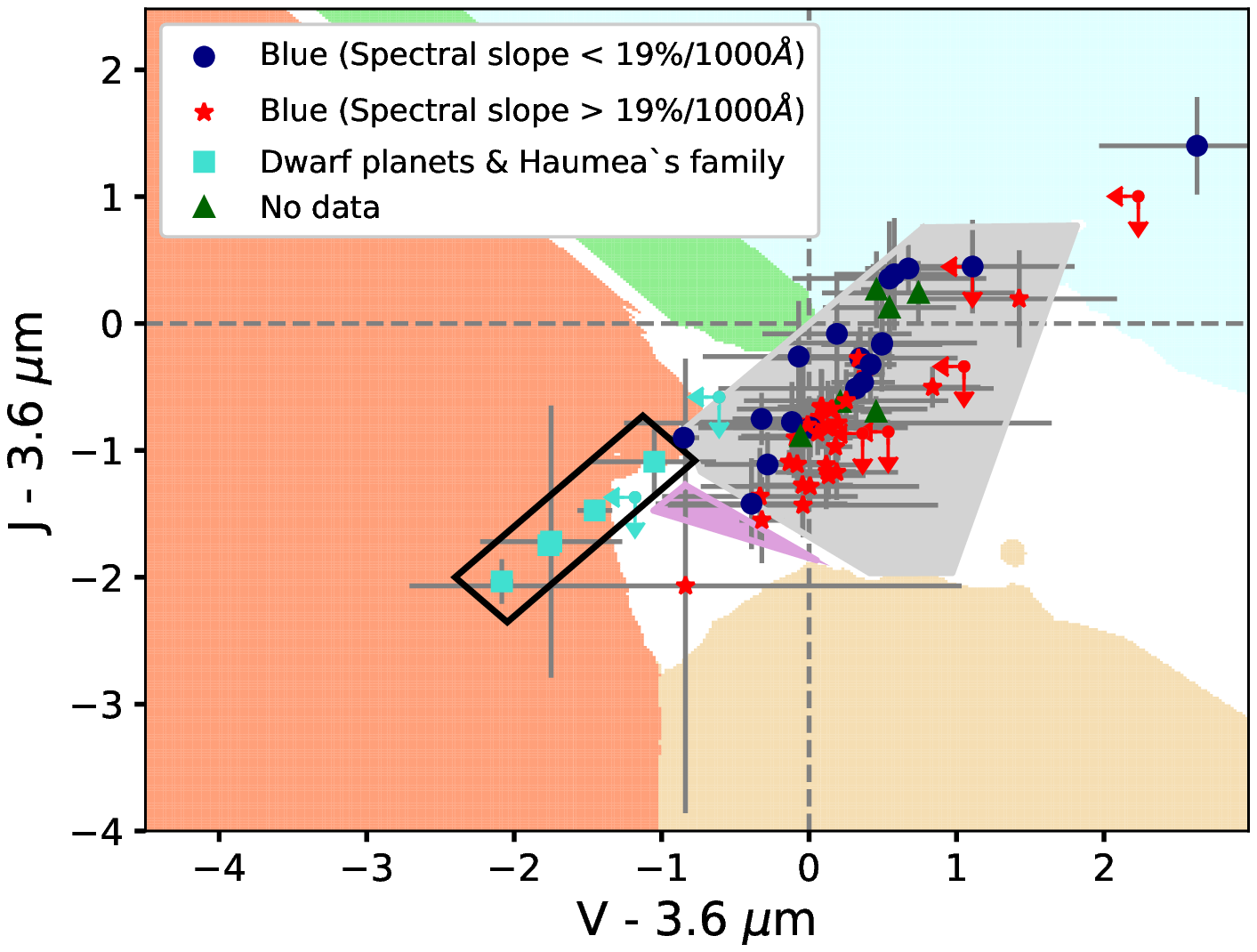}{0.5\textwidth}{(d)}}

  \caption{\label{fig:Taxonomy_CC}\textbf{Upper panels:} Compositional clock as a function of taxonomic classifications. (a) Taxonomy developed by \cite{Barucci2005b} and \cite{Belskaya2015}. (b) Taxonomy developed by \cite{Fraser2012} and \cite{Lacerda2014}. \textbf{Bottom pannels:} $V-3.6\,\mu$m versus $J-3.6\,\mu$m diagrams as a function of taxonomy classifications. (c) Taxonomy developed by \cite{Barucci2005b} and \cite{Belskaya2015}. (d) Taxonomy developed by \cite{Fraser2012} and \cite{Lacerda2014}. The group of objects with extremely high albedo claimed by both taxonomic theories are located within the black oval.}
\end{figure*}

The paucity of spectroscopic data compelled different authors to establish taxonomic classes in order to interpret the different surfaces found in the trans-Neptunian region. Two different taxonomies can be found in the literature. One was first proposed by \cite{Barucci2005}, who proposed four different taxonomic groups: neutral objects (BB), two intermediate slightly red types (BR and IR), and the reddest objects (RR). This taxonomy was updated more recently by \cite{Belskaya2015}, where the BB taxonomic group was dived between objects with high and low albedo (BBb and BB, respectively). The other system, first proposed by \cite{Fraser2012} and developed in \cite{Lacerda2014} presented a different perspective, in which two main classes are defined: the red one includes objects with higher albedos and redder colors, and the blue one includes objects with lower albedos and less red colors. Furthermore, \cite{Lacerda2014} proposed two other groups within the latter taxonomy classification in order to distinguish between dwarf planets and Haumea's family. The taxonomy proposed by \cite{Fraser2012} is also discussed in \cite{Schwamb2019}. 

The most interesting diagrams to plot by taxonomy are the compositional clock ($3.6\,\mu{\rm m}-4.5\,\mu{\rm m}$ versus $K-3.6\,\mu\,{\rm m}$), and the $V-3.6\,\mu$m versus $J-3.6\,\mu$m diagrams (figure \ref{fig:Taxonomy_CC}). In both cases, we see how blue \citep{Lacerda2014} and neutral \citep{Belskaya2015} objects, depending on the reference, fall within regions where our models indicate the presence of high percentages of water and/or silicates, which have generally more neutral spectroscopic slopes. In contrast, redder objects are found in regions where models indicate the presence of organic materials, which have redder spectral slopes. In panels (c) and (d) figure \ref{fig:Taxonomy_CC}, we can also highlight the group of objects that are distinct from the rest of TNOs due to their extremely high albedo. This group can be seen where models with pure water, \met, and \nol\ are located, which have the common characteristic of presenting high albedos.

We want to emphasize that our data can generate a compositional context for the different taxonomic classifications. For instance, IR objects are consistent with having a mixture of similar proportions of silicates and complex organics and having non-zero \wat. Also, most RR \citep{Belskaya2015} and red \citep{Lacerda2014} objects are consistent with being composed of material with smaller proportions of silicates and both rich in organics, and having a non-zero \wat-ice content. While the latter was known for part of our sample from VNIR spectroscopy, the former was only inferred due to their red colors. Our measurements provide a high level of confidence that this red visible color is imparted by organic materials such as tholins \cite[e.g.,][]{Khare1993,Materese2014,Materese2015}, which produce absorption bands beyond 2.2 $\mu$m.

\section{Diagrams by dynamical classes}
\label{sec:dynamical_classes}

\begin{figure*}
\gridline{\fig{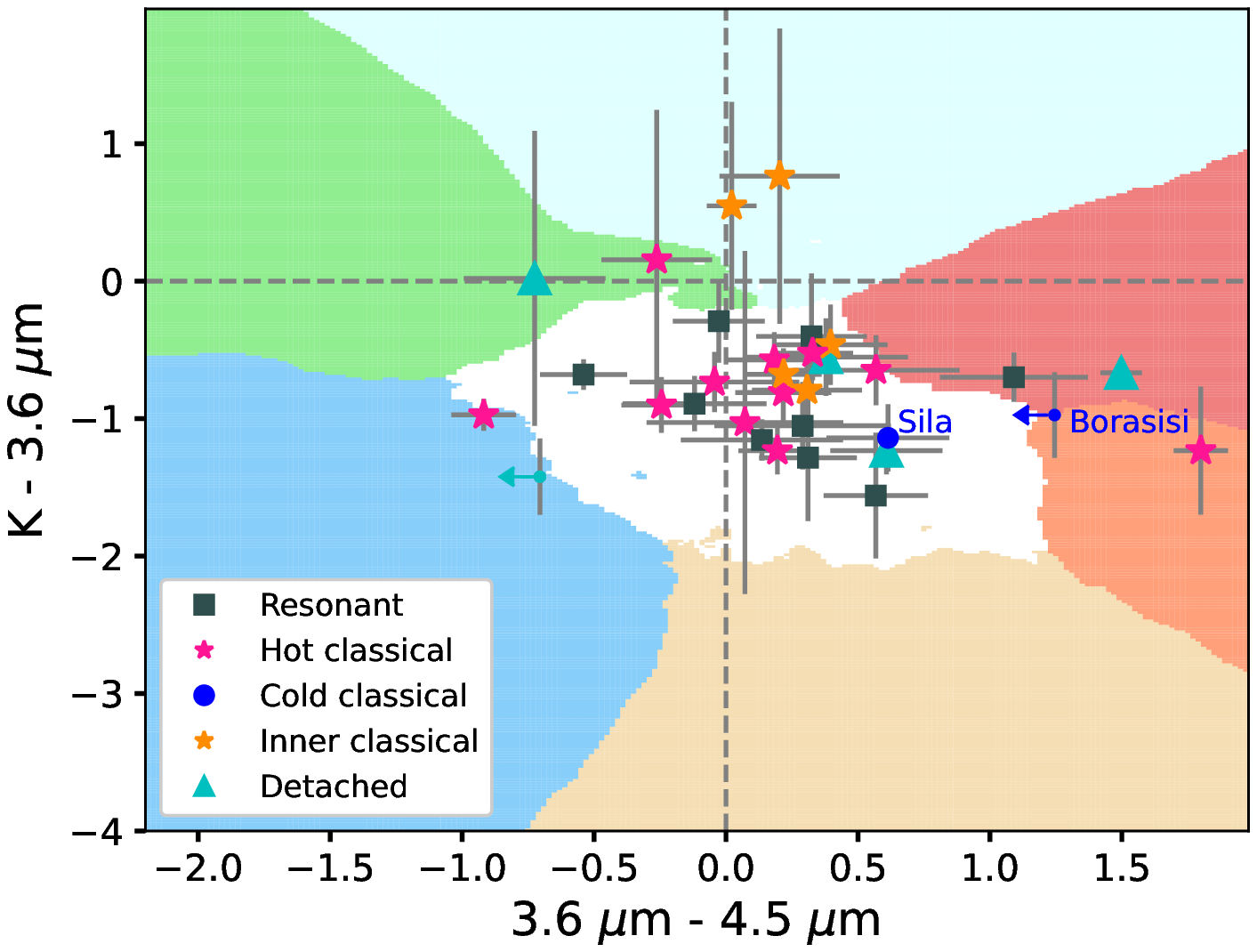}{0.5\textwidth}{(a)}
\fig{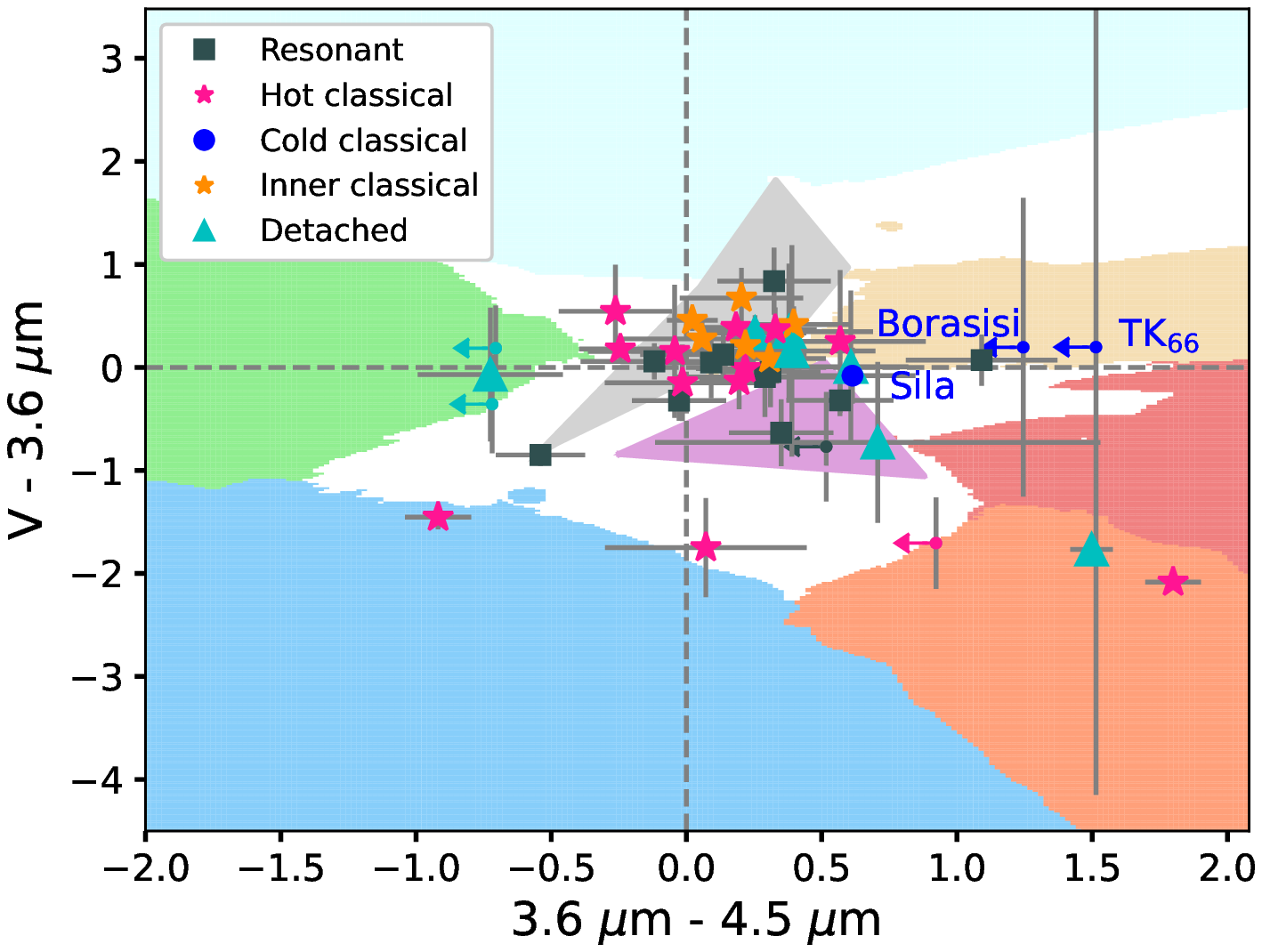}{0.5\textwidth}{(b)}}
\gridline{          \fig{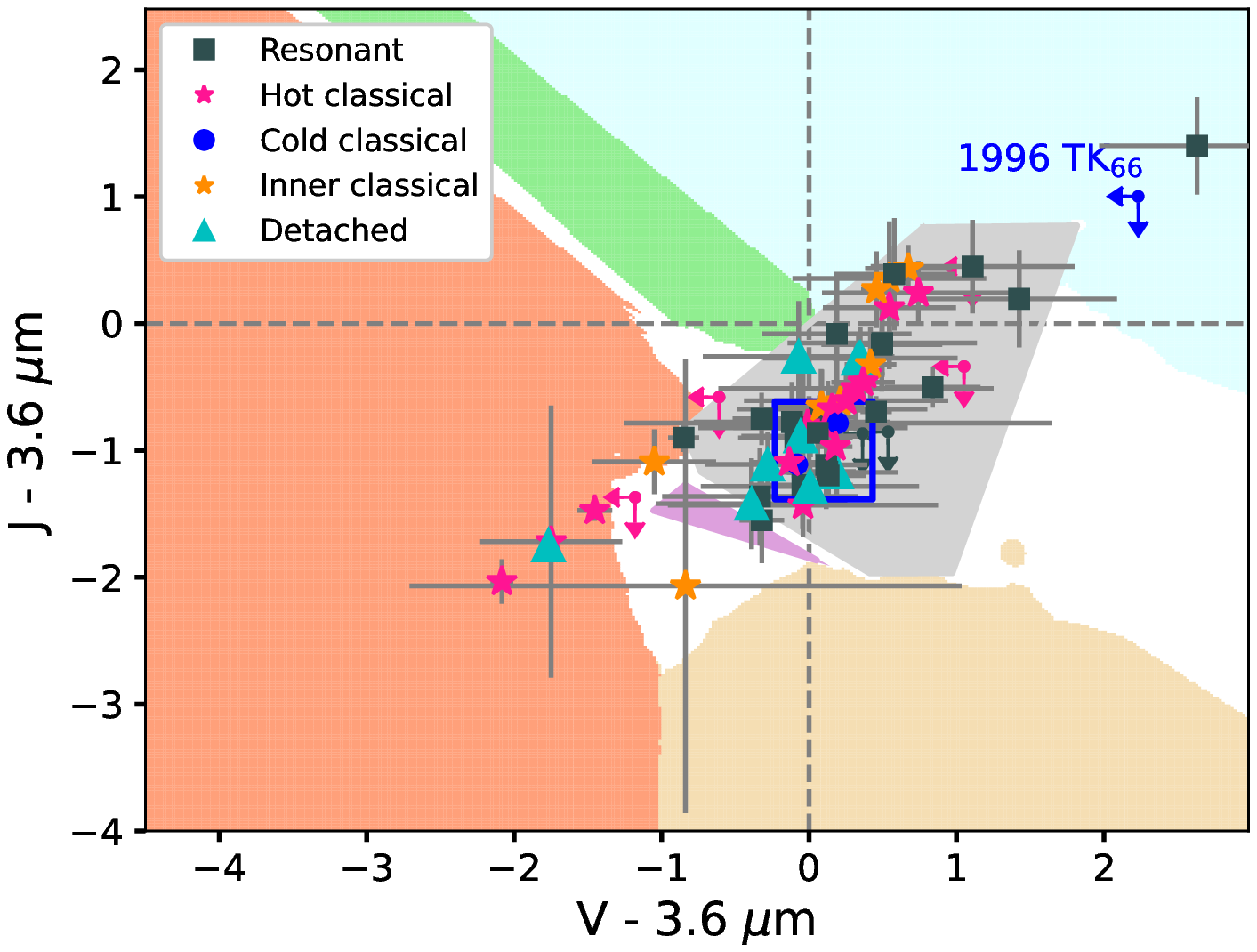}{0.5\textwidth}{(c)}}
      \caption{\label{fig:dynamical_classes} Compositional clock as a function of dynamical classes. Green squares depict resonant objects, turquoise triangles depict detached objects, pink stars depict hot classical objects and blue circles depict cold classical objects. Asterisks represent objects for which $3.6-4.5\,\mu$m color is an upper limit, and the asterisks color is matched to the dynamical class. (c) The blue square has been plotted to show the position of Borasisi and Sila-Nunam, which are the right and left blue circles within the square, respectively.}
\end{figure*}

As mentioned in section \ref{sec:introduction}, one might expect that different dynamical classes could experience different physical processes due to different past and present environments and therefore, exhibit at least slightly different surface composition. As expected, figure \ref{fig:dynamical_classes} shows that there is no distinction between detached, resonant and hot classical objects.

In particular, all of our inner classical belter objects fall firmly in with the majority of TNOs, even the low-$i$ inner belters. This fact thus adds additional support to the hypothesis based on orbital dynamics \citep{Kavelaars2009,Petit2011} that the inner belt classical ($a<39.4$~au classical objects) TNOs are entirely a captured hot population, and the low-$i$ members are just the low-inclination tail of the hot distribution. This hypothesis was confirmed in optical colours by \citet{Tegler2016} and we extend this into the infrared. Thus, the cold classical population appears not to exist today for $a<42.5$~au where it begins (between 39.4 and 42.5 au the $\nu_8$ secular resonance rapidly removes all low-$i$ TNOs).

It is highly desirable to examine the IR features of the cold classical population, as this population is thought to have formed and remain in the same location throughout their life-time \citep{Kavelaars2009, Parker2010}; it is thus a probe of the formation conditions at this distance \citep{Petit2011}.
If any TNO population region should look spectrally distinct, this is the population to study. Unfortunately, we only have measurements for four cold classical objects, from which only three have colors to be represented in our diagrams. However, we can provide some limited statements. Panel (a) in figure \ref{fig:dynamical_classes} provides colors for one cold classical (Sila-Nunam) and an upper limit in the color $3.6\,\mu{\rm m}-4.5\,\mu{\rm m}$ for another object (Borasisi). Panel (b) provides colors for Sila-Nunam and an upper limits in the color $3.6\,\mu{\rm m}-4.5\,\mu{\rm m}$ for two objects (Borasisi and 1996 TK$_{66}$). In these two panels, all cold classical objects appear within the same region, however, note that only Sila-Nunam provides measured colors, as the other two only have $4.6 \mu{\rm m}$ upper limits. Panel (c) provide colors for Sila-Nunam and Borasisi and upper limits in both colors for 1996 TK$_{66}$. In this panel, both objects with colors are located close to each other, and only the one with upper limits is in a region detached from them. Furthermore, agreement appears to exist between all the diagrams regarding the interpretation of the composition of those objects, namely, the colors are consistent with the existence of \nol\ on their surfaces. This is especially true in the case of Sila-Nunam, which has accurate measurements in the three diagrams. This strongly motivates future work to obtain IR colours for a sample of cold classical objects; we tentatively hypothesize that they will be found to have features related to the existence of \nol\ on their surfaces and appear distinct from the other TNO populations (with some overlap in some colours with large objects like Eris and MakeMake due to atmospheric physics).

Interestingly, \cite{Grundy2020} show that the spectra of Arrokoth (also a cold classical object), taken by LEISA on board of New Horizons, has clear \nol\ absorption bands, but do not show convincing evidence for \wat. As mentioned above, our models are consistent with Sila-Nunam containing \nol. In order to explain Arrokoth's composition, \cite{Grundy2020} propose that temperatures at the formation location of cold classical objects would have been low enough that volatile CO and CH$_4$ could freeze onto dust grains in the cold mid-plane of the nebula (where the sunlight was blocked), enabling production of CH$_3$OH and perhaps also destruction of H$_2$O. Once the dust and gas are dissipated from the nebula, the \met\ volatilizes due to the high equilibrium temperature, leaving only \nol. Thus, the cold population should show signatures of \nol\ on their surfaces, either in spectra or color. It has been shown by different authors that cold classicals are both brighter (a characteristic of \nol\ on the spectrum), and redder (as the tholin material produced by the irradiation of \nol) than other populations \cite[e.g.,][]{Brucker2009,Vilenius2014}. \cite{Grundy2020}'s hypothesis could be tested with new color measurements of cold classicals, but we will need to wait for JWST to achieve the desired sensitivity at the specific wavelengths in which \nol\ absorption bands can be detected for such a faint population.

\begin{table*}
    \centering
     \caption{Summary of the different compositions found using the color-color diagrams and taking into account the spectra published in the literature, as described in this work. Statistics are given according to the compositional clock.}
    \label{tb:summary}
    \begin{tabular}{|p{7.5cm}|p{7.5cm}|}
    \hline
         Composition & Objects\\
         \hline\hline
        Dominated by \wat & Haumea  \\
        $>50\%$ of \wat & Quaoar, Orcus\\
        Dominated by \met & Eris, Makemake\\
        Dominated by silicates & Salacia, 1998 SN$_{165}$\\
Presence of {\nol} and/or \met & Sedna, Sila-Nunam, 1999 DE$_{9}$, 2002 VE$_{95}$, 2002 TX$_{302}$.\\
        \hline
      \multicolumn{2}{|c|}{$86\%$ of the sample presents colors consistent with the presence of \wat-ice} \\
        \multicolumn{2}{|c|}{$80\%$ of the sample presents colors consistent with the presence of complex organics} \\
                \multicolumn{2}{|c|}{$93\%$ of the sample presents colors consistent with the presence of amorphous silicates} \\
      \multicolumn{2}{|c|}{$23\%$ of the objects in our sample has or may has {\met} and/or {\nol}} \\
      \multicolumn{2}{|c|}{Only smaller objects are dominated by silicates} \\
      \hline
    
    \end{tabular}
   
\end{table*}

\section{Conclusions}
\label{sec:conclusions}

We present a new method to study the surface composition of small solar system bodies. Using VNIR colors, together with specific filters beyond 2.2 $\mu$m, we have built color-color diagrams in which different materials occupy different regions of the diagram. Using these color-color diagrams, we are able to study very faint objects for which spectroscopic techniques would be either very expensive in time or impossible to carry out. Specifically, the compositional clock can discern compositions that are consistent with mixtures that require small amounts of \wat\ and other ices such as \diox, \met, \nol. The compositional clock also provides a high degree of confidence for the presence of complex organic materials such as tholins. The diagram $V-36\,\mu$m versus $J-3.6\,\mu$m also support this conclusion. A summary of the compositional interpretation by this method can be found in table \ref{tb:summary}.

From the compositional clock, we found that most of the TNOs within our sample ($73\%$), which includes detached, resonant and classical objects, have colors consistent with surfaces mainly composed of a mixture of \wat-ice, complex organics and amorphous silicates. $86\%$ of the sample have signatures consistent with water ice on their surface and $23\%$ have or may have {\met} and/or {\nol}. Also, $80\%$ have colors consistent with the presence of complex organics. Using other diagrams, we notice that only smaller objects seem to have colors that indicate surfaces dominated by silicates. In agreement with other authors, we also noticed that Haumea's family members and dwarf planets have a peculiar composition when compared with other TNOs. We are not able to distinguish very clearly between {\met} and {\nol} using the compositional clock or the other diagrams. Observations with specific/narrower filters should be carried out in order to be able to distinguish between these two components (as IRAC filters are very wide passbands).

There is currently a lack of measurements for cold classical and detached objects at $4.5\,\mu$m due to the faintness of these specific classes. In this regard, JWST will be advantageous for observing these objects. JWST will have a set of filters specifically for the detection of the different materials discussed in this work (figure \ref{fig:sinthetic_models}), see \cite{Kalirai2018}. Thus, JWST will enable similar studies with much fainter objects and will provide additional filters for more specific detections. This capability will be extremely useful to constrain the surface composition of objects within the trans-Neptunian region.

\section*{Acknowledgements}

E. Fern\'andez-Valenzuela acknowledges support from the 2017 Preeminent Postdoctoral Program (P$^3$) at UCF and the ``Earth and Space Based Studies in Support of NASA Space Missions'' under the Space Research Initiative (SRI) Program at FSI. BG and CvL ackowledge funding support from NSERC. N. P.-A. acknowledges funding support from the Spitzer Space Telescope, operated by the Jet Propulsion Laboratory, California and from the SRI/FSI project ``Digging-up Ice Rocks In The Solar System''. TM has received funding from the European Union Horizon 2020 Research and Innovation Programme, under Grant Agreement no 687378, as part of the project ``Small Bodies Near and Far'' (SBNAF). This work is based on observations made with the Spitzer Space Telescope, which is operated by the Jet Propulsion Laboratory, California Institute of Technology under a contract with NASA. Support for this work was provided by NASA. We acknowledge Ra\'ul Carballo-Rubio and the anonymous referees for providing useful comments that helped improve this manuscript.

\bibliography{mybibfile_Spitzer}{}
\bibliographystyle{aasjournal}
\nocite{*}

\newpage

\appendix

\section{Individual analysis of the sample}
\label{ap:sample_composition}

In the following, we analyze and provide the surface composition for each object individually. All proportions for different materials are calculated using the diagram $K-3.6\,\mu{\rm m}$ versus $3.6\,\mu{\rm m}-4.5\,\mu{\rm m}$, also named the compositional clock (unless otherwise indicated in the text):

\begin{itemize}

    \item The spectra of 1996 GQ$_{21}$ was studied by \cite{Doressoundiram2003}, who found no water detection. Later, \cite{Barkume2008} obtained a new spectrum claiming a detection of water at the $3\sigma$ level, which was later supported by \cite{Barucci2011}. \cite{Barkume2008} also suggested the presence of \nol. Our measurements for this object provide only an upper limit for the color index $3.6\,\mu{\rm m}-4.5\,\mu{\rm m}$. However, this limit constrains the region in which the object is localized within the diagram, eliminating the possibility of having other ices but water. Nonetheless, it is possible that large amounts of water could be hiding the {\nol}, as in the case for Orcus (see section \ref{sec:test}). 
    
    \item 1998 SN$_{165}$ has no spectrum published in the literature. Its location in the figure \ref{fig:water_detections} indicates that the surface of this object is dominated by silicates, with a small probability of the presence of water ice (taking into account the error bars). We obtain a proportion of $90\pm20\%$ silicates, and $10\pm10\%$ \wat, and no organics. Note that this object is an inner-belt classical object with a heliocentric ecliptic inclination of $i=4.6^\circ$; we compute a free inclination (with respect to the local forced plane) of 3.5$^\circ$. The fact that this object appears similar to hot classicals despite a low $i$ provides evidence for Section \ref{sec:dynamical_classes}'s argument for a ``hot only'' inner classical belt.
    
    \item 1999 DE$_9$ has been reported by several authors to have tentative water ice bands in its spectrum \citep{Jewitt2001,Brown2007b,Barkume2008}. Counter to this, \cite{Alvarez-Candal2007} published a spectrum with no indications of water related bands, however, they mentioned that the absorption bands that the other authors found were placed at 2\micron{}, a region of the spectrum that they had to remove due to the strong atmospheric absorption. Nonetheless, this object appears in a region of the compositional clock where there is no presence of water ice but {\met}. In fact, 1999 DE$_9$ was reported by \cite{Jewitt2001} to have features near 1.4 and $2.25\,\mu$m similar to what is found in the centaur Pholus, for which they interpreted the presence of solid \nol\ on its surface (although, they claim that the spectrum was not good enough for them to definitively make this identification). Our measurements are consistent with the detection of {\nol} on its surface. We obtain a composition of $20\pm10\%$ \wat, $60\pm10\%$ \nol, and $20\pm10\%$ organics.

\item 1999 OJ$_4$ for which we only provide measurements at 3.6 $\mu$m. However, inspecting its spectrophotometric measurements, this object is red and with and absorption band at 3.6 $\mu$m with respect to the $K$-band. As no absorptions are found at VNIR wavelengths, and due to its deep slope, such an absorption band should be due to complex organics. Considering the diagram $V-3.6\,\mu{\rm m}$ versus $J-3.6\,\mu{\rm m}$, we obtain a composition of $40\pm50\%$ \wat, $0\pm30\%$ silicates, and $60\pm50$ organics. Note that this object is an inner-belt classical object with a heliocentric ecliptic inclination of $i=4.0^\circ$ for which we compute a free inclination of 2.3$^\circ$.

    \item The published spectra for 2000 GN$_{171}$ have presented conflicting interpretations about the presence of water. \cite{deBergh2004} found an absorption band at $\sim1.6$ \micron{} that they reported to be related to water. However, \cite{Alvarez-Candal2007} obtained the spectrum of this object where no water bands were present. Later, \cite{Barkume2008} modeled a new spectra of 2000 GN$_{171}$ that included $10\%$ of water. Finally, \cite{Guilbert2009} presented another spectrum that is in agreement with the one published by \cite{Alvarez-Candal2007}. In the compositional clock, the position of this object corresponds to a composition of $20\pm20\%$ \wat, $40\pm20\%$ silicates, and $40\pm20\%$ organics.
    
     \item 2000 PE$_{30}$ has no spectrum published in the literature. the position of this object in the diagram is consistent with a composition of $20\pm10\%$ \wat, $60\pm20\%$ silicates, and $30\pm10\%$ organics.
          
     \item We only provide upper limit measurements of 2001 QT$_{322}$ at 3.6 and 4.5 $\mu$m. However, this object appears similar to hot classicals despite a low $i$ ($1.8^\circ$, with a free inclination of 2.4$^\circ$). This object provides evidence for Section \ref{sec:dynamical_classes}'s argument for a ``hot only'' inner classical belt.
     
    \item \cite{Doressoundiram2005} and \cite{Barkume2008} both obtained the spectrum of 2002 AW$_{197}$ and both reported a very small fraction of water. However, \cite{Guilbert2009} also observed this object, and suggested that the band found at 2\micron{} is related to incomplete removal of telluric features. Its position in the compositional clock corresponds to a composition of $20\pm10\%$ \wat, $60\pm10\%$ silicates, and $20\pm10\%$ organics.
    
    \item 2002 KX$_{14}$ was studied by \cite{Barkume2008} and \cite{Guilbert2009} with no apparent detection of \wat-ice bands in the spectra. Considering the diagram $3.6\,\mu{\rm m}-4.5\,\mu{\rm m}$ versus $V-3.6\,\mu{\rm m}$, we obtain a composition of $20\pm10\%$ \wat, $60\pm10\%$ silicates, and $20\pm10$ organics, in agreement with previous reports. Note that this object is an inner-belt classical object with a heliocentric ecliptic inclination of $i=0.4^\circ$; we compute a free inclination (with respect to the local forced plane) of 2.7$^\circ$. The fact that this object appears similar to hot classicals despite a low $i$ provides evidence for Section \ref{sec:dynamical_classes}'s argument for a ``hot only'' inner classical belt.
    
    \item 2002 TC$_{302}$ has been reported to tentatively have water ice \citep{Barkume2008,Barucci2011}. \cite{Stansberry2008} also suggested that this object has very little fresh ice on its surface. Our results agree with both conclusions, with a composition of $20\pm10\%$ \wat, $30\pm20\%$ silicates, and $50\pm20$ organics. For a mixture of \wat-\nol-organics, we obtain $40\pm20\%$, $30\pm10\%$, and $30\pm30$, respectively; however, those amounts of \wat{} should have been detected in its spectrum. Also, the visible colors of this object are very red (see appendix \ref{ap:plot_individual_objects}), indicating that higher presence of complex organics are more likely.

    \item The published spectra for 2002 UX$_{25}$ has presented conflicting interpretations. \cite{Barkume2008} suggested a small fraction, $6\%$, of water on its surface. Later, \cite{Barucci2011} obtained a new spectrum and reported no water bands. However, its position correspond to a composition of $20\pm10\%$ \wat, $50\pm10\%$ silicates, and $30\pm10$ organics, similar to 2002 AW$_{197}$. 

    \item 2002 VE$_{95}$ has a strong detection of \nol\ \citep{Barucci2006} and our measurements are in agreement with this detection. This object was discussed in section \ref{sec:test}. We obtain a composition of $20\pm10\%$ \wat, $40\pm10\%$ \nol, and $40\pm20\%$ organics.

     \item 2003 QA$_{92}$ only presents measurements at 3.6 $\mu$m. The lack of infrared data prevent us to provide the existence or not of absorption bands. Note that this object is an inner-belt classical object with a heliocentric ecliptic inclination of $i=3.4^\circ$ for which we compute a free inclination of 2.4$^\circ$.
	
 \item 2004 EW$_{95}$ was studied by \cite{Seccull2018}. They demonstrated that its composition is ``consistent'' with a C-type asteroid and the spectrum present a clear feature produced by hydrated, iron-rich silicates. We obtain a composition of 
Considering the diagram $V-3.6\,\mu{\rm m}$ versus $J-3.6\,\mu{\rm m}$, we obtain a composition of $10\pm10\%$ \wat, $80\pm10\%$ silicates, and $10\pm10$ organics.
	
    \item 2004 NT$_{33}$ has been studied by several authors who are in agreement that there is no detection of water \citep{Barkume2008,Barucci2011}. We obtain a composition of $20\pm40\%$ \wat, $80\pm60\%$ silicates, and $0\pm20\%$ organics. Additionally, the presence of CO$_2$ on its surface is also a possibility. CO$_2$ has been previously detected on the surface of Iapetus (Saturn's moon), where \wat-ice and complex organics coexist \citep[e.g.,][]{Palmer2008,Pinilla-Alonso2011}. CO$_2$ could be originated as a byproduct of the interaction of these two materials.
    
    \item 2004 TY$_{364}$ has been reported to tentatively have water ice \citep{Barkume2008,Barucci2011}. \cite{Barucci2011} measured a depth for the band of $5.8\%$, in agreement with previous works \cite[e.g.,][]{Barkume2008}. Additionally, \cite{Merlin2012} found a band at $2.27\,\mu$m in the spectrum of 2004 TY$_{364}$, which could be associated with methanol. The position of this object in the compositional clock corresponds to a composition of $20\pm10\%$ \wat, $50\pm10\%$ silicates, and $30\pm10$ organics. However, if we considere a mixture of \wat, \nol, and organics, we obtain a proportion of $50\pm10\%$, $30\pm10\%$, and $20\pm10$ for each material, respectively, which increase the amount of \wat\ too much for not have being clearly detected in its spectrum.
        
    \item 2005 RM$_{43}$ has been studied spectroscopically by \cite{Fornasier2009} and \cite{Barucci2011}. No water detection was obtained by \cite{Fornasier2009}, but tentative detection was reported by \cite{Barucci2011}, who model the spectra using up to $\sim40\%$. The position of this object in figure \ref{fig:water_detections} is quite interesting, as this region is the one dominated by CO$_2$. However, error bars are quite large and could easily place the object in a region where the surface would be completely dominated by water. Nonetheless, such amount of \wat\ should be detected by VNIR spectroscopy, which is not the case. Other color-color diagrams support the possibility of this object having CO$_2$ (see section \ref{sec:other_cc}). For a mixture of \wat, silicates and organics, we obtain a proportion of $50\pm40\%$, $50\pm40\%$, a 0\% for each material, respectively.
    
    \item 2005 RN$_{43}$ has been reported to not have water ice on its surface by \cite{Barkume2008} and \cite{Guilbert2009}; however its position in figure \ref{fig:water_detections} indicates that this object is composed of $20\pm10\%$ \wat, $50\pm10\%$ silicates, and $30\pm10\%$ organics.

    \item 2007 JJ$_{43}$ was studied by \cite{Gourgeot2015}. They proposed a surface composition of around 50\% of complex organics and up to $6.5\%$ water. Our measurements result in a composition of $40\pm30\%$ \wat, $40\pm20\%$ silicates, and $20\pm10\%$ organics.
    
    \item Borasisi's spectrum was published in \cite{Barkume2008}, they obtained a spectral slope of $28.67\pm3.61\%/100$ nm and no absorption bands. In our data, Borasisi has only upper limit measurements in the color index $3.6\,\mu {\rm m}-4.5\,\mu {\rm m}$ and, due to its position in the compositional clock, no strong conclusions can be drawn about the composition of this object. However, the color index $K-3.6\,\mu {\rm m}$ indicates that there is an absorption band that may be either related to ices or complex organics. Also, considering the diagram $V-3.6\,\mu{\rm m}$ versus $J-3.6\,\mu{\rm m}$, we obtain a composition of $30\pm50\%$ \wat, $40\pm30\%$ silicates, and $30\pm20\%$ organics.
    
    \item The spectrum of Huya has been studied by \cite{Licandro2001,Jewitt2001} and \cite{Fornasier2004} with the conclusion that no water ice is observed. Nonetheless, \cite{Jewitt2001} suggested the possibility of a wide absorption band near $2.0\,\mu$m that could be related to water. In figure \ref{fig:water_detections}, the position of this object indicates a proportion of $40\pm20\%$ \wat, $30\pm10\%$ silicates, and $30\pm10\%$ organics.

    \item \cite{Barucci2008,Delsanti2010} and \cite{Demeo2010} suggested the presence of methane and their irradiated products on the surface of Orcus. We obtain a composition of $70\pm10\%$ \wat, $20\pm10\%$ silicates, and $10\pm10\%$ organics. Due to the large amount of water ice found in this object, we are unable to detect methane or the irradiated products with our method, and therefore we do not exclude \met\ as part of its composition. \cite{Demeo2010} also suggested the possibility of the presence of {\diox}, and due to the position of this object in figure \ref{fig:ice_detections}, close to the region dominated by {\diox}, we also support this possibility. 

    \item The spectrum of Quaoar has been studied in detail by \cite{Dalle-Ore2009} using VNIR spectroscopic and IRAC data, who reported the presence of H$_2$O ice and \met{}. Our measurements agree with the results from \cite{Dalle-Ore2009}, and we report a composition of $60\pm10\%$ \wat, $20\pm10\%$ silicates, and $20\pm10\%$ organics. The high percentage of water prevents from clearly detecting the presence of \met. Quaoar could be then a good reference for objects with similar spectra.

    \item \cite{Schaller2008} reported that the fraction of water ice in the surface of Salacia is consistent with zero, which is in a very good agreement with our measurements (see figure \ref{fig:color_index}). Its position in the compositional clock suggests that the surface of Salacia is dominated by silicates and depleted from ices. We obtain a composition of $10\pm10\%$ \wat, $90\pm20\%$ silicates, and no organics.
    
    \item The spectrum of Sila-Nunam has been studied by \cite{Grundy2005}, where no water bands were found. Also, \cite{Barucci2011}'s reanalysis of the spectrum published by \cite{Grundy2005} is in agreement. \cite{Grundy2005} reported a neutral spectrum with no strong evidence for \wat\ or \met, although they noticed a dip around $2.33\,\mu$m that may arise from absorption by an organic ice. Our measurements indicate the presence of \met\ or \nol\ on the surface of Sila-Nunam. Due to the small size of the binary system \cite[around 340 km;][]{Vilenius2012}, it is unlikely that this object possesses \met\ on its surface, being more realistic to think that its position in figure \ref{fig:ice_detections} is due to \nol. We obtain a composition of $30\pm10\%$ \wat, $50\pm10\%$ \nol, and $20\pm10\%$ organics.
    
    \item Varda's spectrum has been studied by \cite{Barucci2011}, whose results are in good agreement with its position in the compositional clock. We obtain a composition of $20\pm10\%$ \wat, $60\pm10\%$ silicates, and $30\pm10\%$ organics.
        
    \item 2001 UR$_{163}$, and 2004 GV$_9$ are both located within the pink circle in figure \ref{fig:water_detections}. The spectra of these two were studied by \cite{Barkume2008,Barucci2011,Guilbert2009}. We obtain a composition of $20\pm10$ \wat, $60\pm20\%$ silicates and $20\pm20$ organics for 2001 UR$_{163}$, and $20\pm10$ \wat, $60\pm20\%$ silicates and $20\pm10$ organics for 2004 GV$_9$.

    \item G!k\'un$||$'h\`omd\'im\`a was studied by \cite{Barucci2011}, who reported water detection. We can analyze this object considering the diagram and $V-3.6\,\mu{\rm m}$ versus $3.6\,\mu{\rm m}-4.5\,\mu{\rm m}$, for which we obtain a composition of $20\pm50$ \wat, $10\pm20\%$ silicates and $70\pm50$ organics.

\end{itemize}

The following objects have no spectrum published in the literature. Considering the diagram $V-3.6\,\mu{\rm m}$ versus $J-3.6\,\mu{\rm m}$, we obtain a surface composition as follows:

\begin{itemize}
\item 2000 GP$_{183}$: $10\pm10\%$ \wat, $80\pm10\%$ silicates, and $10\pm10\%$ organics.
\item 2000 QL$_{251}$: $10\pm10\%$ \wat, $90\pm10\%$ silicates, and $10\pm10\%$ organics.
\item 2001 CZ$_{31}$: $10\pm10\%$ \wat, $80\pm10\%$ silicates, and $10\pm10\%$ organics.
\item 2001 QJ$_{181}$: no \wat, $90\pm10\%$ silicates, $10\pm10\%$, and organics.
\item 2002 CY$_{224}$: no \wat, $90\pm10\%$ silicates, $10\pm10\%$, and organics.
\end{itemize}

\newpage
\section{Spectrophotometric measurements plotted for each object}
\label{ap:plot_individual_objects}

In this appendix we present the spectrophotometric measurements of our sample. When available, we also plotted the spectrum or the spectrum model (references are indicated on the plots). The figures are ordered by provisional designation in ascending order, followed by the named objects in alphabetical order.

\includegraphics[width=0.5\columnwidth]{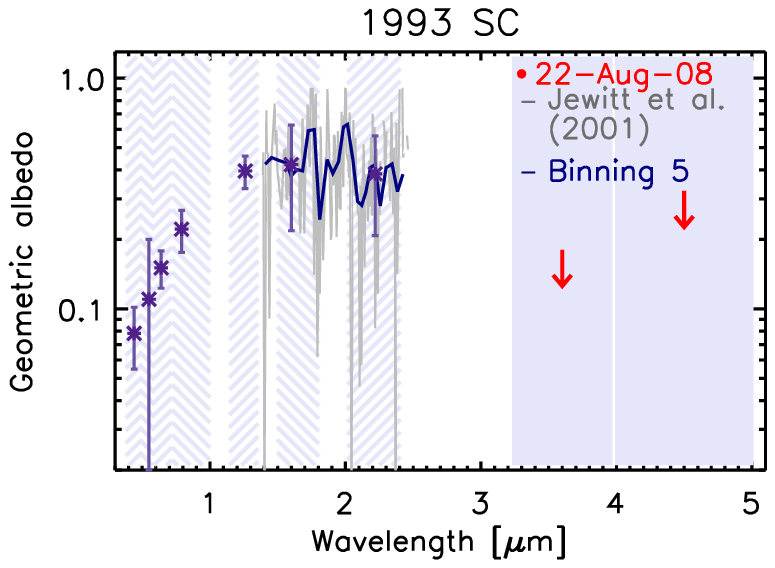}
  ~
     \includegraphics[width=0.5\columnwidth]{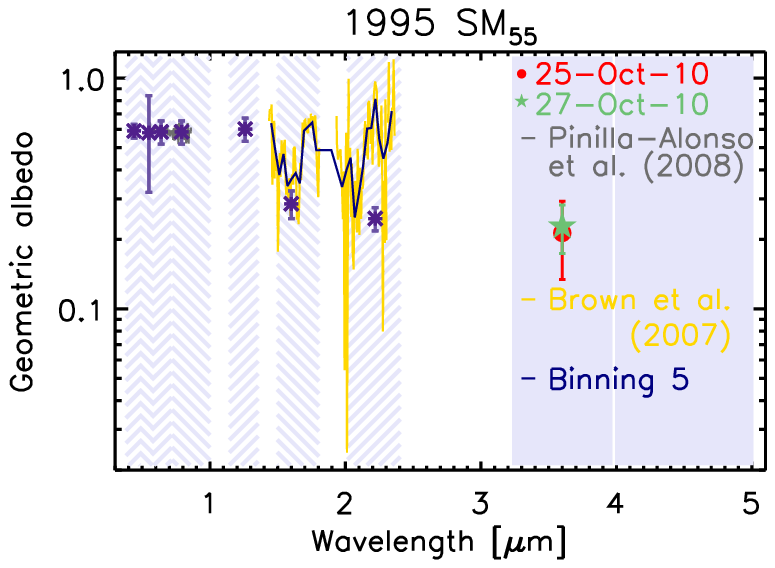}      
 ~
      \includegraphics[width=0.5\columnwidth]{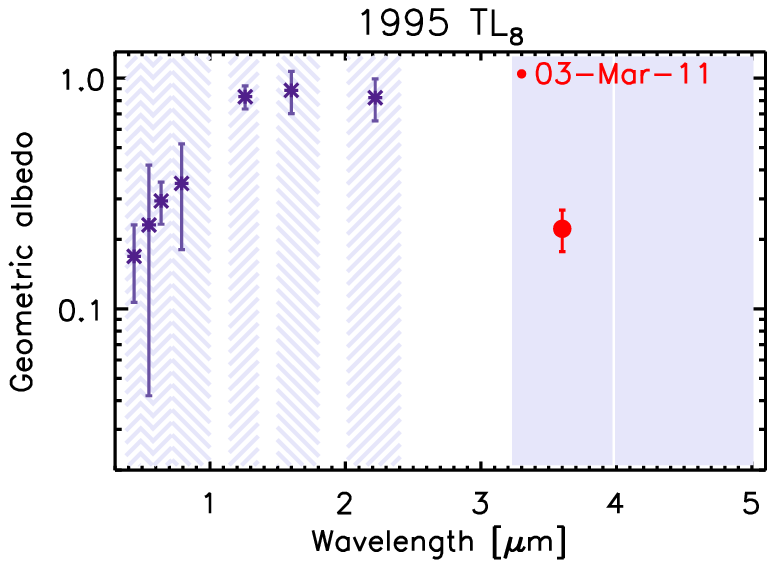}
~
        \includegraphics[width=0.5\columnwidth]{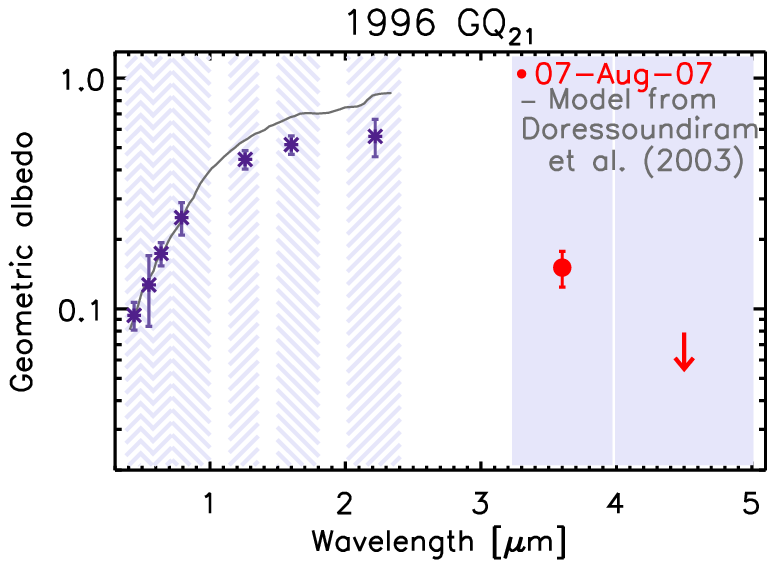}
~
        \includegraphics[width=0.5\columnwidth]{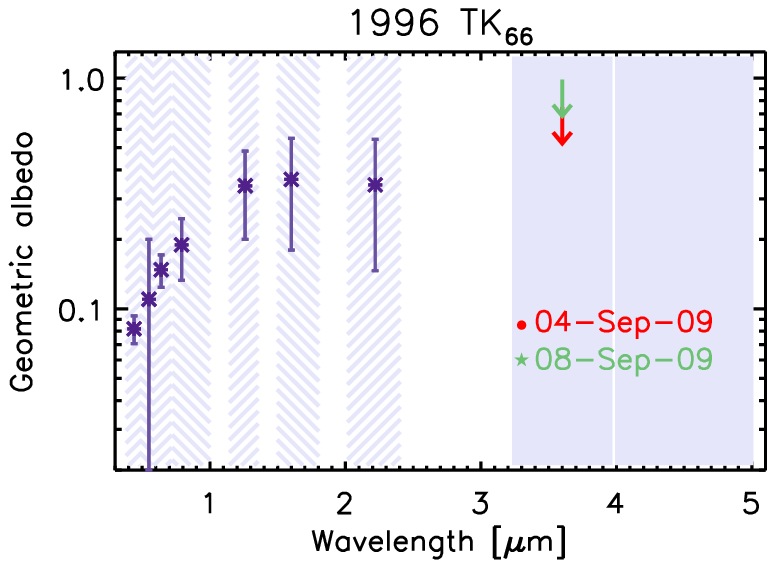}
~
        \includegraphics[width=0.5\columnwidth]{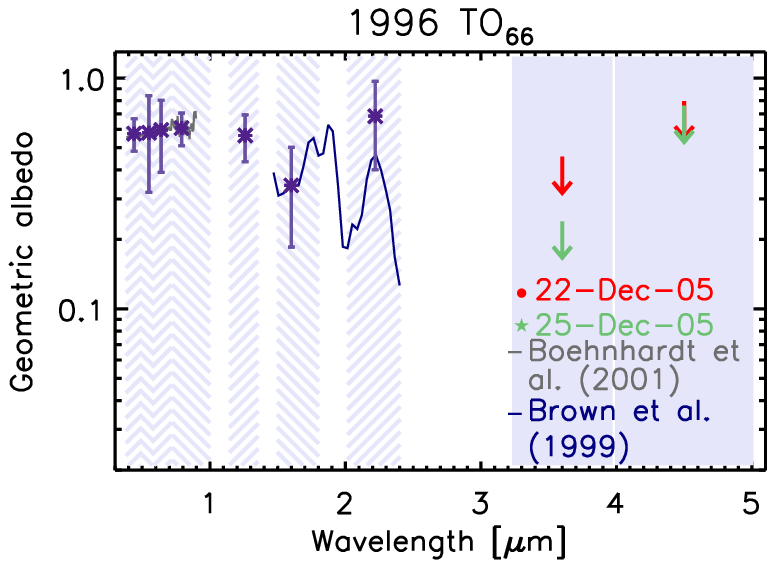}
~
        \includegraphics[width=0.5\columnwidth]{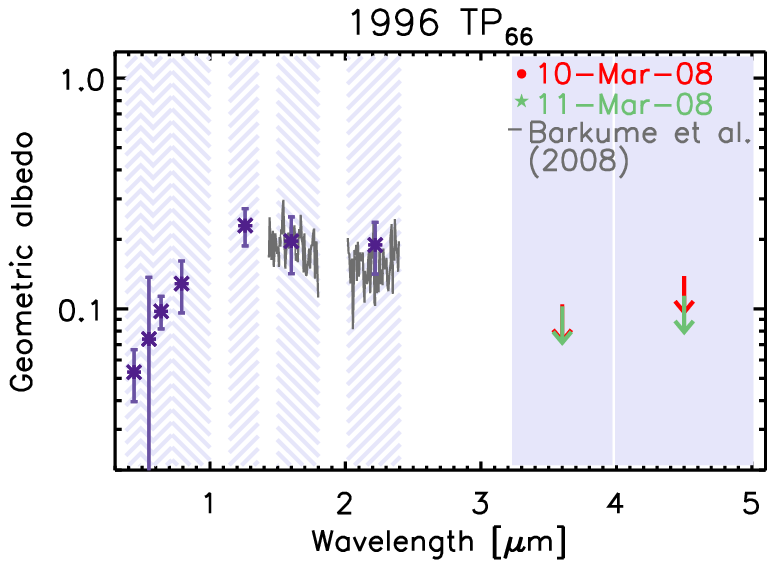}
  ~
        \includegraphics[width=0.5\columnwidth]{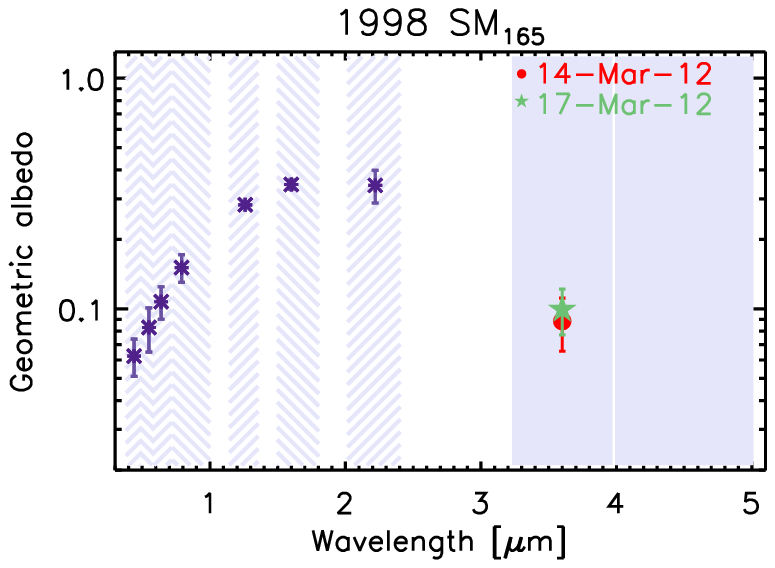}
~
        \includegraphics[width=0.5\columnwidth]{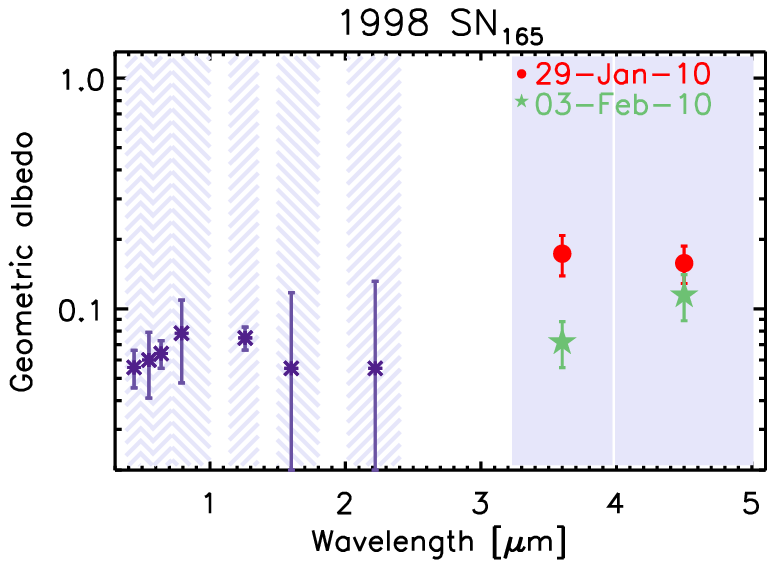}
  ~
        \includegraphics[width=0.5\columnwidth]{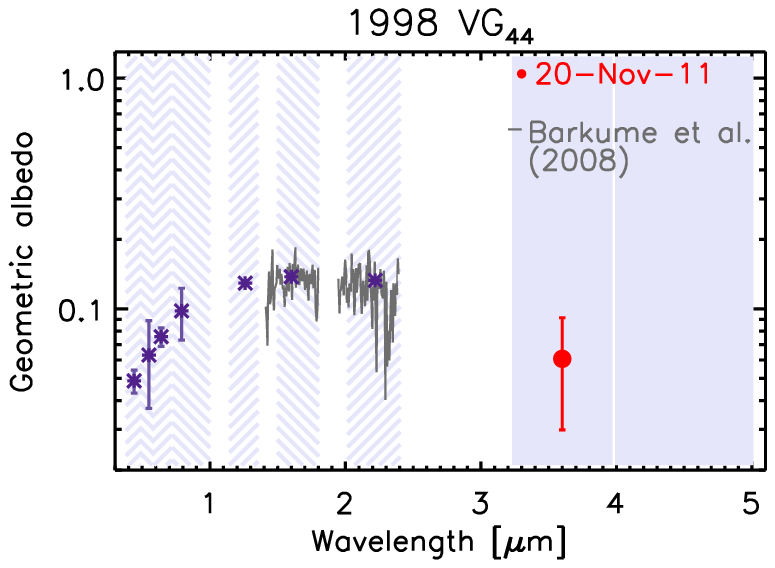}
~
        \includegraphics[width=0.5\columnwidth]{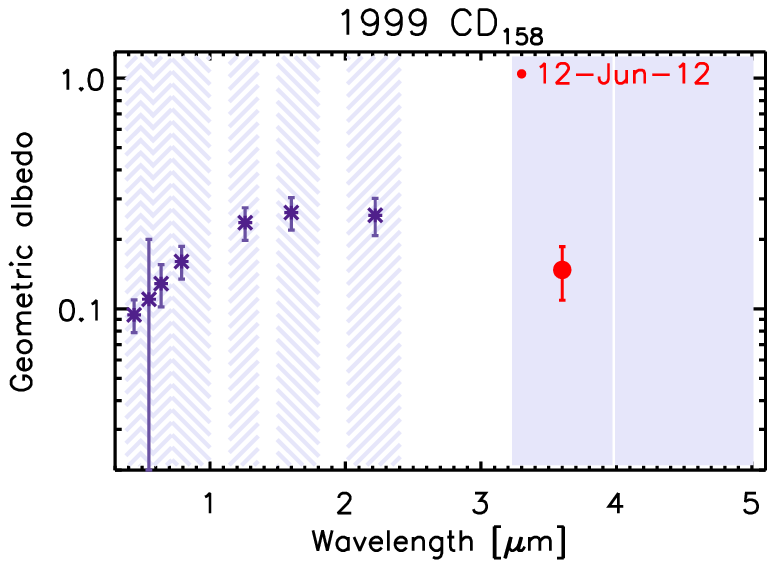}
    ~
        \includegraphics[width=0.5\columnwidth]{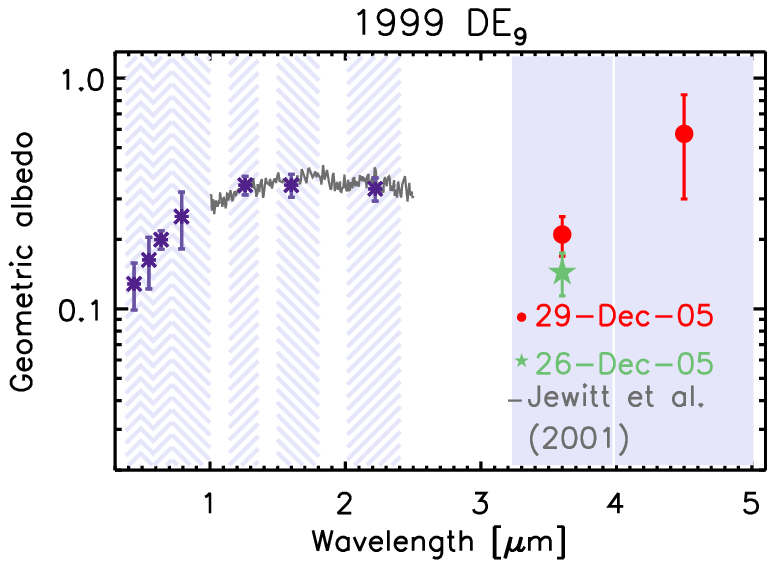}
~
        \includegraphics[width=0.5\columnwidth]{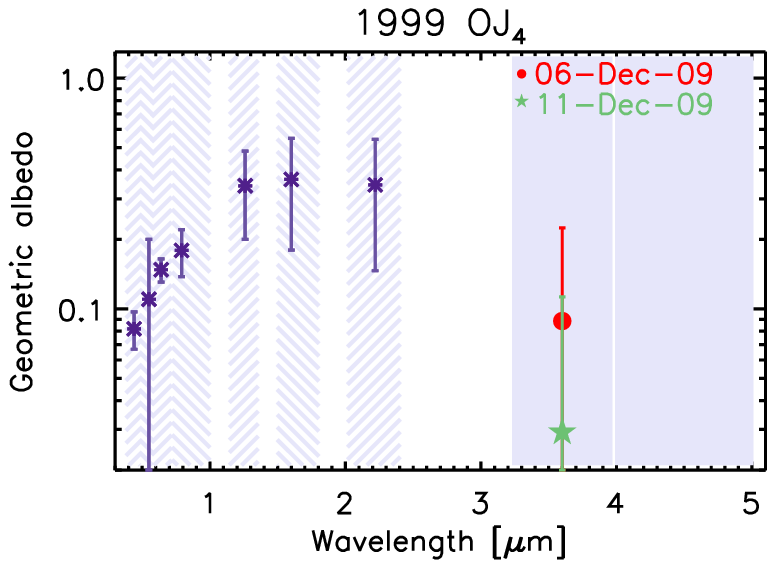}
~
        \includegraphics[width=0.5\columnwidth]{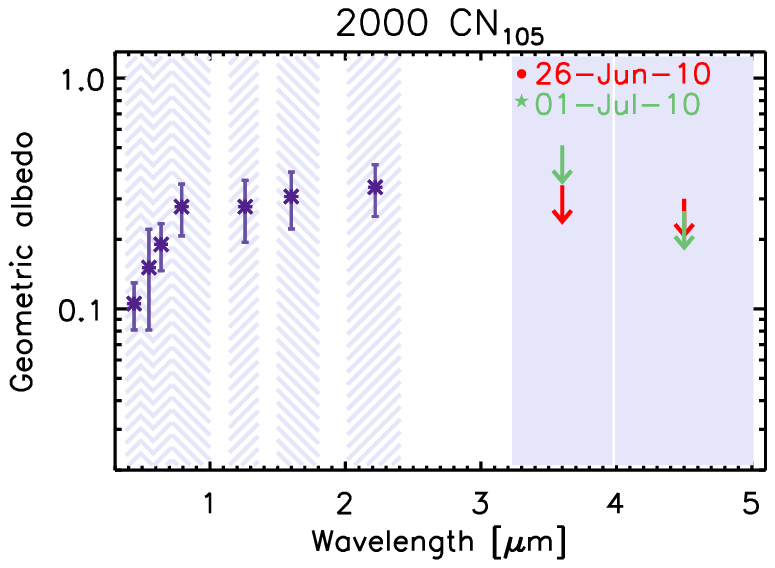}
    ~
        \includegraphics[width=0.5\columnwidth]{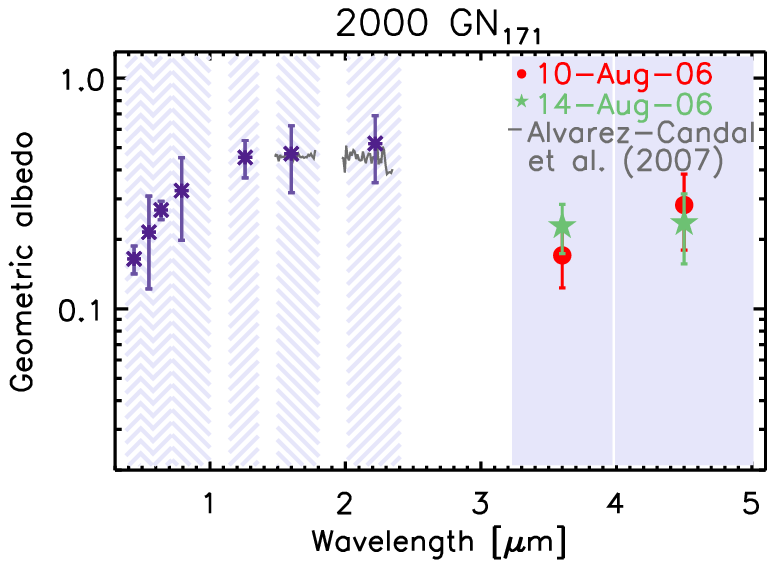}
~
      \includegraphics[width=0.5\columnwidth]{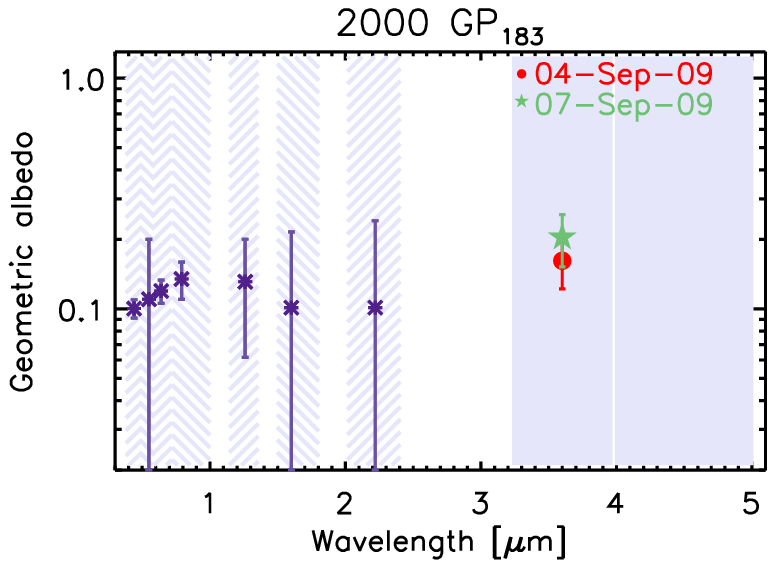}
~
        \includegraphics[width=0.5\columnwidth]{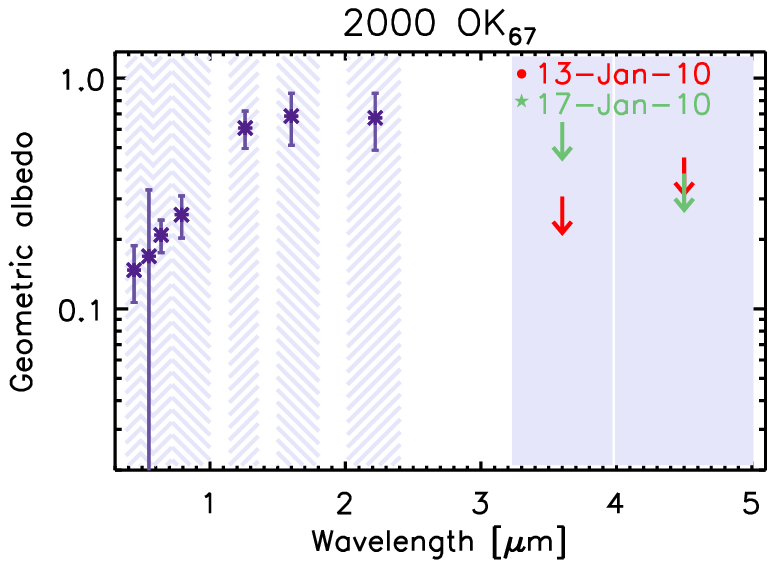}
~
      \includegraphics[width=0.5\columnwidth]{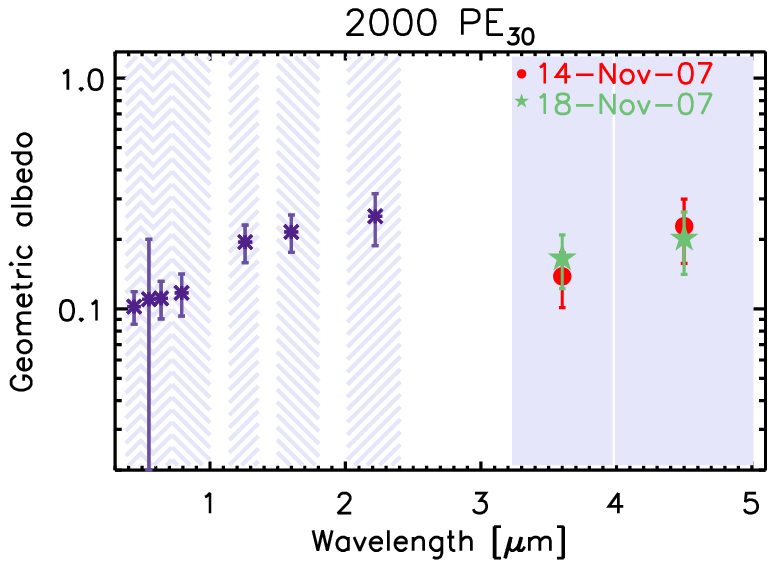}
~
        \includegraphics[width=0.5\columnwidth]{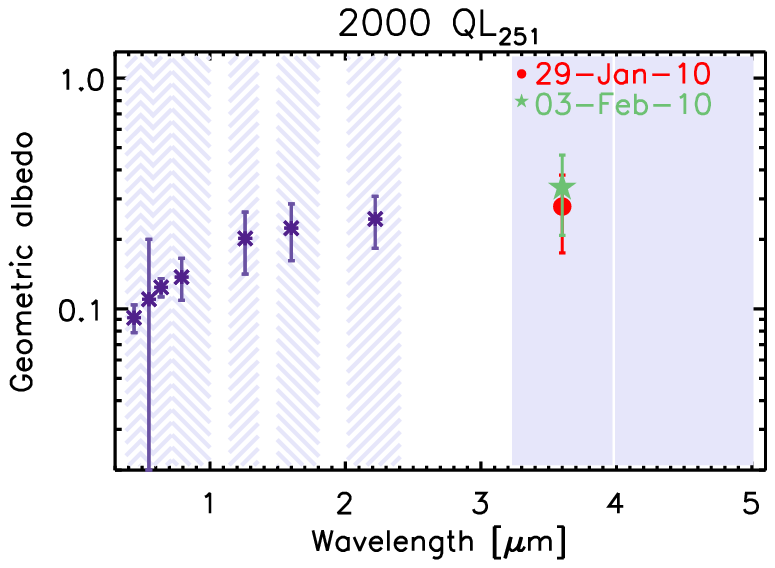}
 ~
        \includegraphics[width=0.5\columnwidth]{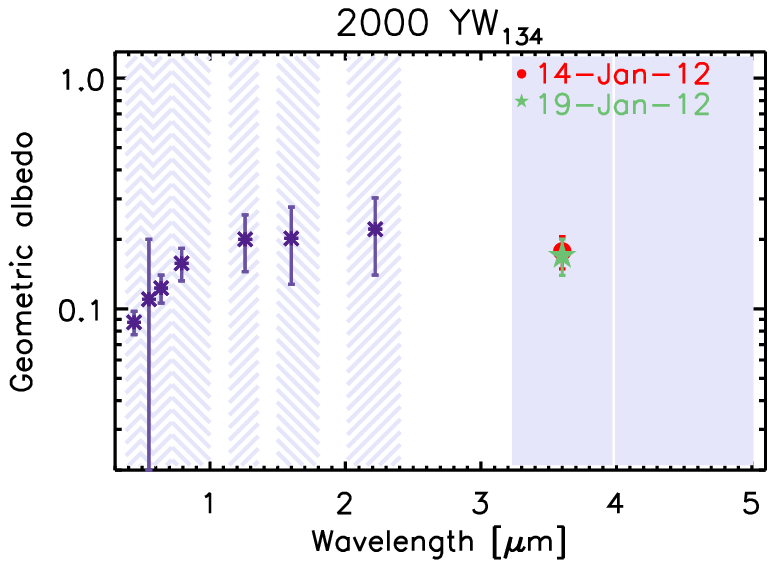}
~
      \includegraphics[width=0.5\columnwidth]{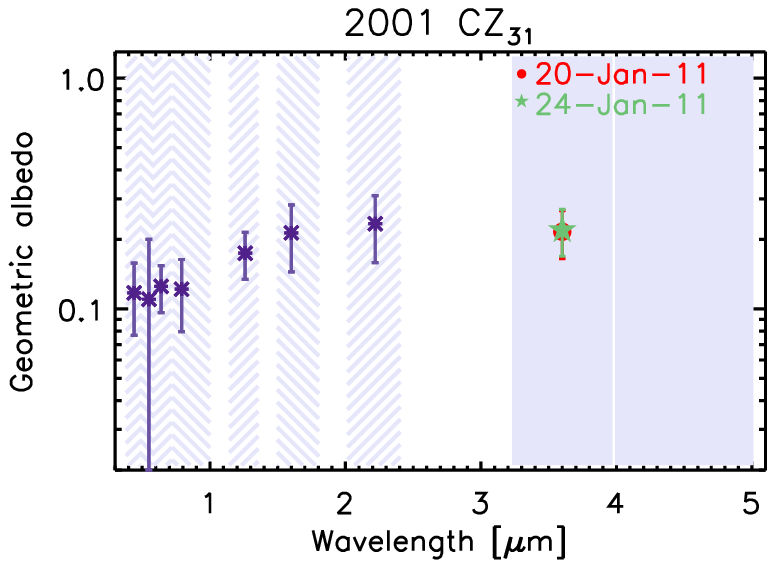}
 ~
        \includegraphics[width=0.5\columnwidth]{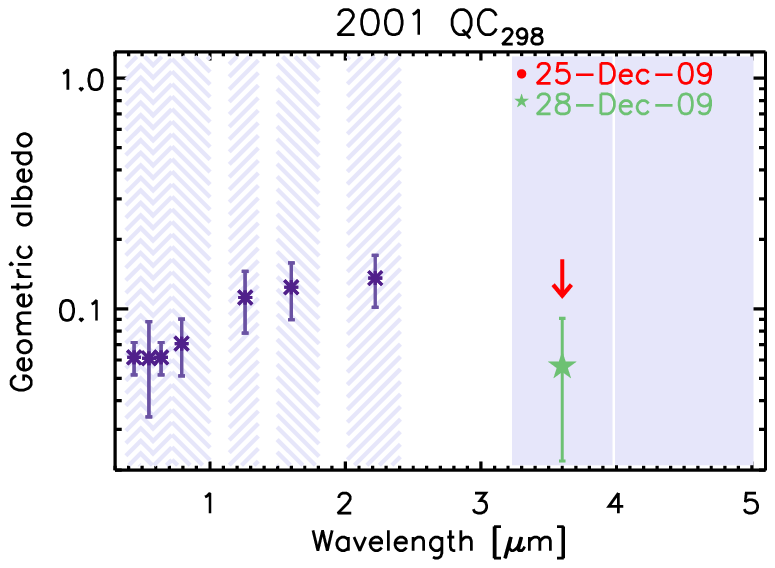}
~
        \includegraphics[width=0.5\columnwidth]{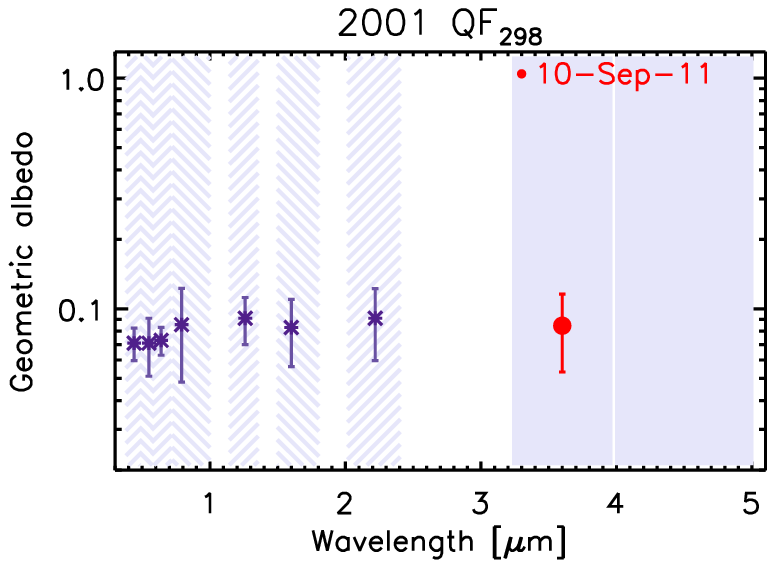}
~
      \includegraphics[width=0.5\columnwidth]{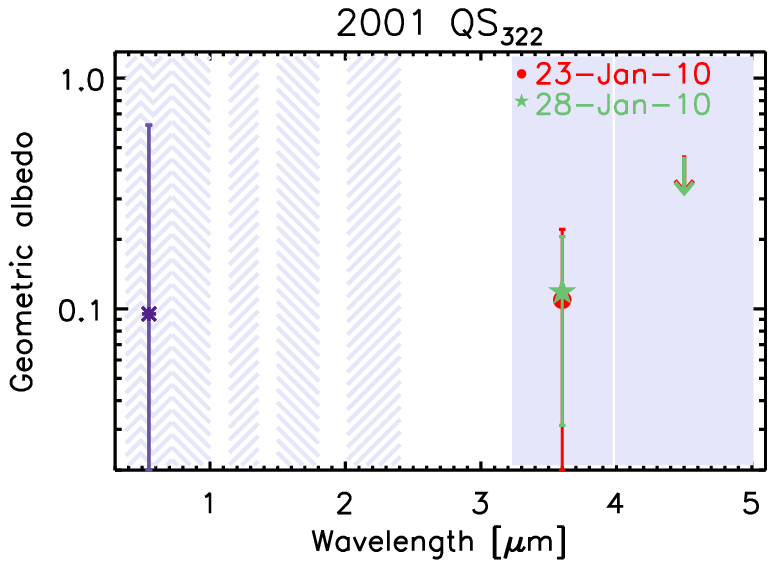}
~
        \includegraphics[width=0.5\columnwidth]{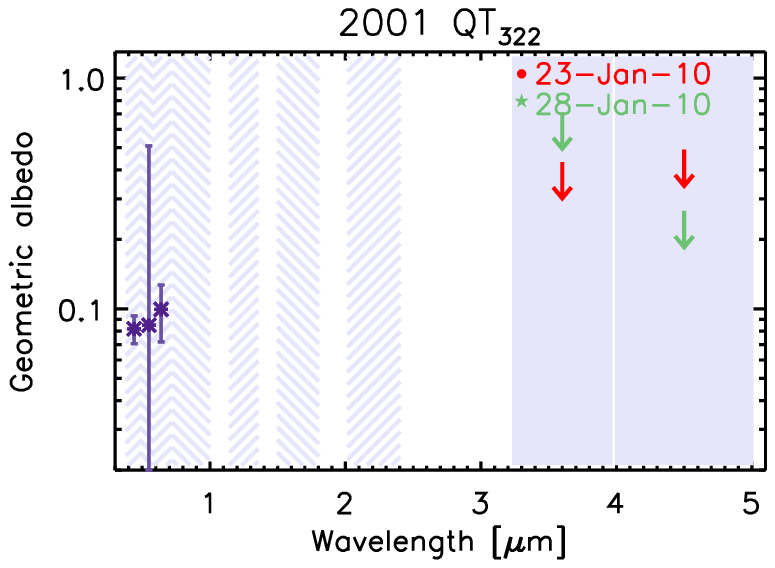}
  ~
        \includegraphics[width=0.5\columnwidth]{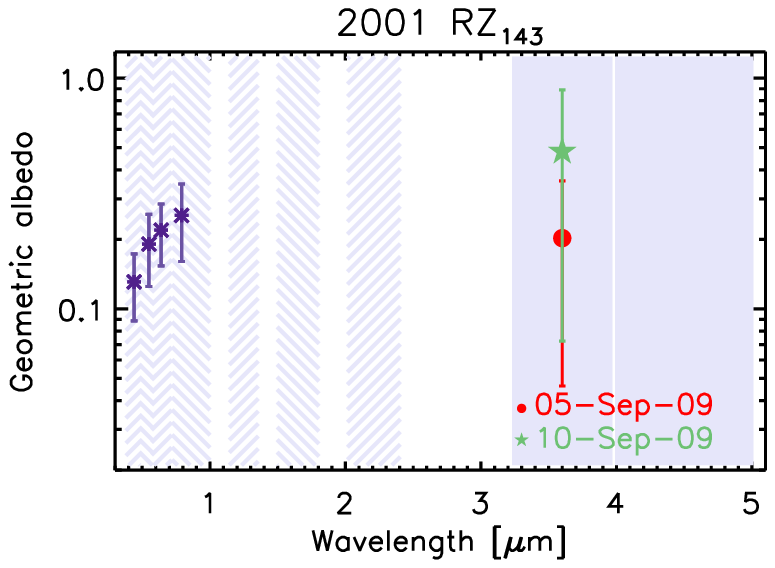}
~
      \includegraphics[width=0.5\columnwidth]{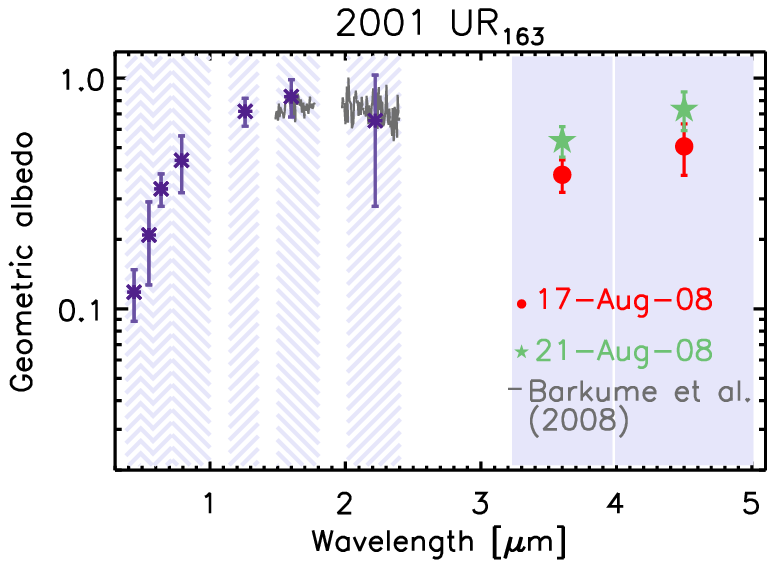}
  ~
        \includegraphics[width=0.5\columnwidth]{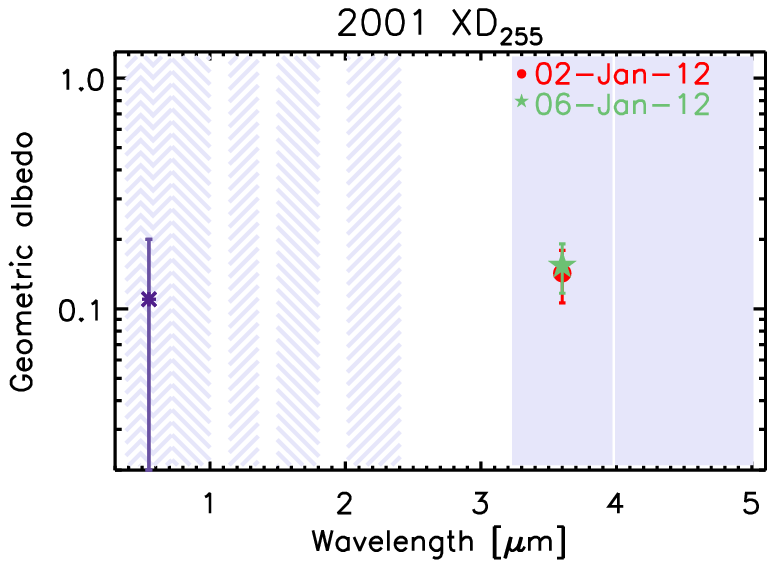}
~ 
  \includegraphics[width=0.5\columnwidth]{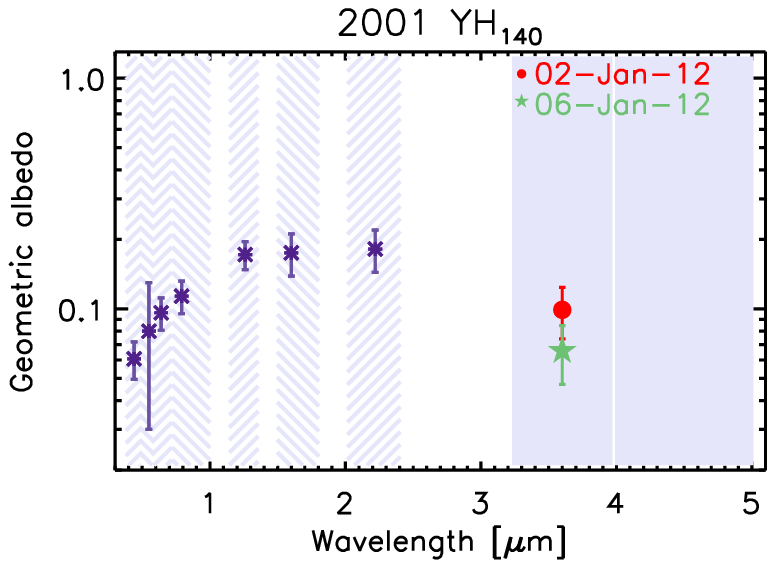}
~
      \includegraphics[width=0.5\columnwidth]{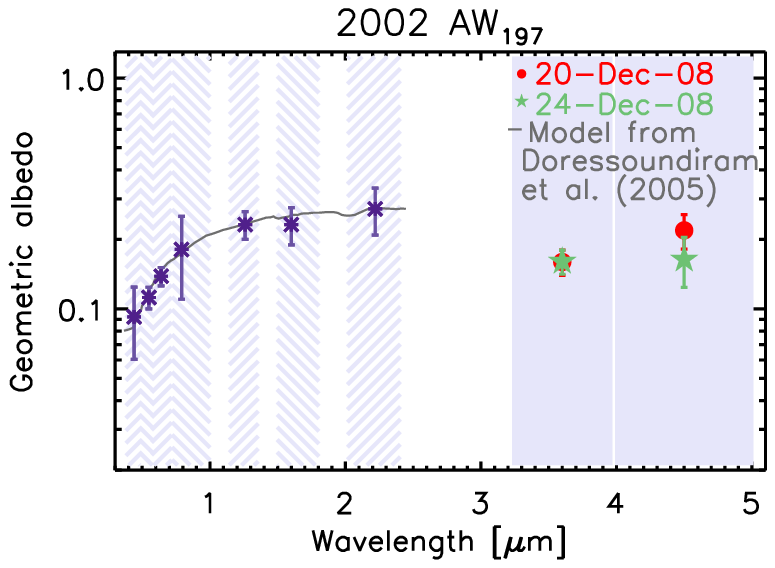}
~
      \includegraphics[width=0.5\columnwidth]{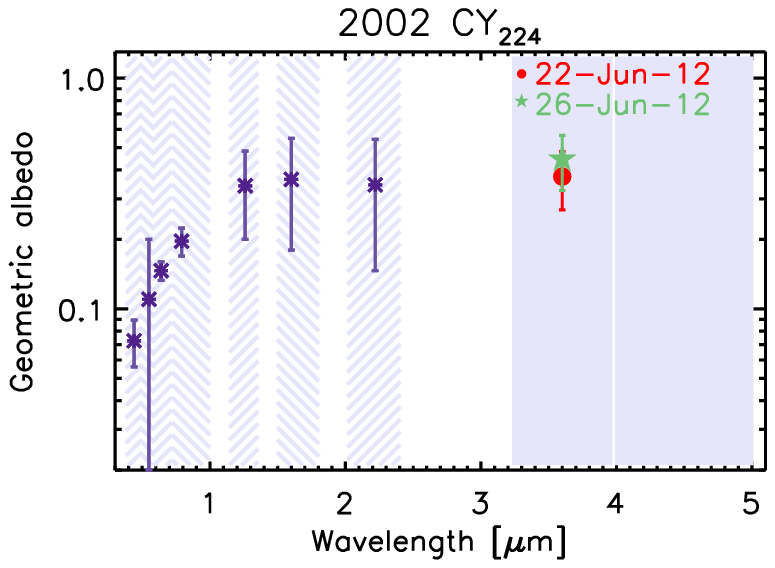}
  ~
        \includegraphics[width=0.5\columnwidth]{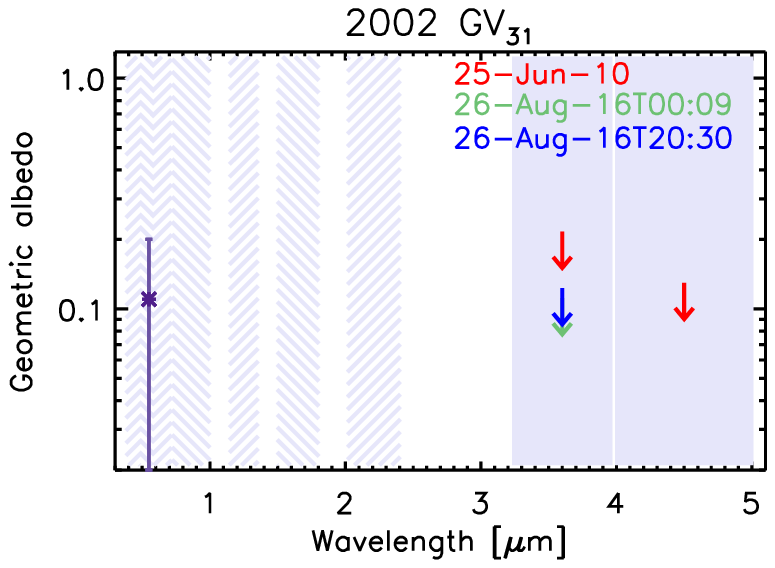}
~
        \includegraphics[width=0.5\columnwidth]{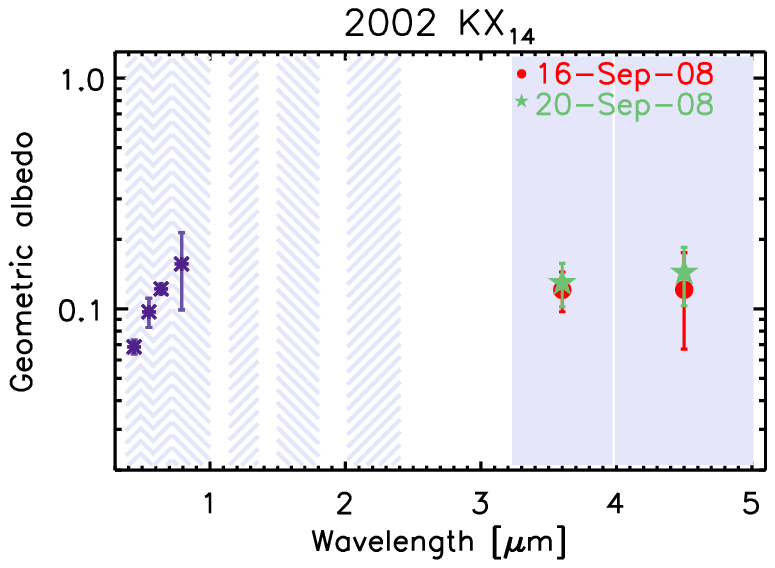}
~
      \includegraphics[width=0.5\columnwidth]{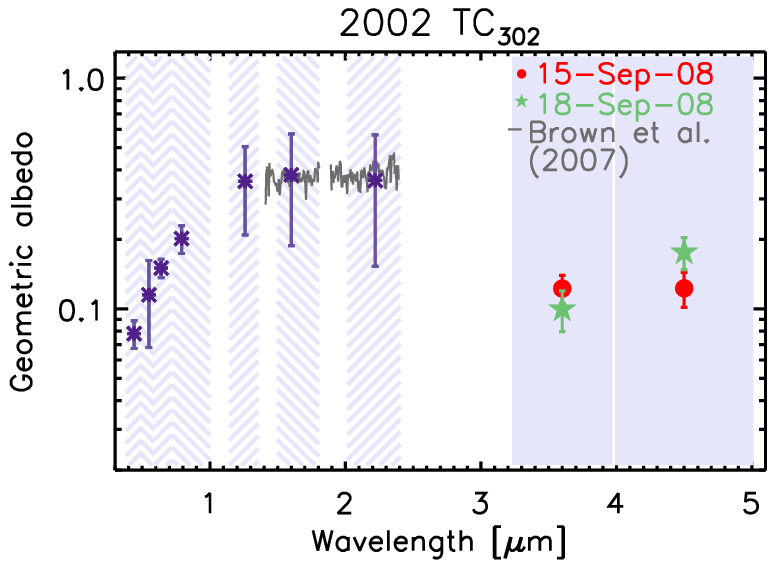}
~
        \includegraphics[width=0.5\columnwidth]{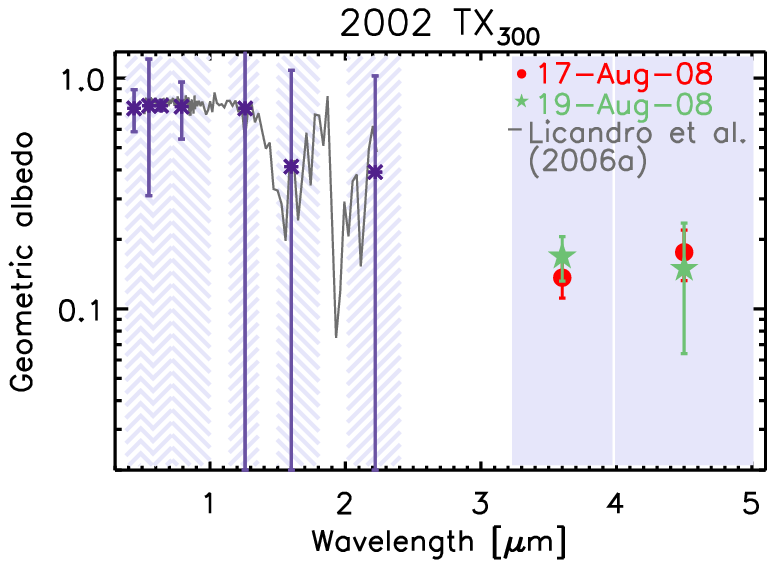}
 ~
        \includegraphics[width=0.5\columnwidth]{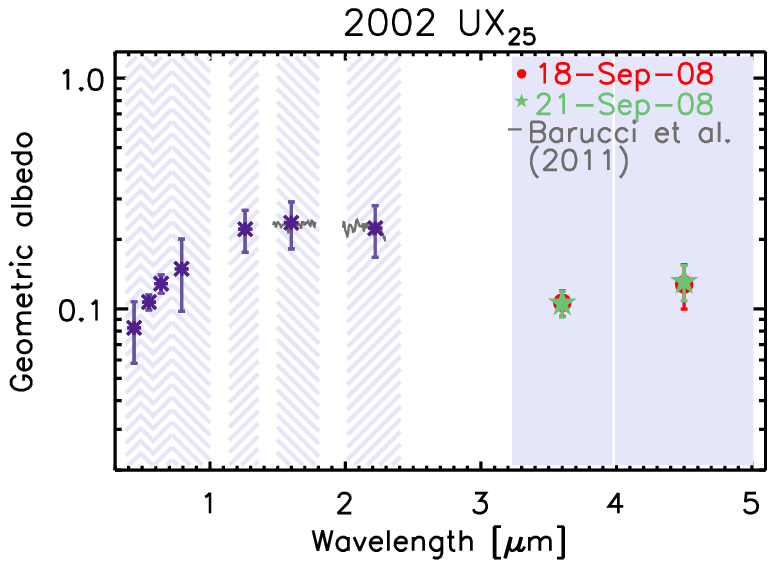}
~
      \includegraphics[width=0.5\columnwidth]{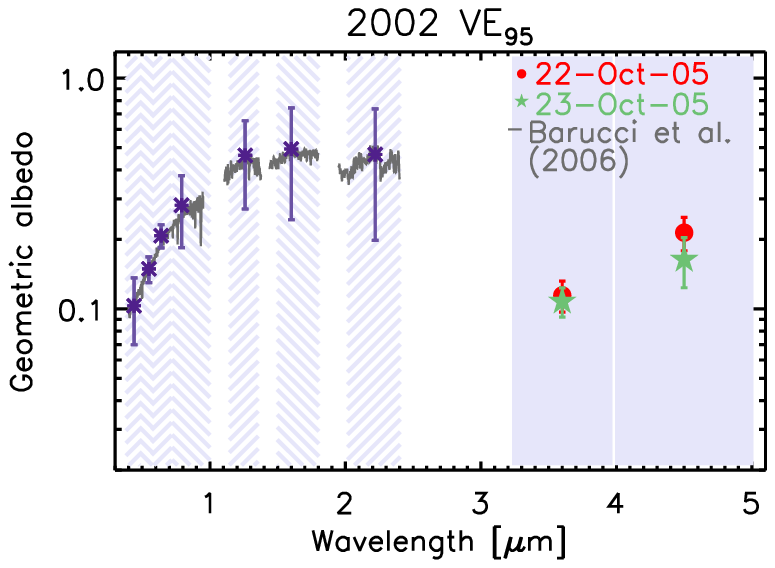}
    ~
        \includegraphics[width=0.5\columnwidth]{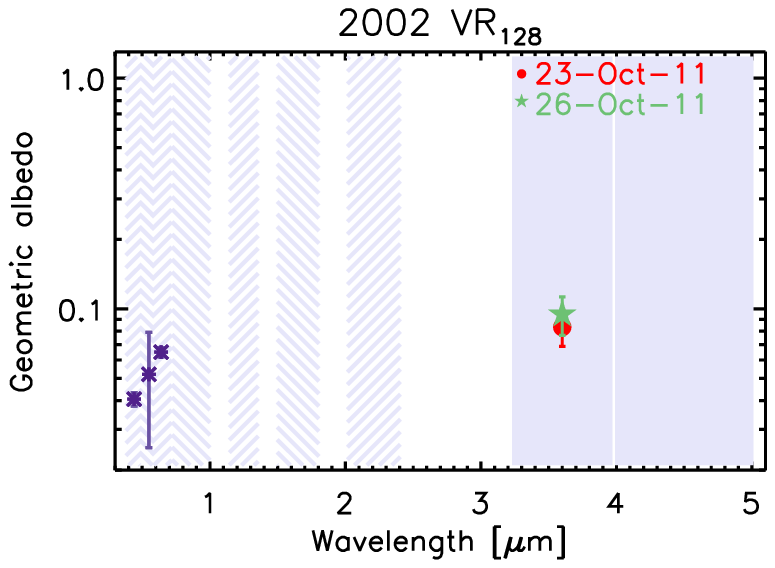}
~
        \includegraphics[width=0.5\columnwidth]{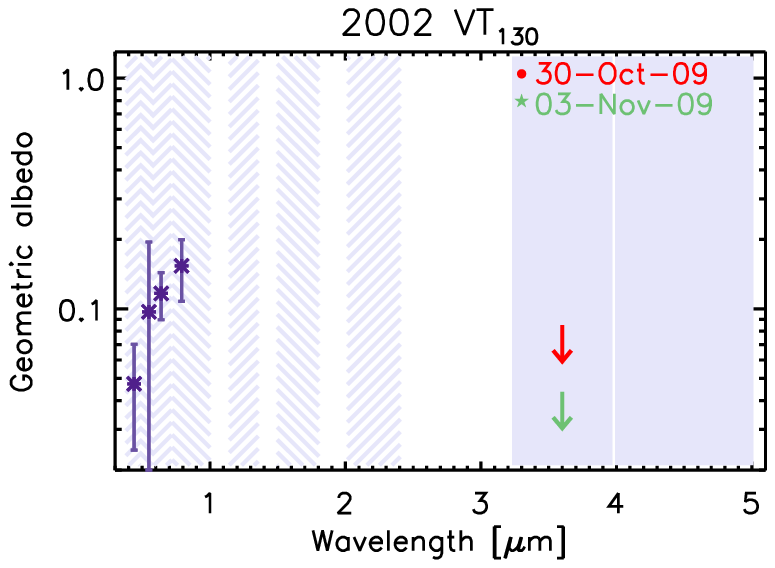}
~
      \includegraphics[width=0.5\columnwidth]{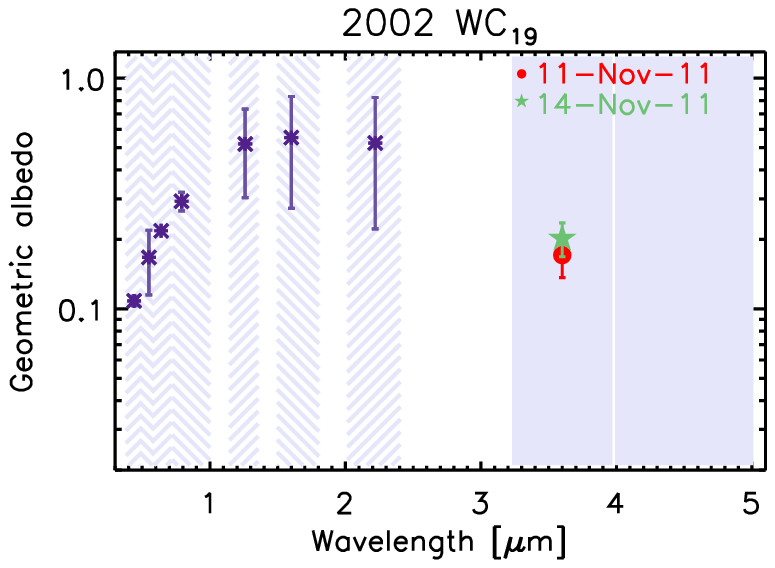}
~
        \includegraphics[width=0.5\columnwidth]{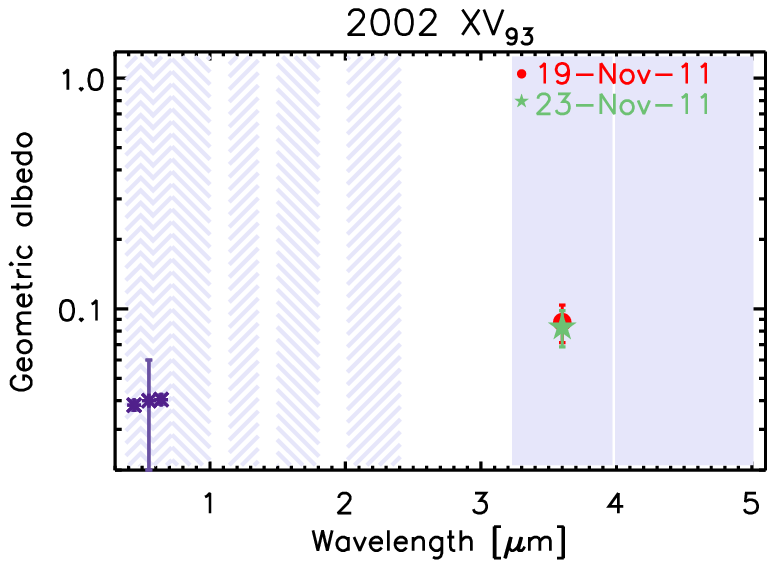}
~
        \includegraphics[width=0.5\columnwidth]{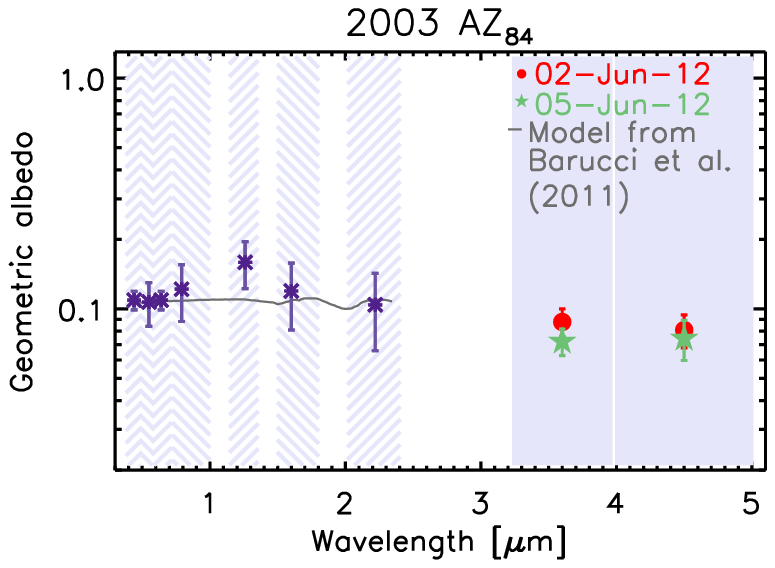}
~
     \includegraphics[width=0.5\columnwidth]{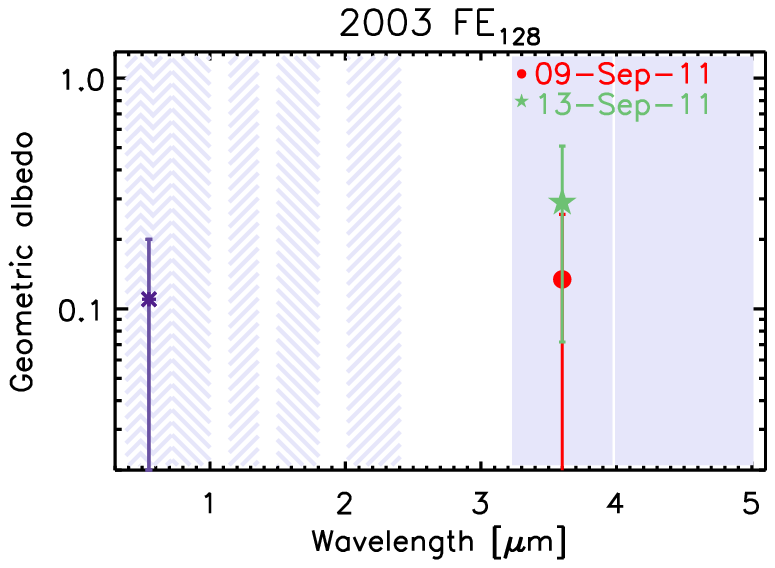}        
~
        \includegraphics[width=0.5\columnwidth]{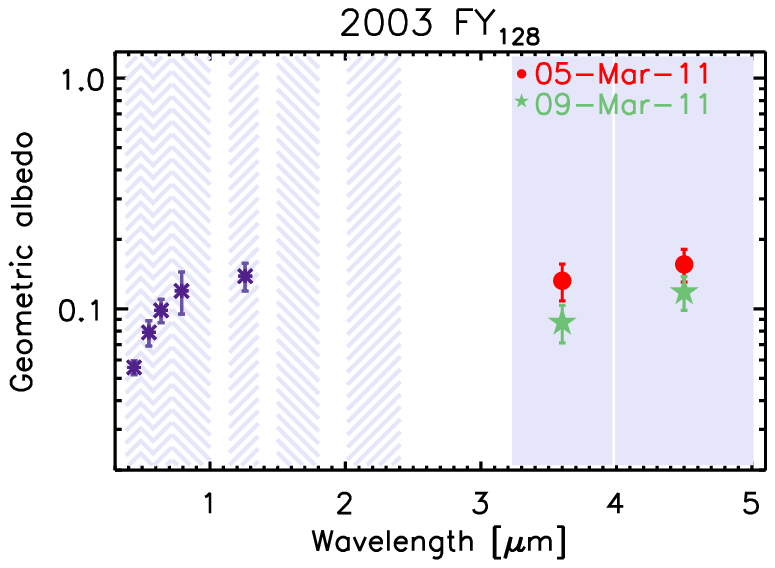}
~
        \includegraphics[width=0.5\columnwidth]{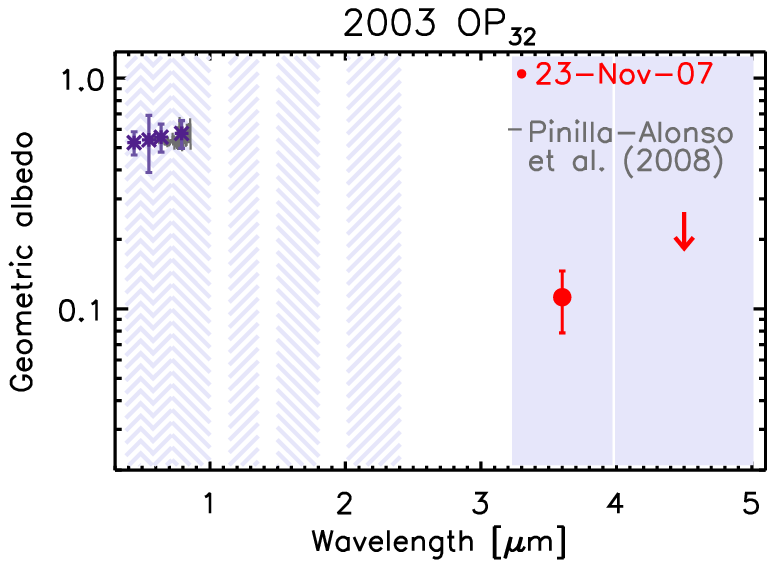}
  ~
      \includegraphics[width=0.5\columnwidth]{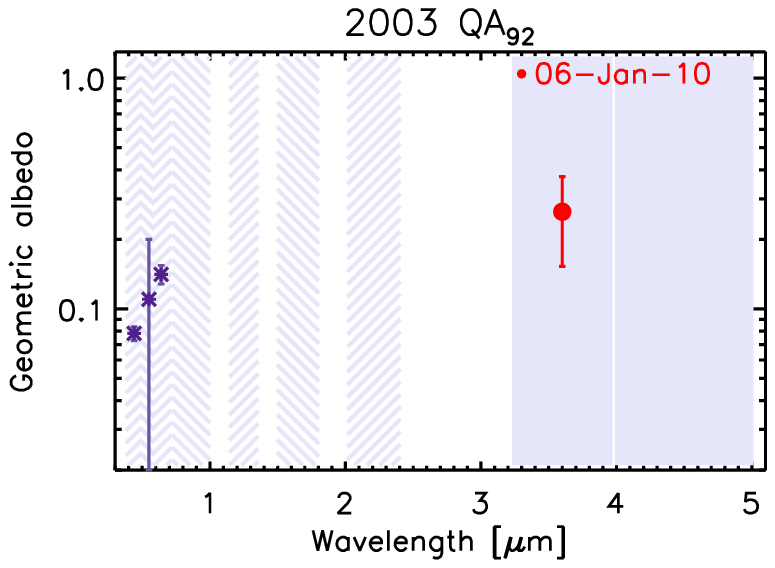}
~
        \includegraphics[width=0.5\columnwidth]{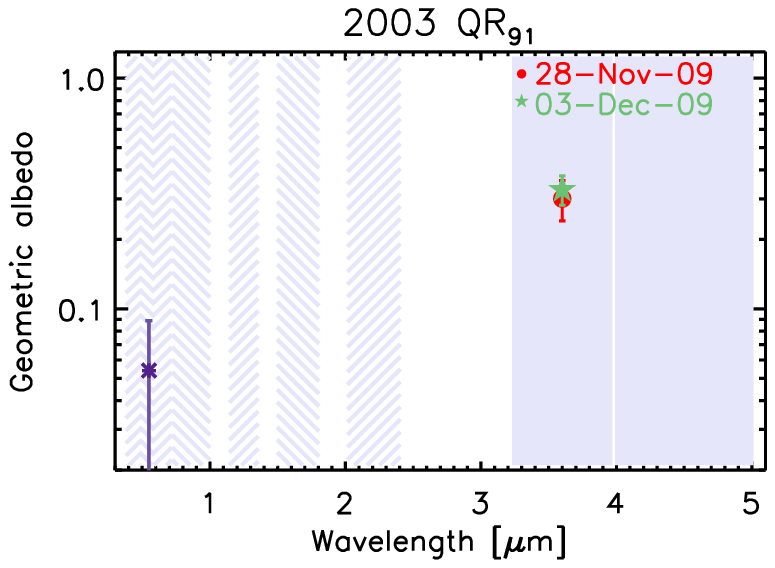}
  ~
        \includegraphics[width=0.5\columnwidth]{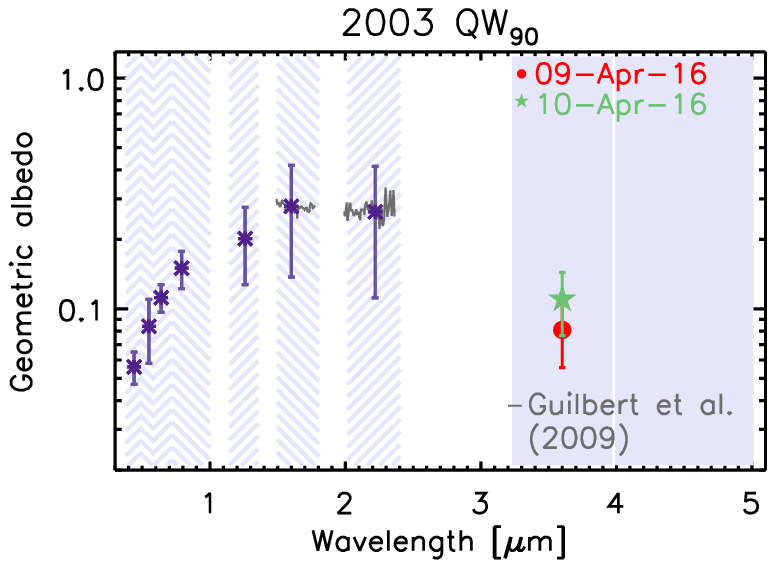}
~
      \includegraphics[width=0.5\columnwidth]{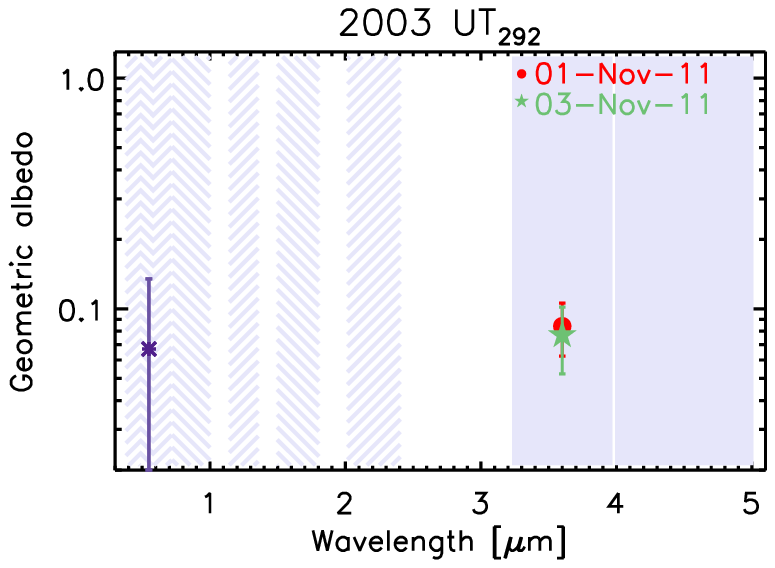}
  ~
        \includegraphics[width=0.5\columnwidth]{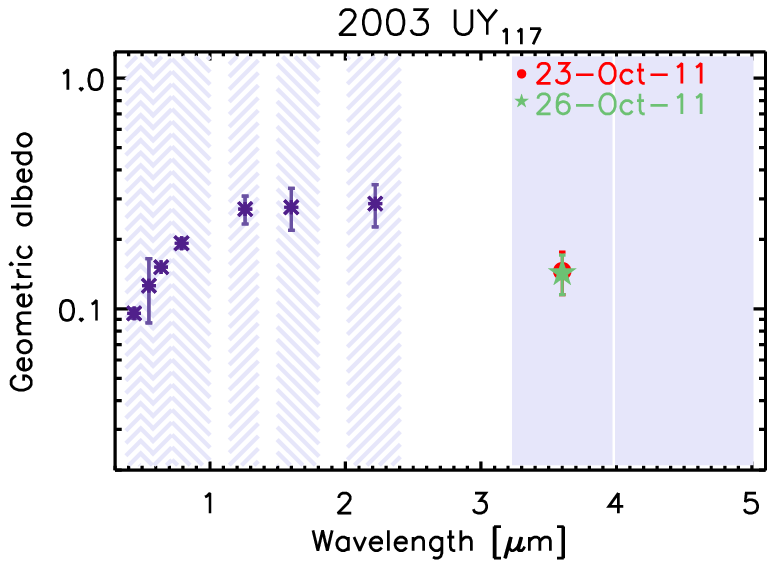}
~
     \includegraphics[width=0.5\columnwidth]{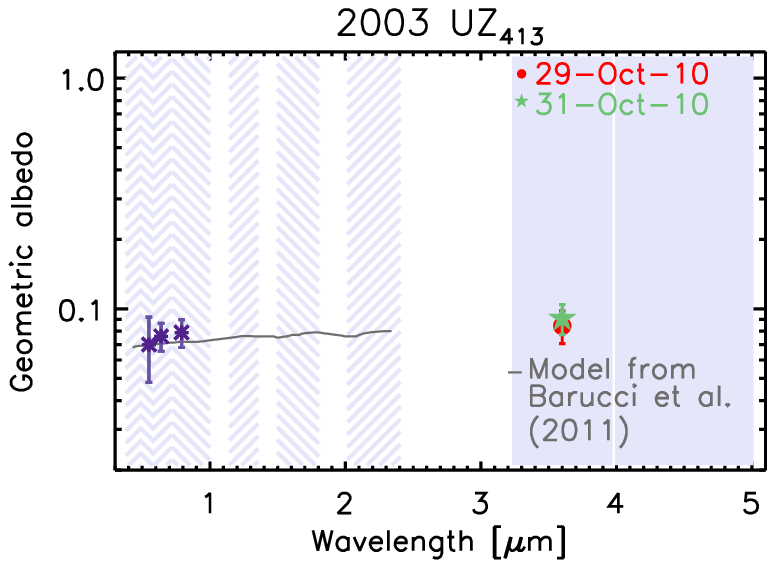}
    ~
        \includegraphics[width=0.5\columnwidth]{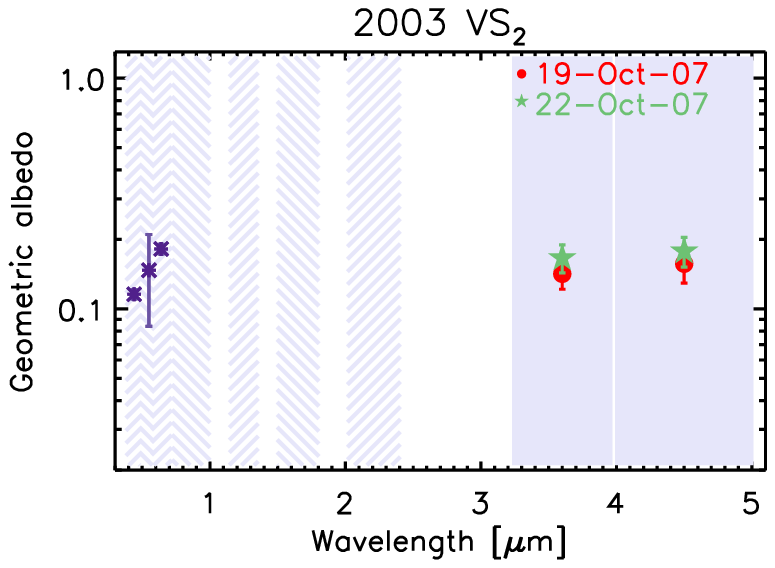}
~
      \includegraphics[width=0.5\columnwidth]{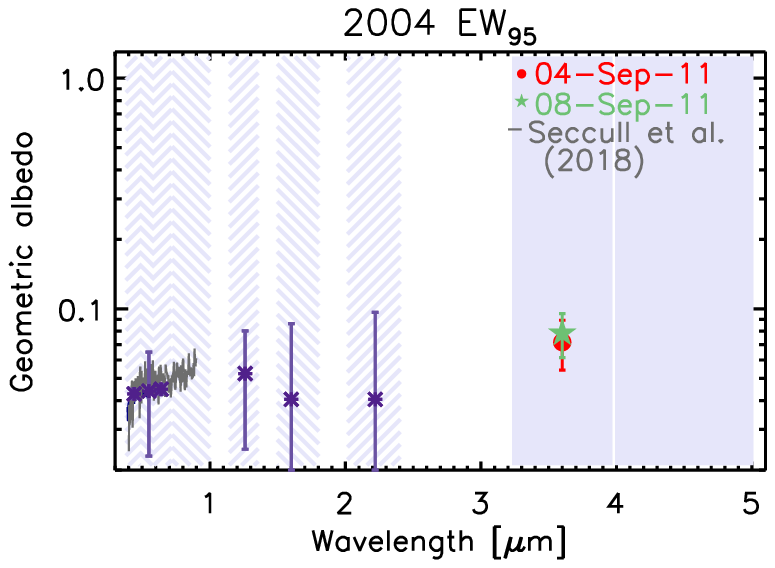}
  ~
        \includegraphics[width=0.5\columnwidth]{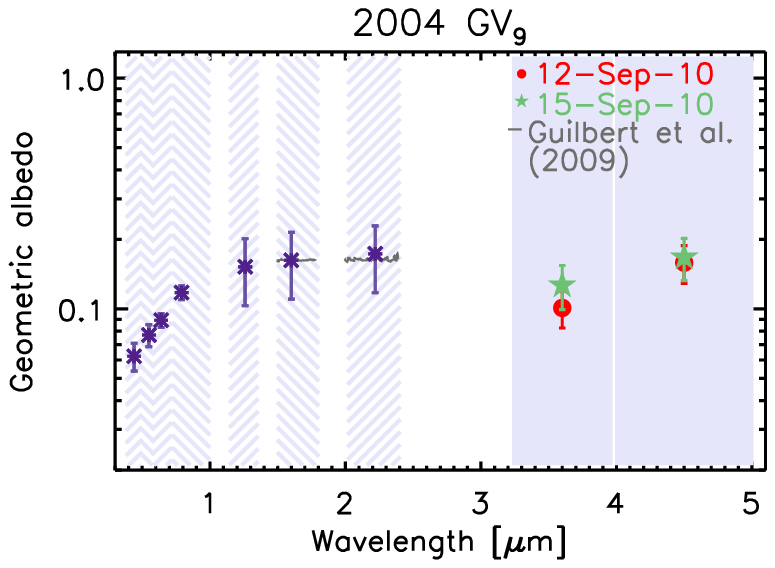}
~
        \includegraphics[width=0.5\columnwidth]{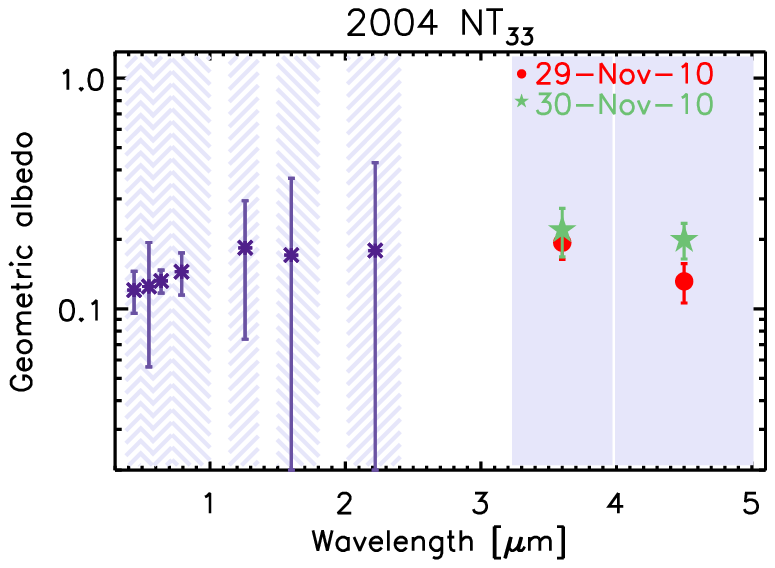}
~
        \includegraphics[width=0.5\columnwidth]{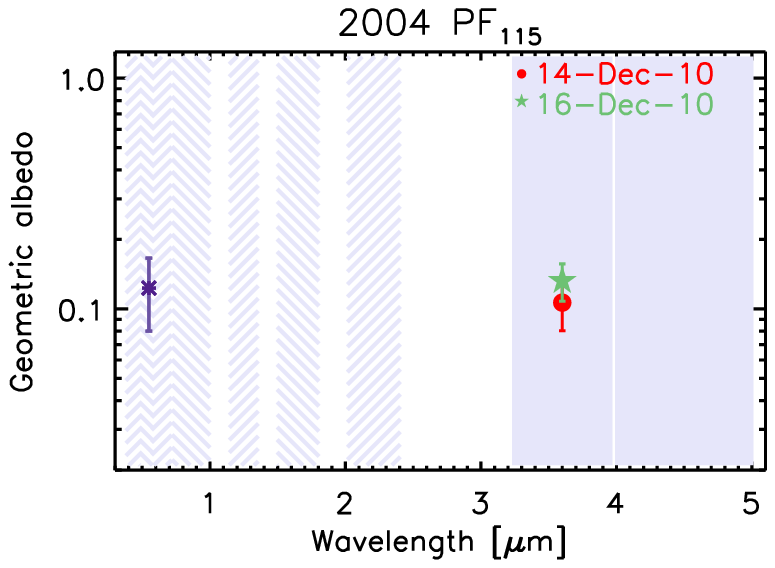}
~
        \includegraphics[width=0.5\columnwidth]{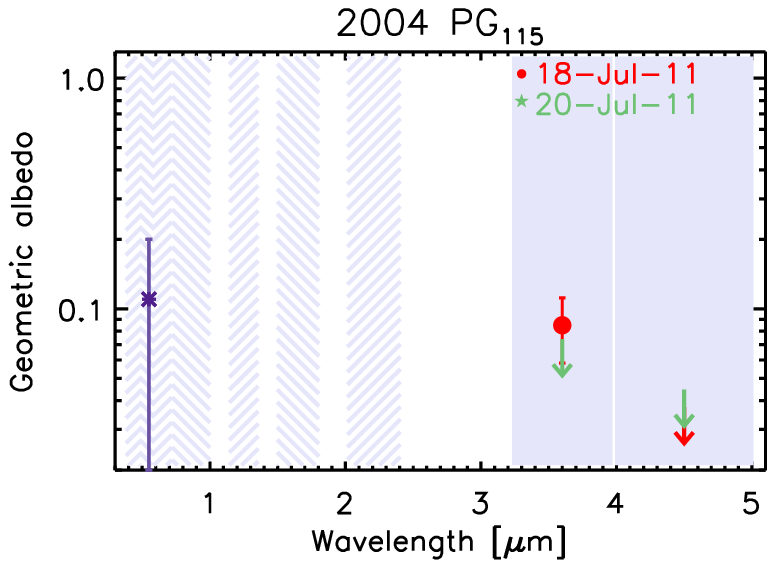}
  ~
        \includegraphics[width=0.5\columnwidth]{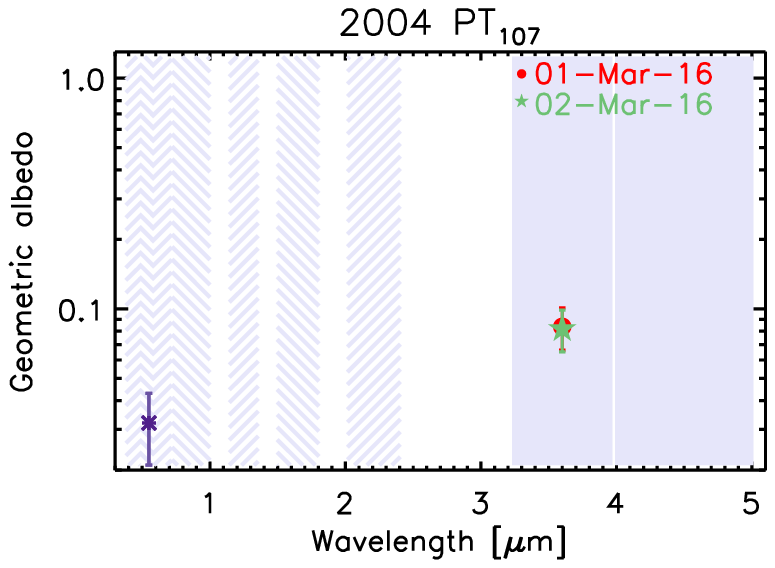}
~
        \includegraphics[width=0.5\columnwidth]{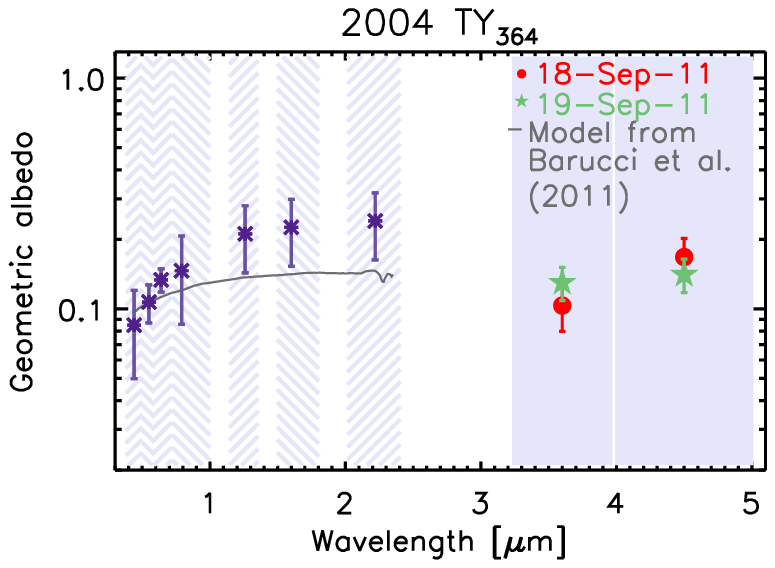}
  ~
        \includegraphics[width=0.5\columnwidth]{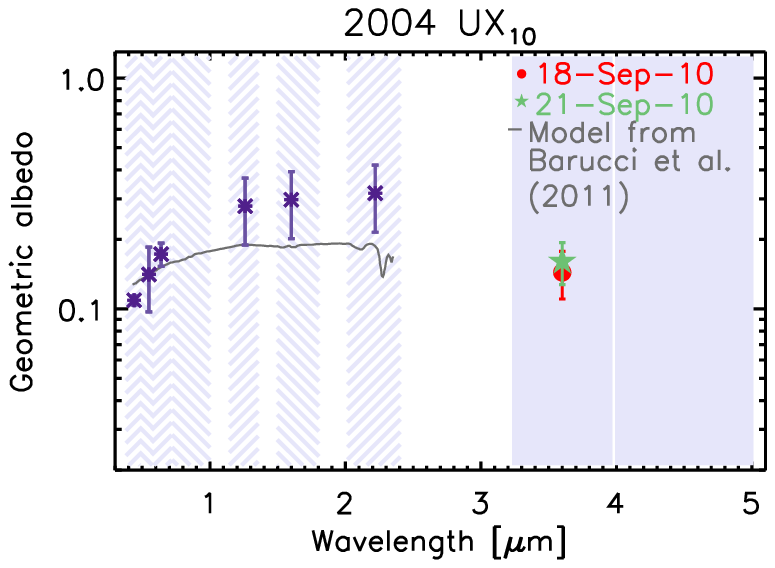}
~
        \includegraphics[width=0.5\columnwidth]{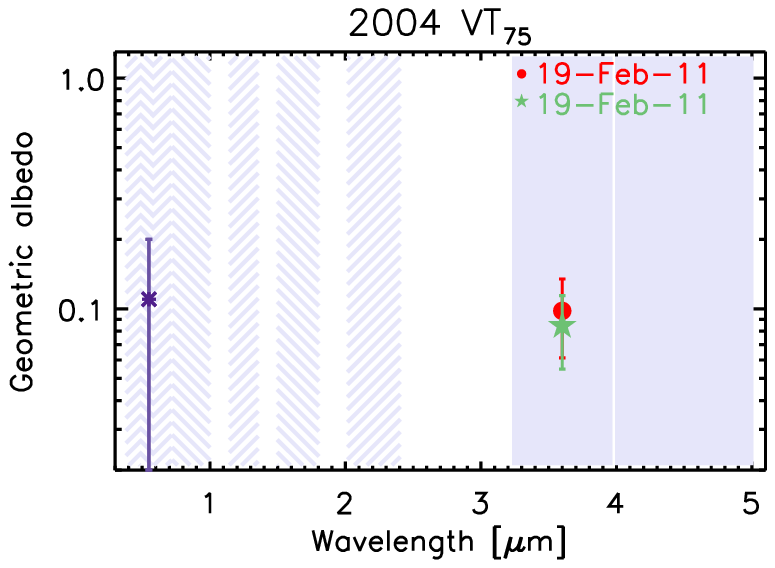}
~
         \includegraphics[width=0.5\columnwidth]{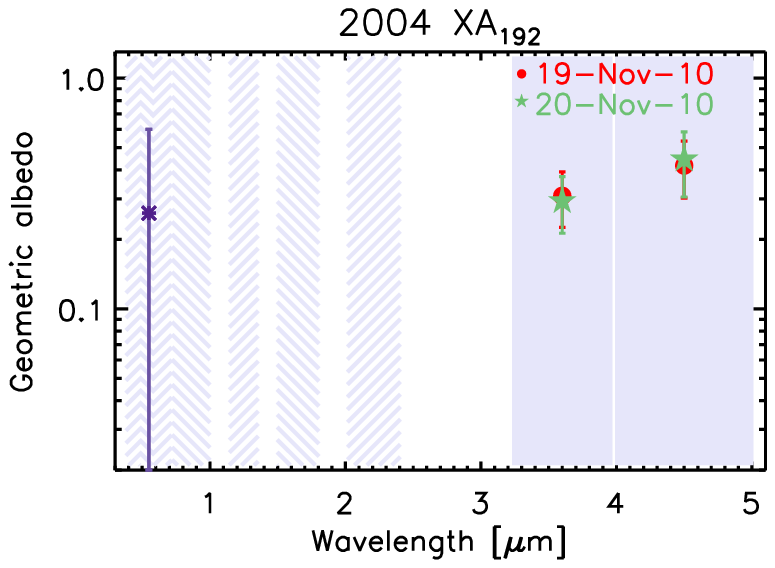}
~
        \includegraphics[width=0.5\columnwidth]{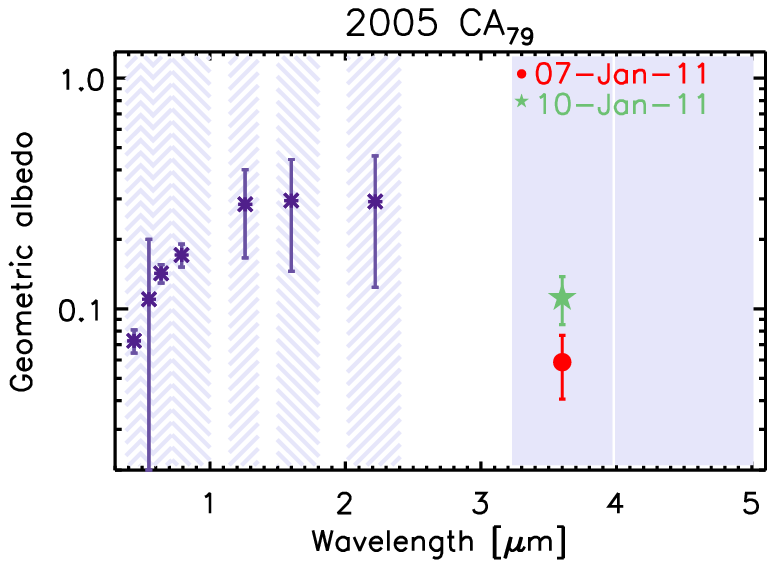}
  ~
          \includegraphics[width=0.5\columnwidth]{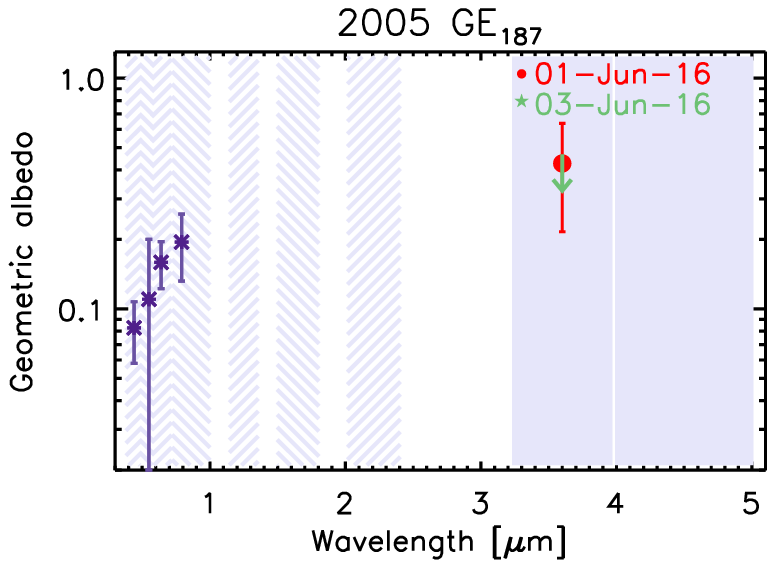}
~
        \includegraphics[width=0.5\columnwidth]{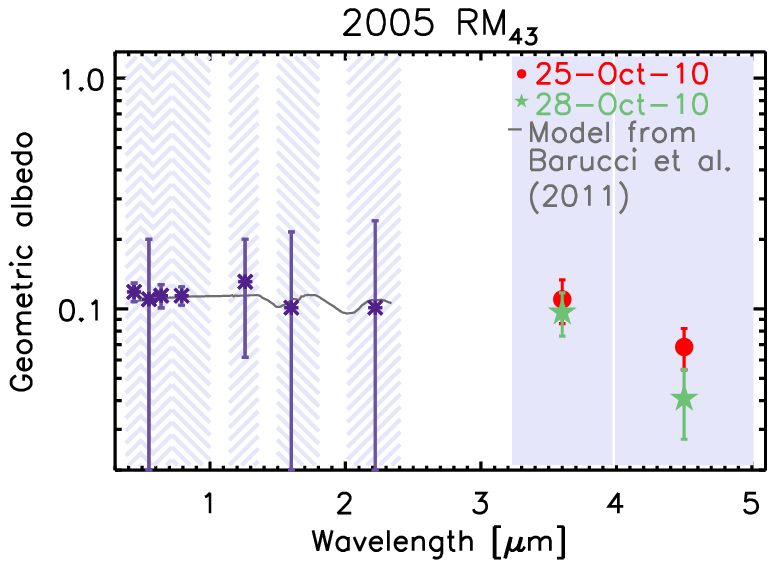}
  ~
         \includegraphics[width=0.5\columnwidth]{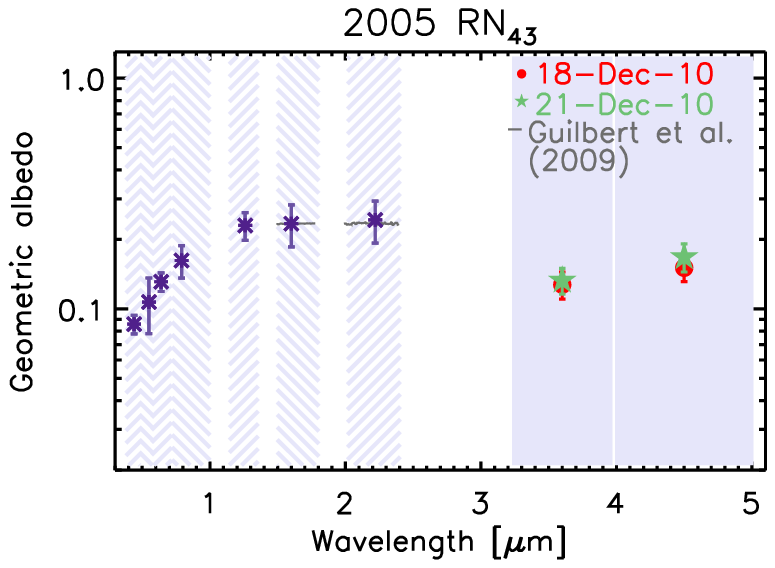}
~
        \includegraphics[width=0.5\columnwidth]{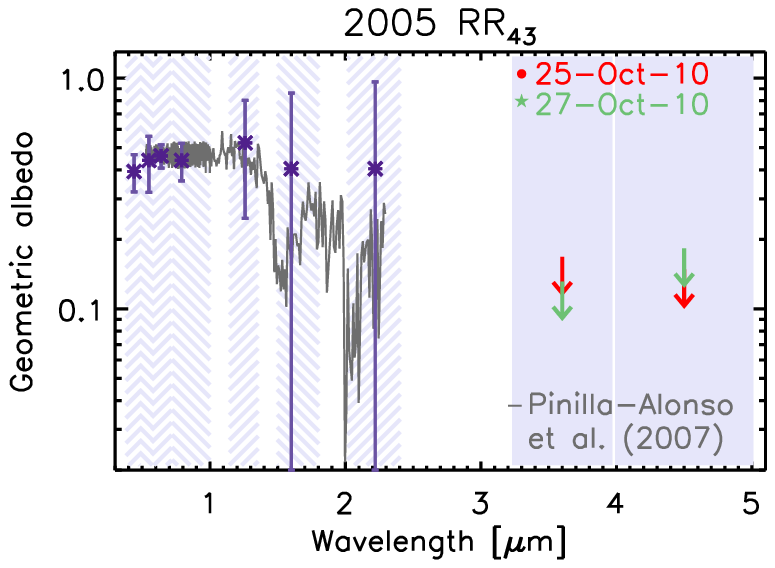}
 ~
        \includegraphics[width=0.5\columnwidth]{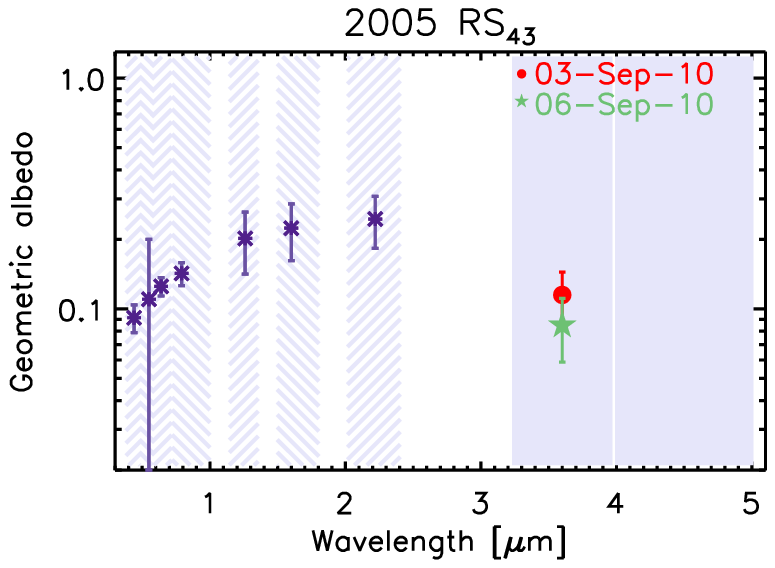}
~
         \includegraphics[width=0.5\columnwidth]{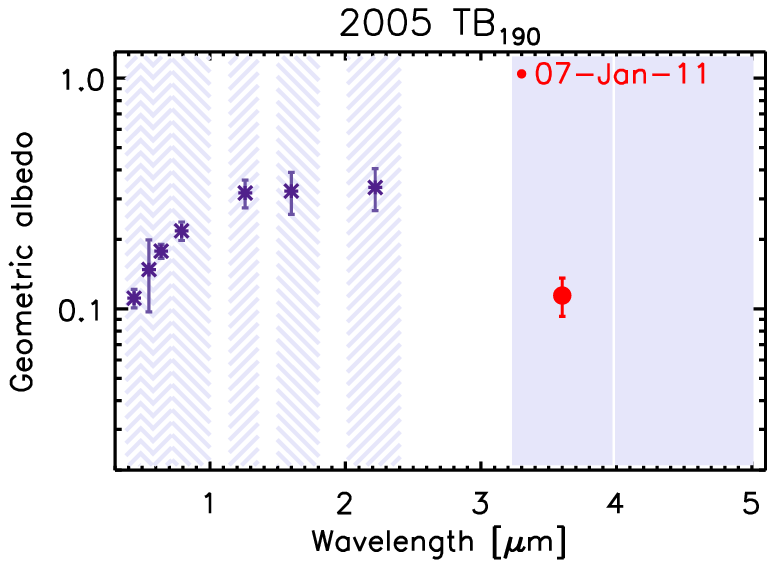}
~
        \includegraphics[width=0.5\columnwidth]{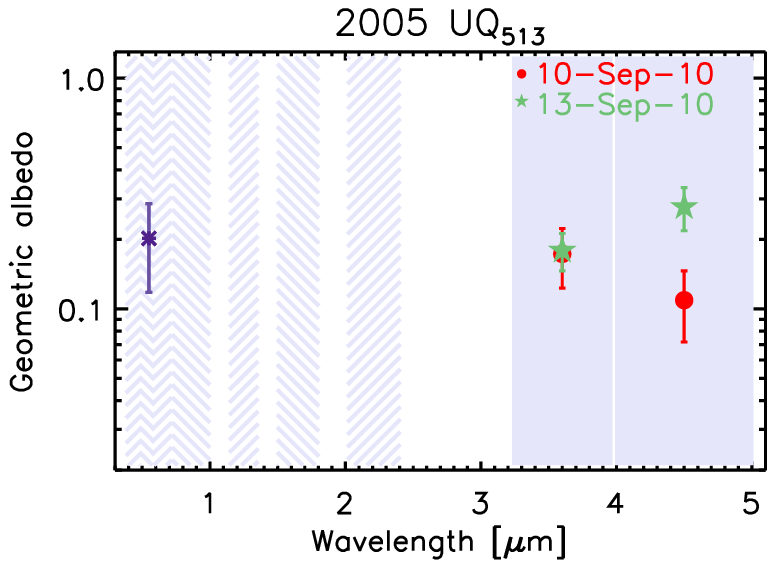}
~
      \includegraphics[width=0.5\columnwidth]{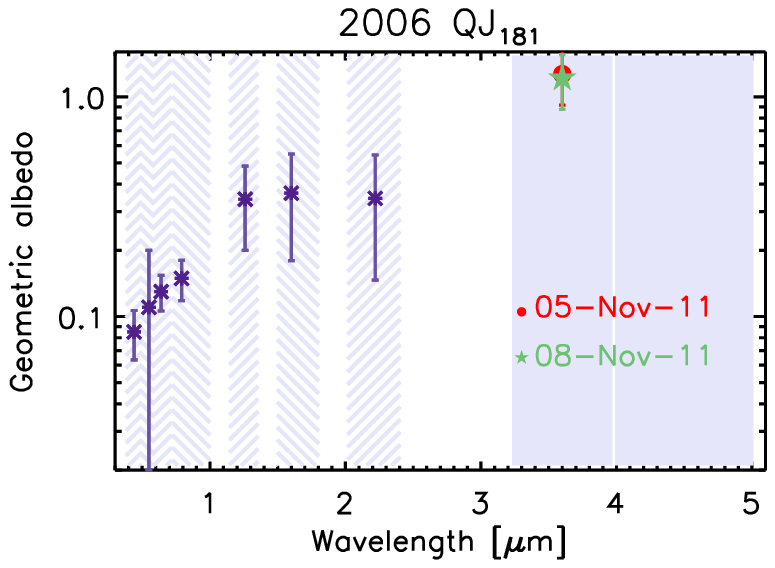}
~
        \includegraphics[width=0.5\columnwidth]{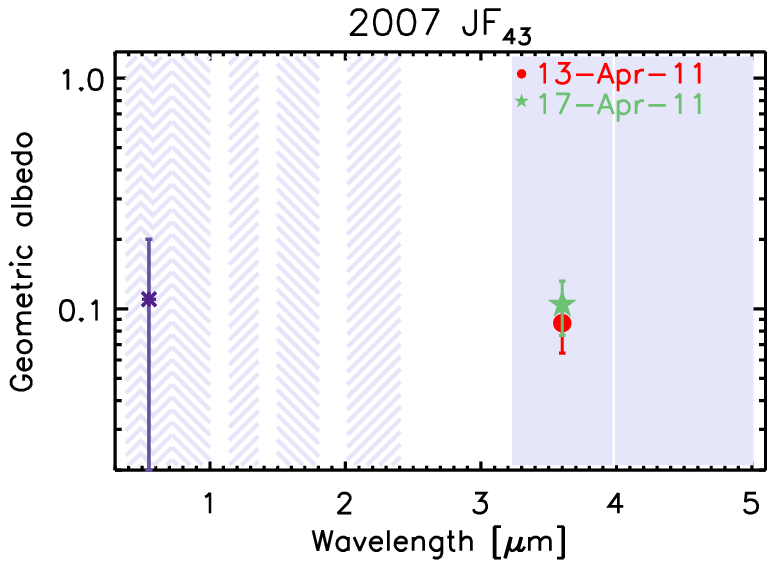}
~
         \includegraphics[width=0.5\columnwidth]{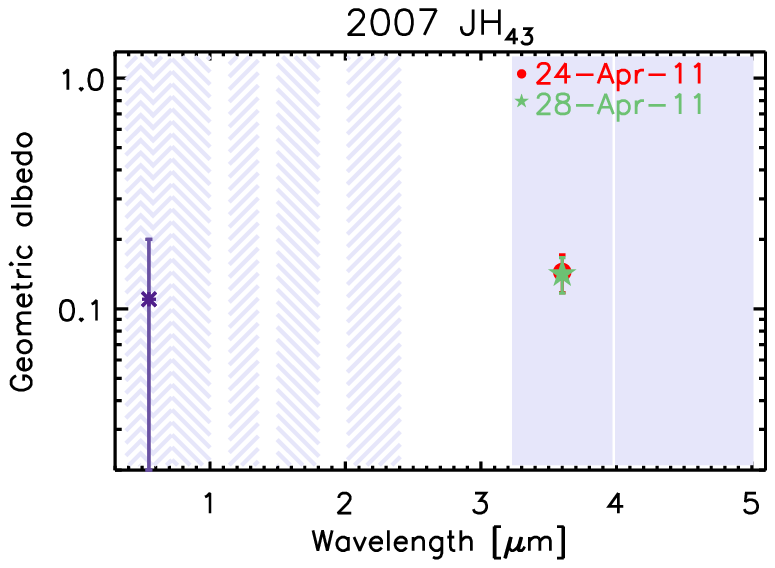}
  ~
        \includegraphics[width=0.5\columnwidth]{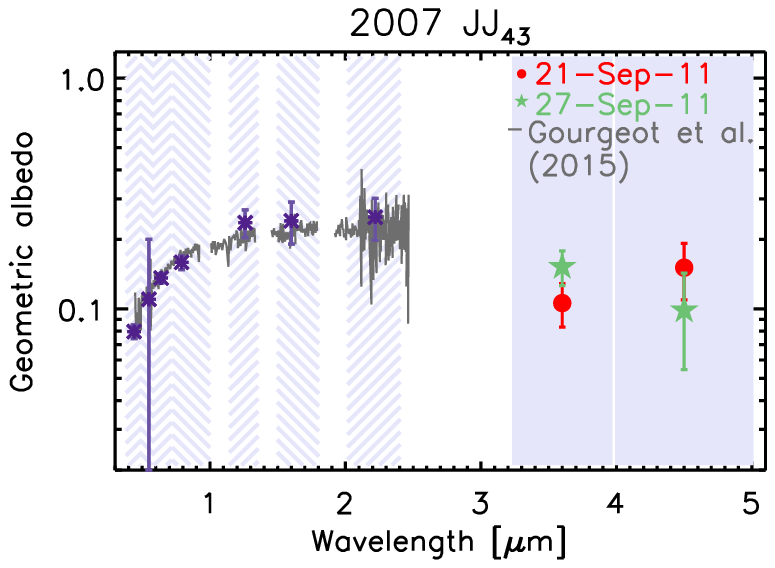}
~
        \includegraphics[width=0.5\columnwidth]{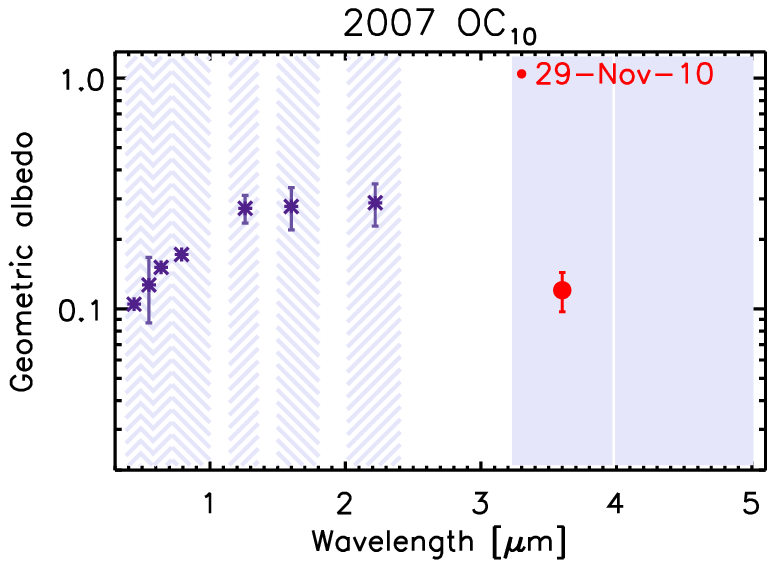}
  ~
        \includegraphics[width=0.5\columnwidth]{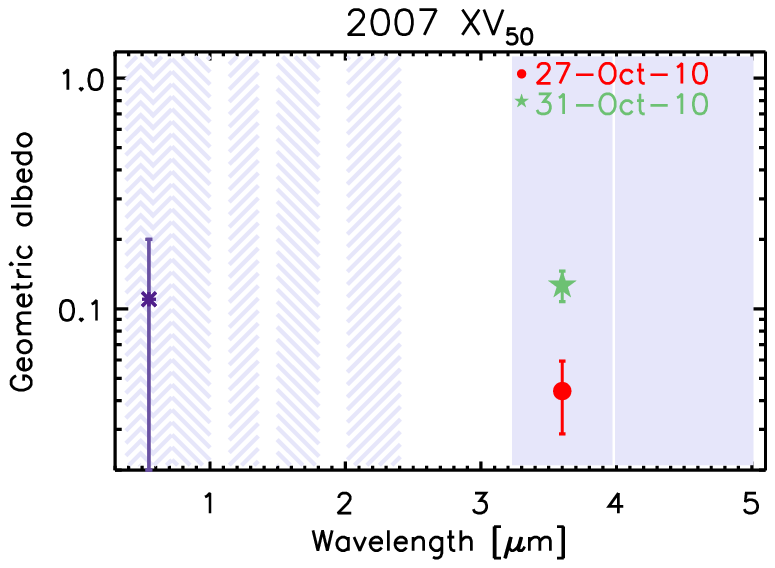}
~
          \includegraphics[width=0.5\columnwidth]{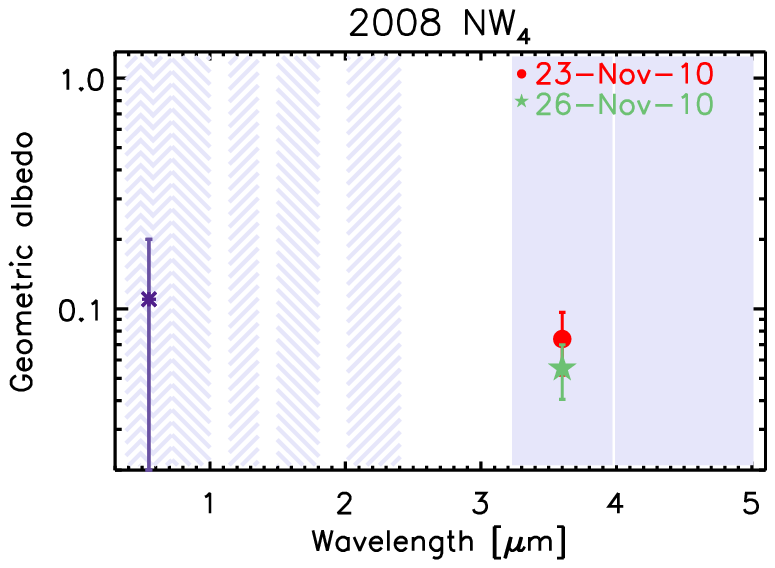}
~
        \includegraphics[width=0.5\columnwidth]{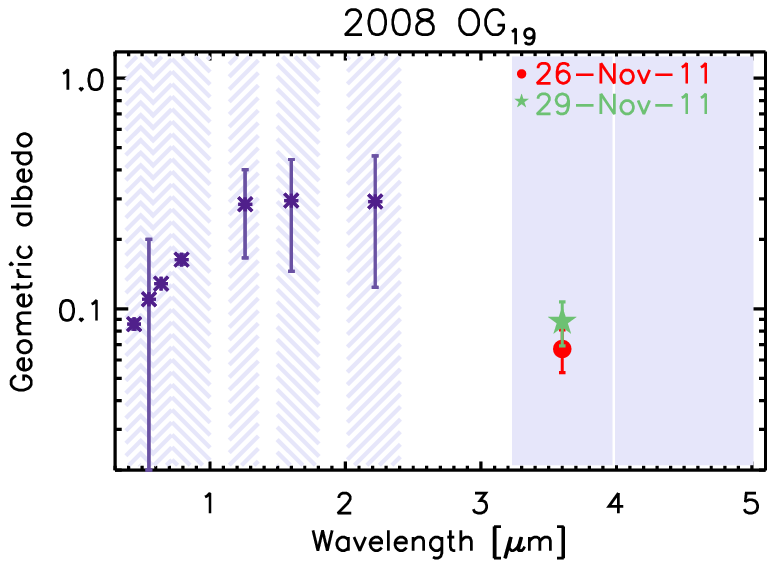}
 ~
        \includegraphics[width=0.5\columnwidth]{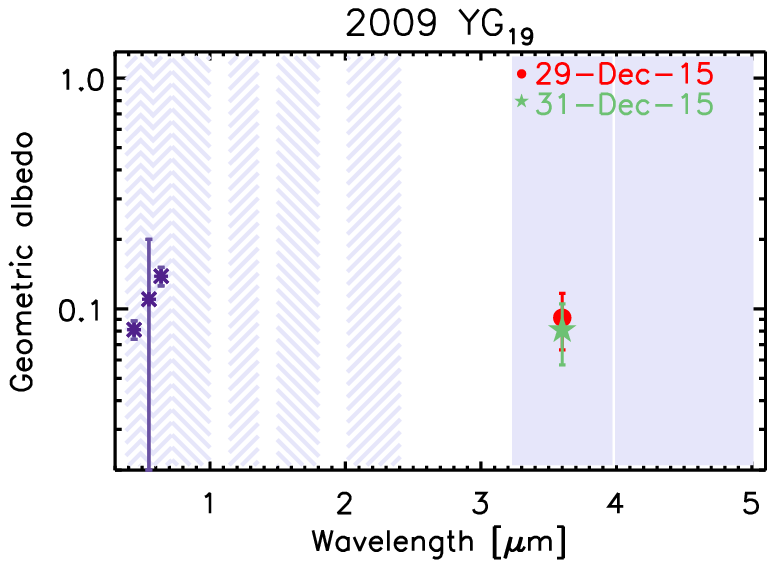}
~
        \includegraphics[width=0.5\columnwidth]{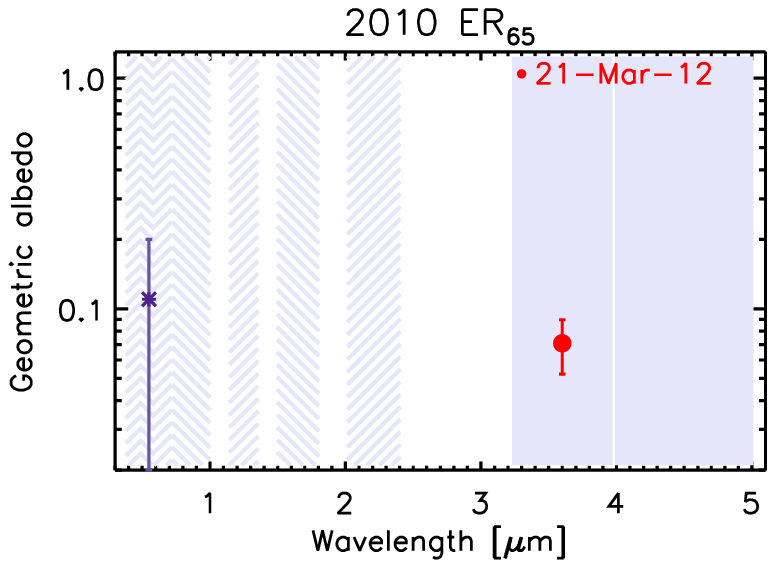}
  ~
        \includegraphics[width=0.5\columnwidth]{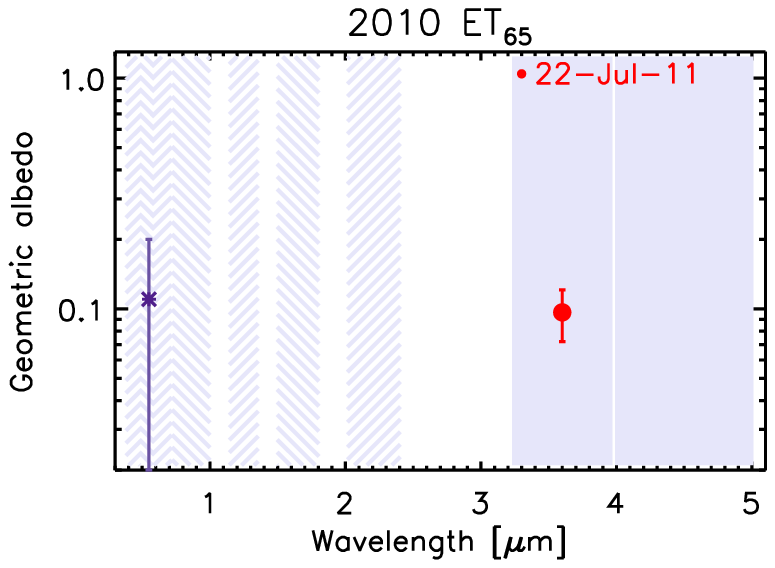}
~
          \includegraphics[width=0.5\columnwidth]{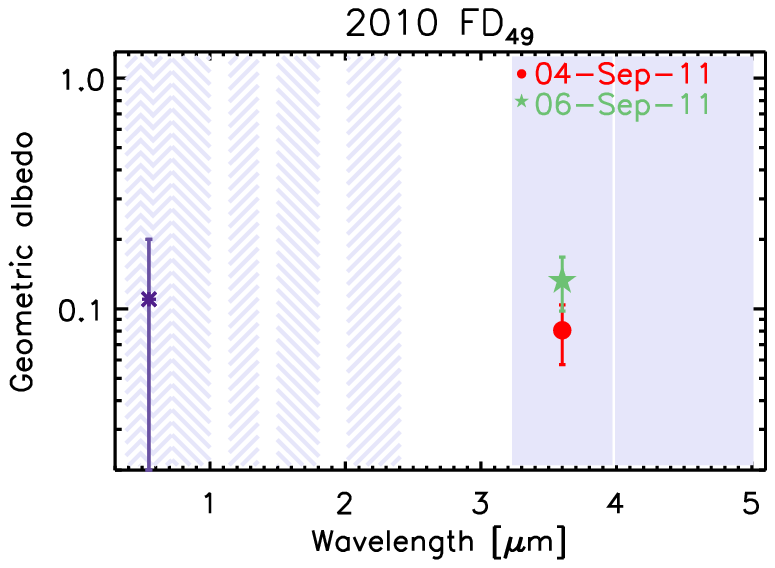}
~
        \includegraphics[width=0.5\columnwidth]{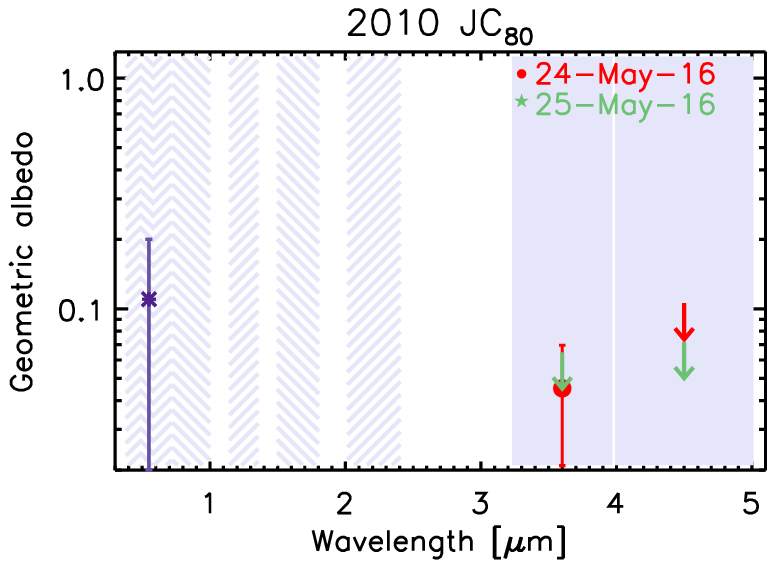}
~
     \includegraphics[width=0.5\columnwidth]{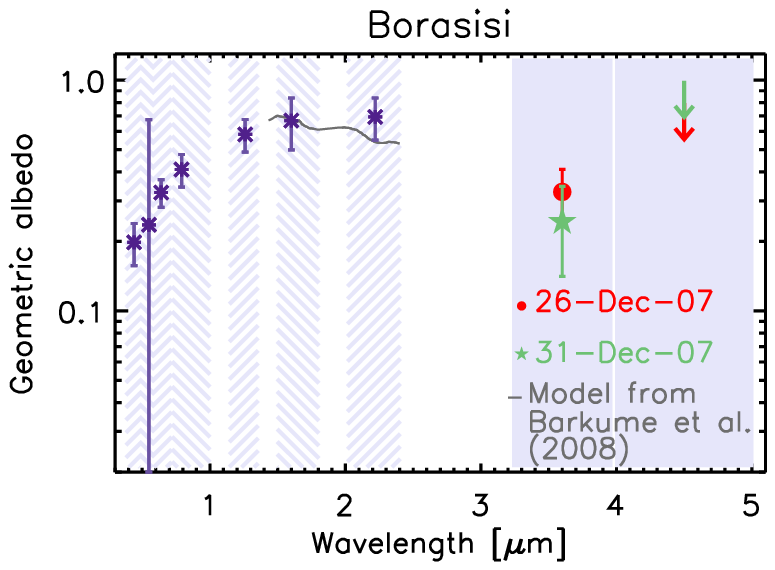}
 ~
          \includegraphics[width=0.5\columnwidth]{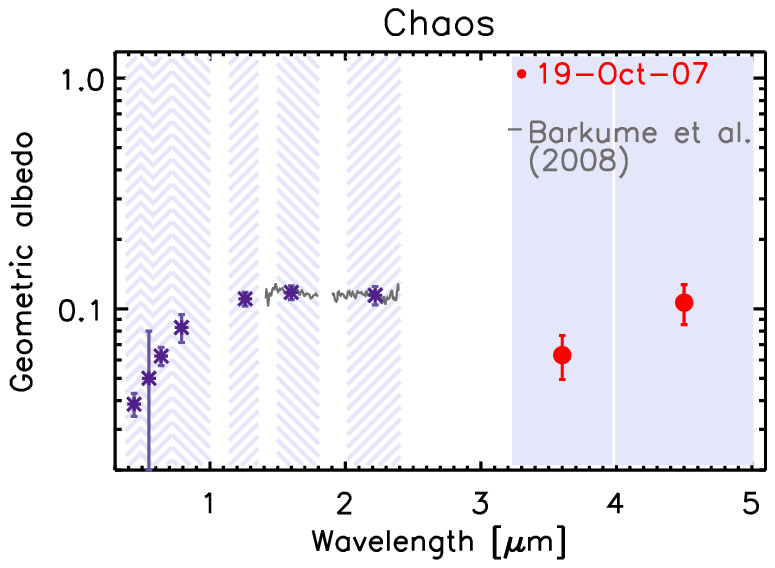}
~ 
                    \includegraphics[width=0.5\columnwidth]{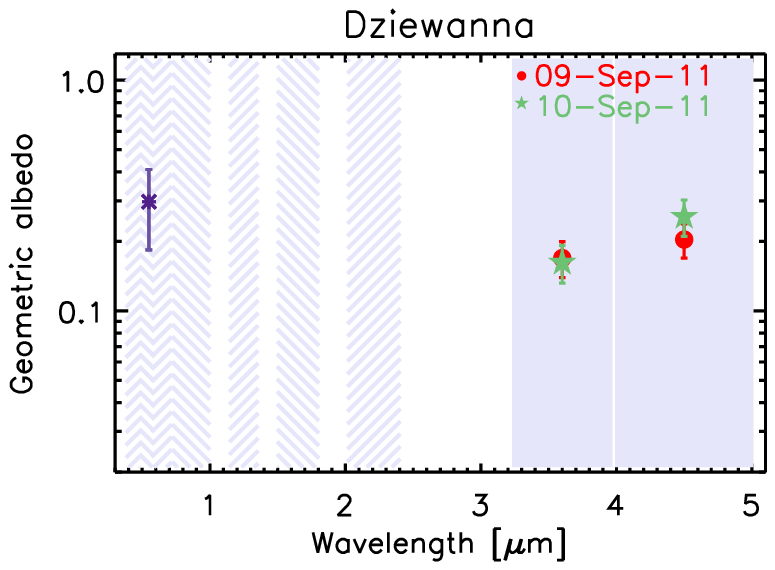}
~
        \includegraphics[width=0.5\columnwidth]{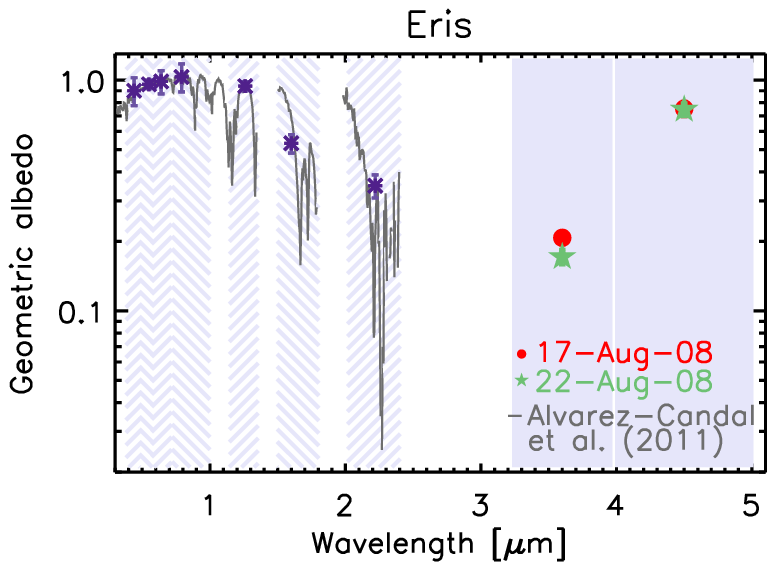}
        ~
        \includegraphics[width=0.5\columnwidth]{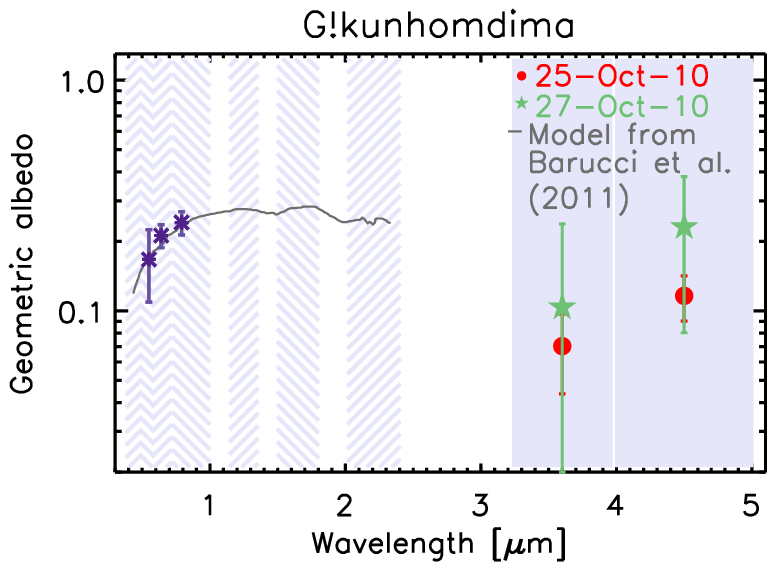}
        ~
                \includegraphics[width=0.5\columnwidth]{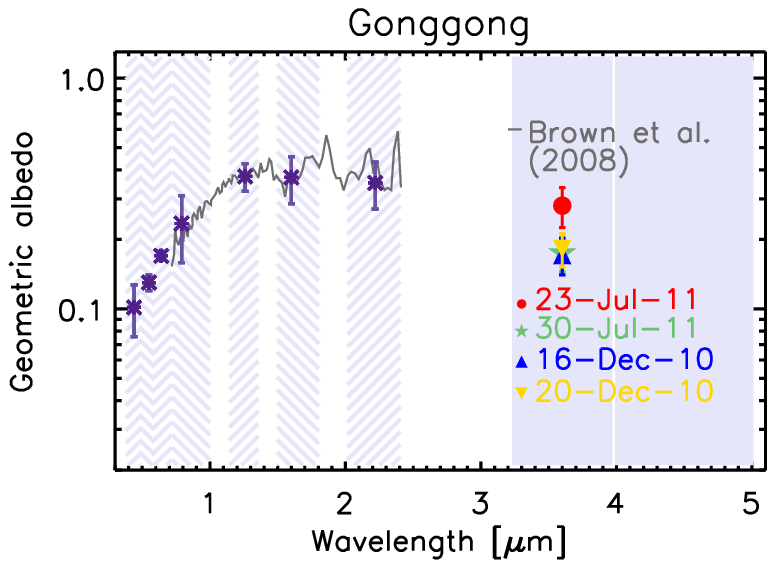}
 ~
        \includegraphics[width=0.5\columnwidth]{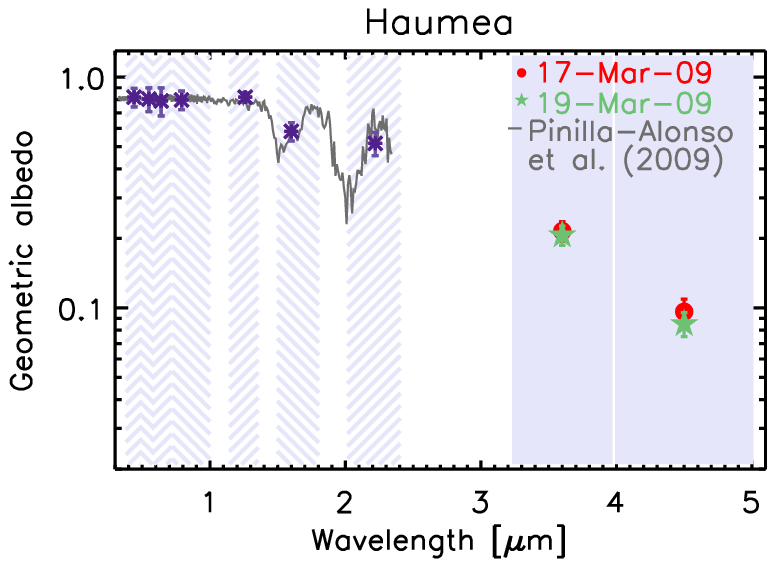}
~
          \includegraphics[width=0.5\columnwidth]{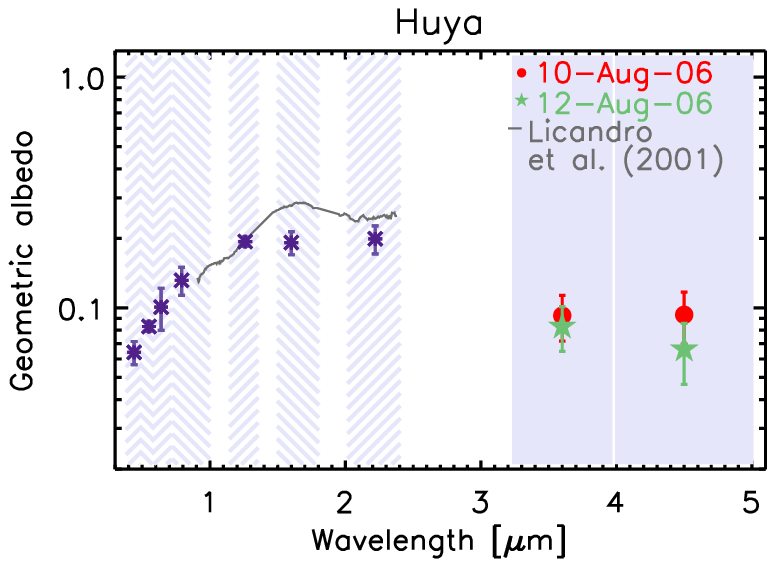}
          ~
         \includegraphics[width=0.5\columnwidth]{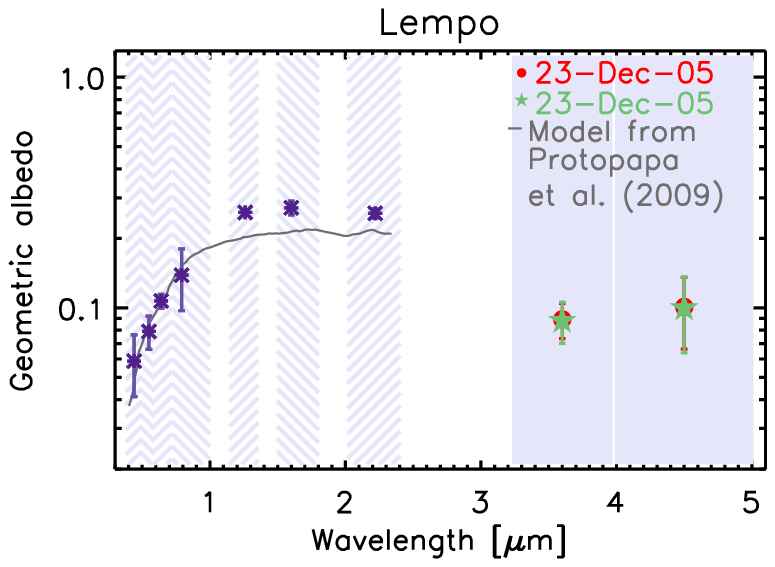}
  ~
        \includegraphics[width=0.5\columnwidth]{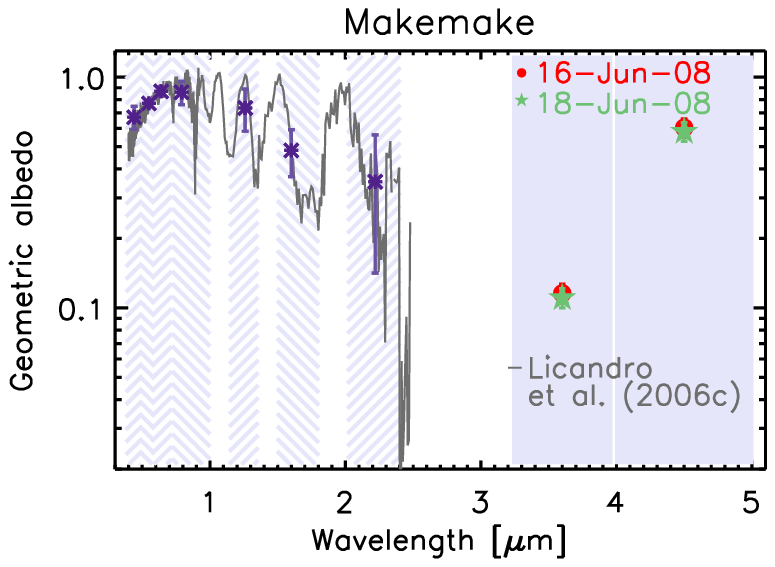}
~
        \includegraphics[width=0.5\columnwidth]{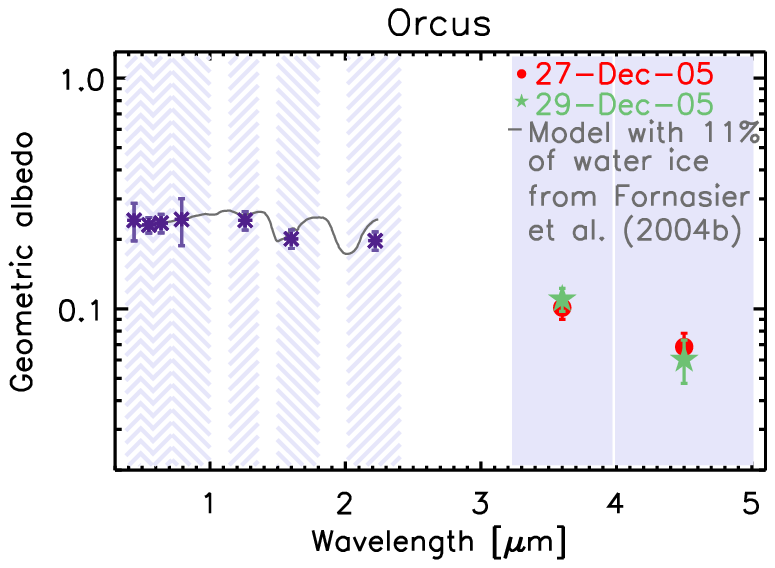}
~
          \includegraphics[width=0.5\columnwidth]{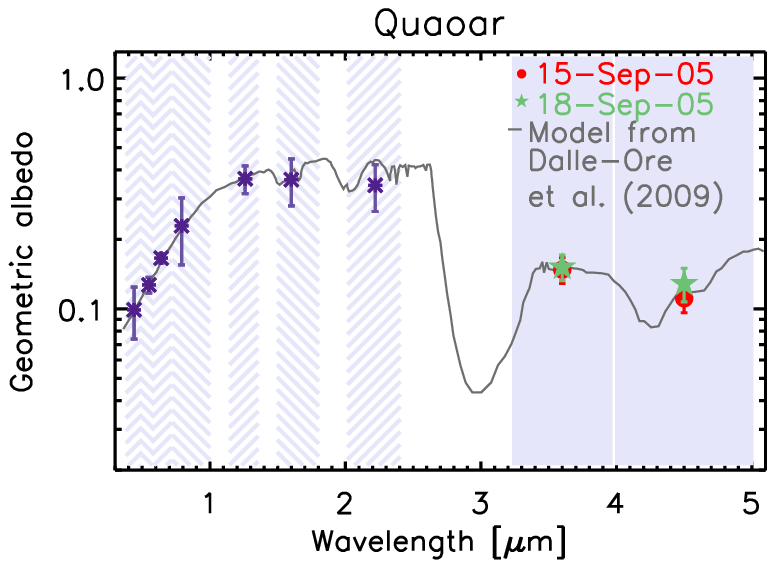}
~
        \includegraphics[width=0.5\columnwidth]{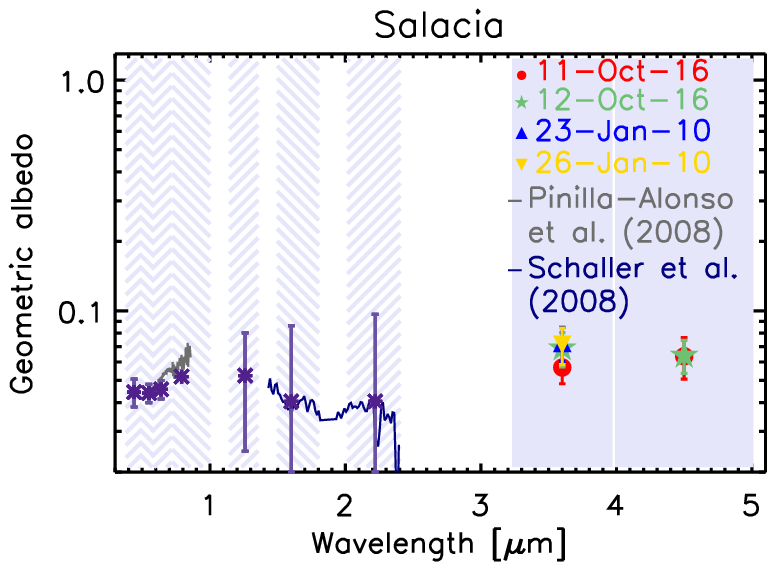}
  ~
        \includegraphics[width=0.5\columnwidth]{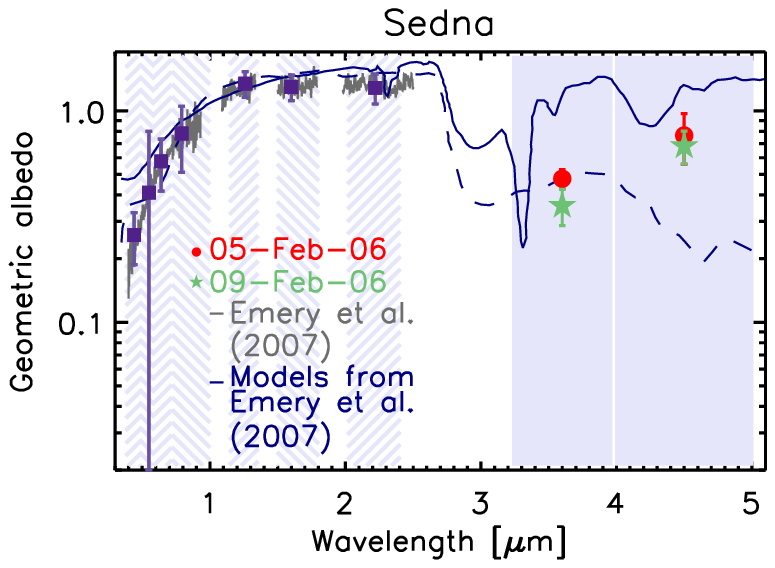}
~
          \includegraphics[width=0.5\columnwidth]{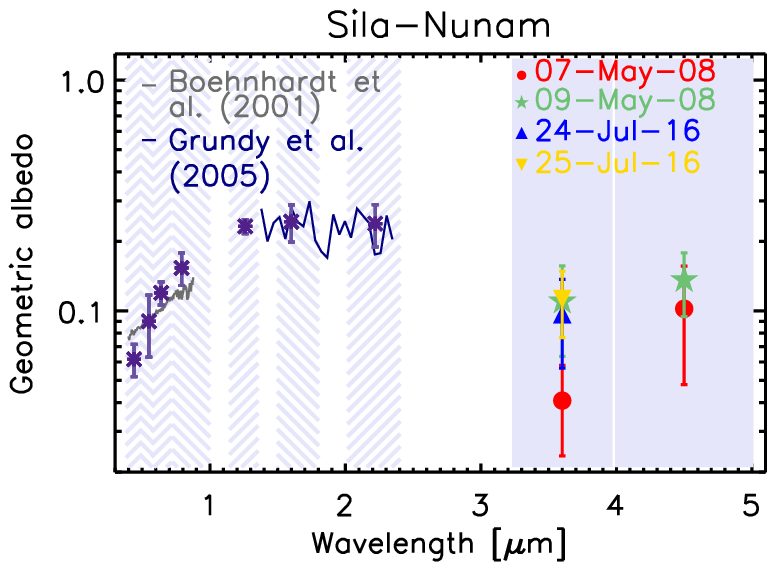}
~
      \includegraphics[width=0.5\columnwidth]{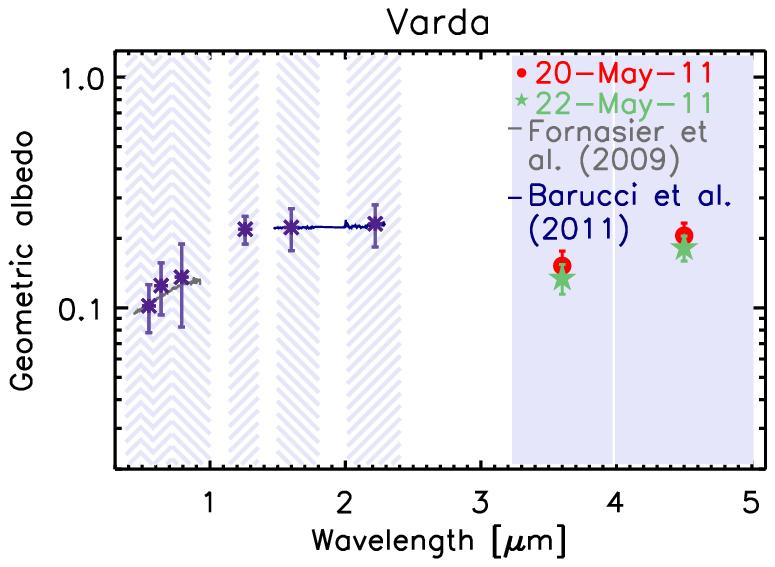}

        \includegraphics[width=0.5\columnwidth]{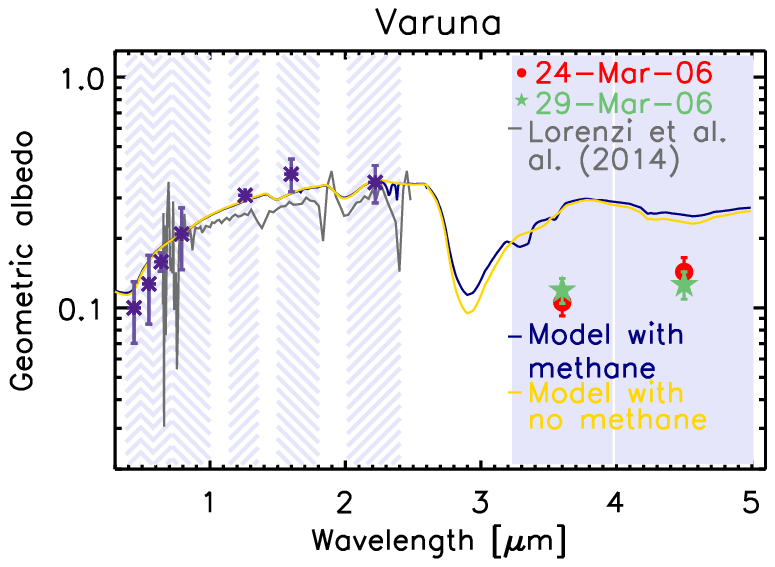}

\newpage
\section{Tables}

\startlongtable
\begin{deluxetable*}{lcccccccc}
\tablecaption{\label{tb:observations}\textbf{Summary of observations.} Abbreviations are defined as follows: geocentric distance ($\Delta$), heliocentric distance ($r_{\rm H}$), phase angle ($\alpha$), flux at 3.6 and 4.5 {\micron} ($F_{\rm 3.6\mu m}$ and $F_{\rm 4.5\mu m}$, respectively), geometric albedo at 3.6 and 4.5 {\micron} ($p_{\rm 3.6\mu m}$ and $p_{\rm 4.5\mu m}$, respectively).}

\tablehead{\colhead{Designation }	&  \colhead{Date}	& \colhead{ $\Delta$	}&  \colhead{$r_{\rm H}$}	&\colhead{$\alpha$   }	&    \colhead{ $F_{\rm 3.6\mu m}$ } & \colhead{$F_{\rm 4.5\mu m}$ }&\colhead{$p_{\rm 3.6\mu m}$  }& \colhead{$p_{\rm 4.5\mu m}$}\\  
		\colhead{	}&\colhead{(DD/MM/YY)}		&\colhead{ (au)	}	&\colhead{  (au)	}	& \colhead{($^{\circ}$)	}&	\colhead{($\mu$Jy)	}		&  \colhead{($\mu$Jy)}&\colhead{ } &	\colhead{ }  }
		
\startdata
1993 SC& 22-Aug-08 & 36.335 & 36.329 & 1.615 & $<1.4$ & $<1.7$ & $<0.2$ & $<0.3$ \\ 
1995 SM$_{55}$  & 25-Oct-10 & 38.109 & 38.572 & 1.372 & 2.6 $\pm$ 0.9 & $-$ & 0.21 $\pm$ 0.08 & $-$ \\ 
& 27-Oct-10 & 38.080 & 38.571 & 1.349 & 2.8 $\pm$ 0.6 & $-$ & 0.23 $\pm$ 0.05 & $-$ \\ 
1995 TL$_{8}$  & 03-Mar-11 & 43.670 & 44.145 & 1.155 & 2.0 $\pm$ 0.3 & $-$ & 0.22 $\pm$ 0.05 & $-$ \\ 
1996 GQ$_{21}$ & 07-Aug-07 & 40.254 & 40.656 & 1.329 & 3.8 $\pm$ 0.5 & 1.3 & 0.15 $\pm$ 0.03 & $<0.08$ \\ 
1996 TK$_{66}$ & 04-Sep-09 & 42.827 & 42.988 & 1.352 & $<1.8$ & $-$ & $<0.75$ & $-$ \\ 
& 08-Sep-09 & 42.761 & 42.988 & 1.337 & $<2.3$ & $-$ & $<1.0$ & $-$ \\ 
1996 TO$_{66}$ & 22-Dec-05 & 46.094 & 46.486 & 1.151 & $<2.0$ & $<2.3$ & $<0.46$ & $<0.79$\\
\enddata

\end{deluxetable*}

\startlongtable
\begin{deluxetable*}{lccclclll}
\tablecaption{\label{tb:physical_properties_compilation}
\textbf{Compilation of the physical properties used in this work for our sample.} Geometric albedo in $V$-band ($p_{\rm V}$), absolute magnitude in $V$-band ($H_{\rm V}$) and diameter ($D$) have been compiled from ``TNOs are cool'' data base. In the case, the object was not study by this project (marked with the symbol $\dag$), we calculated the median value of $p_{\rm V}$ from this data base, with $H_{\rm V}$ from the Minor Planet Center; and we calculated $D$ using the equation (\ref{eq:radius}). For the Haumea family members for which ``TNOs are cool'' project only provided a lower limit (marked with $^{\star}$), we used the procedure but using a median value from those family members with known albedo ($p_{\rm V, Haumea}=0.58\pm0.26$). Taxonomy was taken \cite{Fulchignoni2008}, with the exception of those marked with the symbol $\ddag$, for which was taken from \cite{Belskaya2015}. The spectral gradient was taken from \cite{Peixinho2015}. Binarity was taken from the webpage of \href{http://www2.lowell.edu/users/grundy/tnbs/status.html}{Lowell Observatory}; in this column, Haumea family members are labeled as HH. Dynamical classification (Dyn. Class.) following the work by \cite{Gladman2008}. The orbital behavior (OB) is characterized as secure (S) or insecure (I).}

\tablehead{\colhead{Designation }&  \colhead{$p_{\rm V}$   }          & \colhead{ $H_{\rm V}$ }                 & \colhead{$D$ }                    &  \colhead{Taxonomy   }                &\colhead{ $S'$  }    & \colhead{ Binarity   }       & \colhead{Dyn. Class. } & \colhead{OB}\\
         \colhead{ }                       &     \colhead{ }          &  \colhead{(mag)       }            &  \colhead{(km)   }                 &    \colhead{}                    &  \colhead{(\%/1000)  }      &  \colhead{ }   &    \colhead{   }  &\colhead{}     } 
\startdata
1993 SC$^\dag$& $0.11\pm0.09$ & $7.00\pm0.70$ & $160.0\pm50.0$  & RR & $6.84\pm4.88$ & No & Resonant &S\\ 
1995 SM$_{55}^{\star}$ &$0.58\pm0.26$ & $4.49\pm0.03$ & $220.0\pm17.0$ & BBb? & $1.27\pm2.91$ & No & Inner Classical&S\\ 
1995 TL$_{8}$ &$0.23\pm0.19$ & $5.29\pm0.06$ & $244.0\pm82.0$ & RR & $3.17\pm6.08$ & Si & Hot Classical&I\\ 
1996 GQ$_{21}$ &$0.13\pm0.04$ & $5.20\pm0.30$ & $349.0\pm49.0$ & RR & $4.74\pm3.15$ & No & Detached&I\\ 
1996 TK$_{66}^\dag$ &$0.11\pm0.09$ & $6.10\pm0.60$ & $120.0\pm30.0$ & RR & $8.62\pm4.51$ & $-$& Cold Classical&S\\ 
1996 TO$_{66}^{\star}$ &$0.58\pm0.26$ & $4.81\pm0.11$ & $190.0\pm17.0$ & BBb? & $8.12\pm8.43$ & HH & Hot Classical&S\\ 
1996 TP$_{66}$ &$0.07\pm0.06$ & $7.51\pm0.09$ & $154.0\pm33.0$ & RR & $0.36\pm4.74$ & No & Resonant&S\\ 
1998 SM$_{165}$ &$0.08\pm0.02$ & $6.02\pm0.08$ & $291.0\pm26.0$ & RR & $9.02\pm4.33$ & Si & Resonant&S\\
\enddata

\end{deluxetable*}

\startlongtable
\begin{deluxetable*}{lccccccc}
\tablecaption{\label{tb:colors_compilation}
\textbf{\textbf{Visible and near-infrared colors compilation for our sample.} Visible colors are taken from \cite{Peixinho2015} and infrared colors are taken from \cite{Fulchignoni2008}. For those objects without VNIR colors but taxonomic classification, we took the average color for the correspondent taxonomy ($^{\star}$). See table's footnote and section \ref{sec:data_sample} for exceptions.}}

\tablehead{\colhead{Designation }&  \colhead{$B-V$   }          & \colhead{ $V-R$ }                 & \colhead{$V-I$ }                    &  \colhead{$V-J$   }                &\colhead{ $V-H$  }    & \colhead{ V-K   }      \\
         \colhead{ }                       &     \colhead{(mag)  }          &  \colhead{(mag)       }            &  \colhead{(mag)    }                 &    \colhead{(mag) }                    &  \colhead{(mag)  }      &  \colhead{(mag)  }   } 
\startdata
1993 SC& $1.04\pm0.13$ & $0.70\pm0.08$ & $1.45\pm0.09$  & $2.42\pm0.07$ &$2.82\pm0.21$ & $2.78\pm0.20$ \\ 
1995 SM$_{55}$ & $0.65\pm0.03$ & $0.37\pm0.05$ & $0.70\pm0.05$  & $1.07\pm0.05$ &$0.59\pm0.06$ & $0.49\pm0.05$ \\ 
1995 TL$_{8}$ & $1.01\pm0.16$ & $0.62\pm0.09$ & $1.14\pm0.21$  & $2.42\pm0.05$ &$2.82\pm0.09$ & $2.8\pm0.09$ \\ 
1996 GQ$_{21}$ & $1.00\pm0.06$ & $0.70\pm0.05$ & $1.42\pm0.07$  & $2.39\pm0.04$ &$2.88\pm0.04$ & $3.03\pm0.08$ \\ 
1996 TK$_{66}^{\star}$ & $0.99\pm0.06$ & $0.68\pm0.07$ & $1.28\pm0.13$  & $2.26\pm0.18$ &$2.66\pm0.22$ & $2.66\pm0.25$ \\ 
1996 TO$_{66}$ & $0.68\pm0.07$ & $0.39\pm0.15$ & $0.74\pm0.07$  & $1.\pm0.1$ &$0.79\pm0.2$ & $1.6\pm0.18$ \\ 
1996 TP$_{66}$ & $1.03\pm0.11$ & $0.66\pm0.07$ & $1.29\pm0.11$  & $2.26\pm0.08$ &$2.42\pm0.12$ & $2.44\pm0.11$ \\ 
1998 SM$_{165}$ & $0.98\pm0.08$ & $0.64\pm0.07$ & $1.34\pm0.06$  & $2.36\pm0.01$ &$2.91\pm0.02$ & $2.96\pm0.07$ \\
\enddata
\tablenotetext{\dag}{\cite{Tegler2016}}
\tablenotetext{\ddag}{\cite{Peixinho2015} \& \cite{Belskaya2015}}
\tablenotetext{\S}{\cite{Fulchignoni2008}}
\tablenotetext{\P}{\cite{Perna2013}}
\tablenotetext{\clubsuit}{From its spectrum}
\end{deluxetable*}


\startlongtable
\begin{deluxetable*}{lccccccc}
\tablecaption{\label{tb:albedos_compilation}
\textbf{\textbf{Visible and near-infrared colors compilation for our sample.} Abbreviations are defined as follows: albedo at wavelengths: $B$, $R$, $I$, $J$, $H$ and $K$ ($p_{\rm B}$, $p_{\rm R}$, $p_{\rm I}$, $p_{\rm J}$, $p_{\rm H}$ and $p_{\rm K}$, respectively). Albedo in $V$-band is shown in table \ref{tb:physical_properties_compilation} together with the physical properties of each object.}}

\tablehead{\colhead{Designation }&  \colhead{$p_{\rm B}$   }          & \colhead{ $p_{\rm R}$ }                 & \colhead{$p_{\rm I}$ }                    &  \colhead{$p_{\rm J}$   }                &\colhead{ $p_{\rm H}$  }    & \colhead{$p_{\rm B}$  }}      

\startdata									
1993 SC& $0.08\pm0.02$ & $0.15\pm0.03$ & $0.22\pm0.05$  & $0.40\pm0.06$ &$0.42\pm0.20$ & $0.38\pm0.18$ \\ 
1995 SM$_{55}$ & $0.59\pm0.04$ & $0.59\pm0.07$ & $0.59\pm0.07$  & $0.60\pm0.07$ &$0.29\pm0.04$ & $0.25\pm0.03$ \\ 
1995 TL$_{8}$ & $0.17\pm0.06$ & $0.29\pm0.06$ & $0.35\pm0.17$  & $0.83\pm0.10$ &$0.89\pm0.18$ & $0.82\pm0.17$ \\ 
1996 GQ$_{21}$ & $0.09\pm0.01$ & $0.17\pm0.02$ & $0.25\pm0.04$  & $0.44\pm0.04$ &$0.52\pm0.05$ & $0.56\pm0.10$ \\ 
1996 TK$_{66}$ & $0.08\pm0.01$ & $0.15\pm0.02$ & $0.19\pm0.06$  & $0.34\pm0.14$ &$0.36\pm0.18$ & $0.34\pm0.20$ \\ 
1996 TO$_{66}$ & $0.57\pm0.09$ & $0.60\pm0.21$ & $0.61\pm0.10$  & $0.56\pm0.13$ &$0.34\pm0.16$ & $0.68\pm0.28$ \\ 
1996 TP$_{66}$ & $0.05\pm0.01$ & $0.10\pm0.02$ & $0.13\pm0.03$  & $0.23\pm0.04$ &$0.20\pm0.05$ & $0.19\pm0.05$ \\ 
1998 SM$_{165}$ & $0.06\pm0.01$ & $0.11\pm0.02$ & $0.15\pm0.02$  & $0.28\pm0.01$ &$0.35\pm0.02$ & $0.34\pm0.06$ \\
\enddata

\end{deluxetable*}


\startlongtable
\begin{deluxetable*}{lcccc}
\tablecaption{\label{tb:colors_Spitzer_compilation}
\textbf{Colors used in this work for the different diagrams.} }

\tablehead{\colhead{Designation }&  \colhead{$V-3.6\,\mu$m   }          & \colhead{$J-3.6\,\mu$m }                 & \colhead{$K-3.6\,\mu$m }                    &  \colhead{$3.6\,\mu{\rm m}-4.5\,\mu{\rm m}$ } \\
 \colhead{ } &  \colhead{(mag)   }          & \colhead{(mag) }                 & \colhead{(mag) }                    &  \colhead{(mag) }    }      

\startdata		
1993 SC & $<0.54$ & $<-0.85$ & $<-0.82$  & $<0.64$  \\ 
1995 SM$_{55}$ & $-1.05\pm0.42$ & $-1.09\pm0.26$ & $-0.12\pm0.26$  & $-$  \\ 
1995 TL$_{8}$ & $-0.04\pm0.92$ & $-1.43\pm0.25$ & $-1.42\pm0.32$  & $-$  \\ 
1996 GQ$_{21}$ & $0.19\pm0.41$ & $-1.17\pm0.22$ & $-1.42\pm0.28$  & $<-0.71$  \\ 
1996 TK$_{66}$ & $<2.23$ & $<1.00$ & $<0.99$  & $-$  \\ 
1996 TO$_{66}$ & $<-0.61$ & $<-0.58$ & $<-0.79$  & $<0.93$  \\ 
1996 TP$_{66}$ & $<0.36$ & $<-0.87$ & $<-0.66$  & $<0.21$  \\ 
1998 SM$_{165}$ & $0.13\pm0.25$ & $-1.20\pm0.19$ & $-1.41\pm0.22$  & $-$  \\
\enddata

\end{deluxetable*}


\end{document}